\documentclass[11pt,a4paper]{article}
\usepackage{bbm}
\pdfoutput=1
\usepackage{jheppub}
\usepackage{amsmath}
\usepackage{epsfig}
\usepackage{amssymb}
\usepackage{graphics}
\usepackage[active]{srcltx}
\usepackage{epstopdf}
\usepackage{subfigure}
\usepackage{cancel}

\newcommand{\be}{\begin{equation}}
\newcommand{\ee}{\end{equation}}
\newcommand{\bea}{\begin{eqnarray}}
\newcommand{\eea}{\end{eqnarray}}
\newcommand{\bean}{\begin{eqnarray*}}
\newcommand{\eean}{\end{eqnarray*}}
\newcommand{\nn}{\nonumber \\}

\def\W #1{\widetilde{#1}}

\def\eref#1{(\ref{#1})}

\def\Label#1{\label{#1}%
  \smash{\hbox to0pt{\raise1ex\hbox{\tiny[#1]}\hss}}}

\title{Factorizations for tree amplitudes in the double-cover framework: from gravity to other theories}
\author[a]{Kang Zhou}
\affiliation[a]{Center for Gravitation and Cosmology, College of Physical Science and Technology, Yangzhou University,\\
 No.180, Siwangting Road, Yangzhou, 225009, P.R. China.}
\emailAdd{zhoukang@yzu.edu.cn}
\date{\today}
\abstract{In this paper, we demonstrate that the factorizations for tree amplitudes in the double-cover framework, for various theories, can be generated from the gravity amplitude in the double-cover prescription. Using our method, the factorized formulae for amplitudes of Yang-Mills theory, special Yang-Mills-scalar theory, and bi-adjoint scalar theory, have been derived. The differential operators indicate some non-trivial relationships among factorizations for above theories, as can be understood through four angles which are factorization channels, pole-structures, gauge choices, as well as BCFW recursions.
}


\begin{document}

\maketitle \flushbottom

%%%%%%%%%%%%%%%%%%%%%%%%%%%%%%%%%%%%%%
\section{Introduction}
\label{secintro}
%%%%%%%%%%%%%%%%%

The well-known Cachazo-He-Yuan (CHY) formalism is an elegant new representation of tree-level amplitudes for massless particles in arbitrary space-time dimensions \cite{Cachazo:2013gna,Cachazo:2013hca,Cachazo:2013iea,Cachazo:2014nsa,Cachazo:2014xea}. It provides both intriguing theoretical understanding and novel computational tool for S-matrix elements for a wide range of theories. However, in the original CHY construction, the factorizations, which reflect the unitarity and locality of S-matrix, are deeply hidden. Recently, the factorizations for tree amplitudes in the CHY framework, which are different from the factorizations arise from the traditional Feynman diagram approach, have been realized by the so called double-cover prescription developed by Gomez \cite{Gomez:2016bmv,Cardona:2016bpi,Bjerrum-Bohr:2018lpz,Gomez:2018cqg,Bjerrum-Bohr:2018jqe,Gomez:2019cik}. Using the double-cover prescription, the factorized formulae for amplitudes of Yang-Mills theory (YM), special Yang-Mills-scalar theory (sYMS), as well as non-linear sigma model (NLSM), were obtained in \cite{Bjerrum-Bohr:2018lpz,Gomez:2018cqg,Bjerrum-Bohr:2018jqe,Gomez:2019cik}.

The factorized formula expresses an tree amplitude as
\bea
{\pmb A}=\sum_{\rm channels}\,{\cal A}_L{1\over P^2}{\cal A}_R\,,~~~~\label{fact-form}
\eea
where ${\pmb A}$ is the full on-shell amplitude, ${\cal A}_L$ and ${\cal A}_R$ are two off-shell sub-amplitudes which contain off-shell external legs from the propagator ${1\over P^2}$. Throughout this paper, we use ${\pmb A}$ and ${\cal A}$ to denote on-shell and off-shell amplitudes, respectively.
An interesting observation is the similarity between new factorized formulae for YM and NLSM amplitudes in the double-cover framework \cite{Bjerrum-Bohr:2018lpz,Bjerrum-Bohr:2018jqe}, which are given as
\bea
{\pmb A}^{\epsilon}_{\rm YM}(1,\cdots,n)&=&\sum_{\epsilon^M}\,{\cal A}^{\epsilon}_{\rm YM}({4},\cdots,n,1,\underline{\hat{P}}^{\epsilon^{M}}_{23}){1\over P^2_{23}}{\cal A}^{\epsilon}_{\rm YM}(\underline{2},\underline{\hat{3}},\underline{\hat{P}}^{\epsilon^{M}}_{4:1})\nn
& &+\sum_{i=4}^n\sum_{\epsilon^M}\,{\cal A}^{\epsilon}_{\rm YM}(i+1,\cdots,\underline{\hat{1}},\underline{2},\underline{\hat{P}}^{\epsilon^{M}}_{3:i}){1\over P^2_{i+1:2}}{\cal A}^{\epsilon}_{\rm YM}(\underline{\hat{3}},\underline{4},\cdots,i,\underline{\hat{P}}^{\epsilon^{M}}_{i+1:2})\nn
& &-2\sum_{i=4}^n\sum_{\epsilon^L}\,{\cal A}^{\epsilon}_{\rm YM}(i+1,\cdots,\underline{\hat{1}},\underline{3},\underline{\hat{P}}^{\epsilon^{L}}_{2(4:i)}){1\over P^2_{(i+1:1)3}}{\cal A}^{\epsilon}_{\rm YM}(\underline{\hat{2}},\underline{4},\cdots,i,\underline{\hat{P}}^{\epsilon^{L}}_{(i+1:1)3})\,,~~~~\label{fact-form-YM}
\eea
and
\bea
{\pmb A}_{\rm NLSM}(1,\cdots,n)&=&{\cal A}_{\rm NLSM}(\underline{\bar{4}},\cdots,n,\underline{\bar{\hat{1}}},\underline{\hat{P}}_{23}){1\over P_{23}^2}{\cal A}_{\rm NLSM}(\underline{\bar{2}},\underline{\bar{\hat{3}}},\underline{\hat{P}}_{4:1})\nn
& &+\sum_{i=1}^n\,{\cal A}_{\rm NLSM}(i+1\cdots\underline{\bar{\hat{1}}},\underline{\bar{2}},\underline{\hat{P}}_{3:i}){1\over P_{i+1:2}^2}{\cal A}_{\rm NLSM}(\underline{\bar{\hat{3}}},\underline{\bar{4}},\cdots,i,\underline{\hat{P}}_{i+1:2})\,,~~~~\label{fact-form-NLSM}
\eea
respectively. The meanings of notations will be explained in next sections. Comparing the first and second lines of \eref{fact-form-YM} and \eref{fact-form-NLSM}, one can observe that the factorization channels in these lines are totally the same for two theories, and the sub-amplitudes ${\cal A}_L$
and ${\cal A}_R$ for two theories can be related by simply replacing gluons by scalar particles. This similarity implies that there is an underlying relationship links two factorized formulae together.

To understand such relationship, a natural tool is the set of differential operators proposed by Cheung, Shen and Wen \cite{Cheung:2017ems}, which unifies tree amplitudes of various theories. In the unified web indicated by differential operators, the tree gravitational (GR) amplitude \footnote{In this paper, the gravity theory is understood in a generalized version, i.e., Einstein gravity theory couples to a dilaton and two-forms.} can be transmuted to tree amplitudes of other theories via proper operators which act on kinematic variables. Since the web includes not only YM and NLSM,
but also a variety of other theories, one can expect that the relationship between factorized formulae exist among a wider range of theories.

Motivated by the above idea, in this paper we demonstrate that the factorizations for various theories in the double-cover framework can be generated from the GR amplitude in the double-cover representation, via proper differential operators. More explicitly, suppose the amplitude of theory-$a$ can be factorized as in \eref{fact-form}, and this amplitude can be transmuted to the amplitude ${\pmb A}'$ of theory-$b$ through the differential operator ${\cal O}$ as ${\pmb A}'={\cal O}{\pmb A}$. We factorize the operator ${\cal O}$ as ${\cal O}\cong{\cal O}_L\cdot{\cal O}_R$. Here $\cong$ means operators on two sides are not equal at the algebraic level, but are equivalent to each other when applying to physical amplitudes. The operator ${\cal O}_L$ transmutes ${\cal A}_L$ to ${\cal A}'_L$ and annihilates ${\cal A}_R$, while ${\cal O}_R$ transmutes ${\cal A}_R$ to ${\cal A}'_R$ and annihilates ${\cal A}_L$, where ${\cal A}'_L$ and ${\cal A}'_R$ are off-shell sub-amplitudes for theory-$b$. Then we arrive at
\bea
{\pmb A}'=\sum_{\rm channels'}\,{\cal A}'_L{1\over P^2}{\cal A}'_R\,,~~~~\label{fact-form-2}
\eea
which is the factorized formula for theory-$b$.
Using this method, we can reproduce the factorization for the YM amplitude by applying the differential operator to the GR amplitude. By applying differential operators to the factorized YM amplitude, we also derive the factorizations for sYMS, NLSM, as well as bi-adjoint scalar (BAS) amplitudes. The obtained factorizations for sYMS and NLSM amplitudes coincide with the results obtained in the literature, while the result for the BAS amplitude will be verified through the standard double-cover approach. Although our consideration do not include all theories in the unified web, the effect of all three types of differential operators, ${\cal T}[\alpha_1,\cdots,\alpha_n]$, ${\cal T}_{{\cal X}_{2m}}$ and ${\cal T}[a,b]\cdot{\cal L}$ in \cite{Cheung:2017ems} (the definitions of them will be given in the next section), are discussed in the current work.

The relationships among factorized formulae for different theories can be understood by our method. Firstly, the factorization channels for ${\pmb A}'$ are selected from channels for ${\pmb A}$ by the operator ${\cal O}$. Secondly, the definitions of ${\cal T}[\alpha_1,\cdots,\alpha_n]$ and ${\cal T}_{{\cal X}_{2m}}$ indicate that these two operators will not create or annihilate any pole, thus they transmute physical poles to physical poles, and transmute spurious poles to spurious poles. On the other hand, the definition of ${\cal T}[a,b]\cdot{\cal L}$ indicates the possibility of canceling physical poles, therefore this operator can transmute physical poles to spurious poles. Consequently, the pole-structure of the factorization for ${\pmb A}'$
arise from the pole-structure of ${\pmb A}$ via differential operators. Thirdly, ${\cal A}'_L$ and ${\cal A}'_R$ are off-shell amplitudes in the CHY formula, which depend on the gauge choices. Our method shows that the gauge choices for ${\cal A}'_L$ and ${\cal A}'_R$ are inherited from the gauge choices for ${\cal A}_L$ and ${\cal A}_R$. Finally, terms in the factorized formula can be related to terms in Britto-Cachazo-Feng-Witten (BCFW) recursion relation \cite{Britto:2004ap,Britto:2005fq,Feng:2011np}. Such relation for ${\pmb A}'$ can also be understood from the corresponding relation for ${\pmb A}$ through differential operators.

The factorized formula for an amplitude depend on the gauge choice in the double-cover prescription. The proper gauge choice, which leads to the factorized formulae appear in the relations mentioned above, will also be discussed.

This paper is organized as follows. In section \ref{secreview}, we give a brief review about the necessary background including the CHY construction, the double-cover prescription, and the differential operators which link amplitudes of different theories together. In section \ref{secGRYM}, we illustrate how to get the factorization for the YM amplitude from the GR amplitude in the double-cover representation, and discuss the relations between two theories indicated by differential operators. In section \ref{secYMother}, we consider the factorizations for sYMS, NLSM and BAS amplitudes. In section \ref{secconclu}, we end with a brief summary and discussion. Some details of computation in section \ref{secGRYM} are given in Appendix \ref{phypole}.

%%%%%%%%%%%%%%%%%%%%%%%%%%%%%%%%%%%%%%
\section{Background}
\label{secreview}
%%%%%%%%%%%%%%%%%

For reader's convenience, in this section we rapidly review the CHY construction, double-cover prescription, as well as the differential operators.

%%%%%%%%%%%%%%%%%%%%%%%%%%%%
\subsection{CHY construction}
\label{RV-CHY}
%%%%%%%%%%%%%%%%%%%%%%%%%%%%

In the CHY construction, tree level amplitudes for $n$ massless particles arise from a multi-dimensional contour integral over
the moduli space of genus zero Riemann surfaces with $n$ punctures, ${\cal M}_{0,n}$ \cite{Cachazo:2013gna,Cachazo:2013hca,Cachazo:2013iea,Cachazo:2014nsa,Cachazo:2014xea}. It can be expressed as
\bea
{\pmb A}_n=\int d\mu_n\,{\cal I}_L(k,\epsilon,z){\cal I}_R(k,\W\epsilon,z)\,,~~~~\label{CHY}
\eea
which possesses the M\"obius ${\rm SL}(2,\mathbb{C})$ invariance. The measure is defined as
\bea
d\mu_n\equiv{d^n z\over{\rm vol}\,{\rm SL}(2,\mathbb{C})}{|pqr|\over \prod_{i=1,i\neq pqr}^n\,{\cal E}_i(z)}\,.
\eea
Here the factor $|pqr|$ is given by $|pqr|\equiv z_{pq}z_{qr}z_{rp}$, where $z_{ij}\equiv z_i-z_j$. The scattering equations are given as \footnote{In this paper, we choose $2k_i\cdot k_j$ rather than $s_{ij}$ to define the scattering equations. Two choices are un-equivalent for off-shell amplitudes. The factor $2$ is kept for reproducing the propagator with correct factor in the factorized formulae.}
\bea
{\cal E}_i(z)\equiv\sum_{j\in\{1,2,\ldots,n\}\setminus\{i\}}{2k_i\cdot k_j\over z_{ij}}=0\,.~~~~\label{SCeqS}
\eea
The $(n-3)$ independent scattering equations provide poles which define the contour of integral.
After fixing the ${\rm SL}(2,\mathbb{C})$ symmetry by the Faddeev-Popov method, the measure part becomes
\bea
d\mu_n\equiv{\big(\prod^n_{j=1,j\neq a,b,c}\,dz_j\big)|pqr||abc|\over \prod_{i=1,i\neq p,q,r}^n\,{\cal E}_i(z)}\,.
\eea

The integrand in \eref{CHY} depends on the theory under consideration. For any theory known to have a CHY expression, the corresponding integrand can be separated into two
parts ${\cal I}_L(k,\epsilon,z)$ and ${\cal I}_R(k,\W\epsilon,z)$. The function ${\cal I}_L(k,\epsilon,z)$ depends on $\{k_i\}$, $\{\epsilon_i\}$ and $\{z_i\}$, where $k_i$, $\epsilon_i$ and $z_i$ are the momentum, polarization vector, and puncture location for $i^{\rm th}$
particle, respectively. Correspondingly, ${\cal I}_R(k,\W\epsilon,z)$ depend on $\{k_i\}$, $\{\W\epsilon_i\}$ and $\{z_i\}$, where $\{\W\epsilon_i\}$ is another independent set of polarization vectors. Either of ${\cal I}_L(k,\epsilon,z)$ and ${\cal I}_R(k,\W\epsilon,z)$ are weight-$2$ for each variable $z_i$
under the M\"obius transformation. In Table \ref{tab:theories}, we list integrands for
theories which will be encountered in this paper \footnote{For theories contain gauge or flavor groups, we only show
the integrands for color-ordered partial amplitudes instead of full ones.}.
\begin{table}[!h]
    \begin{center}
        \begin{tabular}{c|c|c}
            Theory& ${\cal I}_L(k,\epsilon,z)$ & ${\cal I}_R(k,\W\epsilon,z)$ \\
            \hline
            GR & ${\bf Pf}'\Psi_n$ & ${\bf Pf}'\W\Psi_n$ \\
            YM & ${PT}_n(\pi)$ & ${\bf Pf}' \Psi_n$ \\
            sYMS & ${PT}_n(\pi)$ & ${\bf Pf}'\Psi_{n-2m,2m;n-2m}{\bf Pf}{\cal X}_{2m}$ \\
            BAS & ${PT}_n(\pi)$ & ${PT}_n(\pi')$ \\
            NLSM & ${PT}_n(\pi)$ & ${\bf det}' A_n$  \\
        \end{tabular}
    \end{center}
    \caption{\label{tab:theories}Integrands for various theories}
\end{table}

We now explain each ingredient appearing in Table \ref{tab:theories} in turn. The Park-Taylor factor ${PT}_n(\pi)$ is defined by
\bea
{PT}_n(\pi)\equiv {1\over z_{\pi_1\pi_2}}{1\over z_{\pi_2\pi_3}}\cdots{1\over z_{\pi_{n-1}\pi_n}}{1\over z_{\pi_n\pi_1}}\,,
\eea
where $\pi$ stands for the permutation of $n$ elements in $\{1,2,\cdots,n\}$.
The $2m\times 2m$ matrices $X_{2m}$ and ${\cal X}_{2m}$ are
\bea
X_{ij}\equiv\begin{cases} \displaystyle \frac{1}{z_{ij}} & i\neq j\,,\\
\displaystyle ~ ~ 0 & i=j\,,\end{cases} \qquad\qquad\qquad\qquad
{\cal X}_{ij}\equiv\begin{cases} \displaystyle \frac{\delta^{I_i,I_j}}{z_{ij}} & i\neq j\,,\\
\displaystyle ~ ~ 0 & i=j\,.\end{cases}
\eea
where $\delta^{I_i,I_j}$ forbids the interaction between particles with different flavors.
The $2n\times2n$ antisymmetric matrix $\Psi_n$ is given by
\bea\label{Psi}
\Psi_n \equiv \left(
         \begin{array}{c|c}
           ~~A_n~~ &  ~~C_n~~ \\
           \hline
           -C^{\rm T}_n & B_n \\
         \end{array}
       \right)\,,
\eea
where the $n\times n$ blocks $A_n$, $B_n$ and $C_n$ are defined through
\bea
& &A_{ij} \equiv \begin{cases} \displaystyle {k_{i}\cdot k_j\over z_{ij}} & i\neq j\,,\\
\displaystyle  ~~~ 0 & i=j\,,\end{cases} \qquad\qquad\qquad\qquad B_{ij} \equiv \begin{cases} \displaystyle {\epsilon_i\cdot\epsilon_j\over z_{ij}} & i\neq j\,,\\
\displaystyle ~~~ 0 & i=j\,,\end{cases} \nn
& &C_{ij} \equiv \begin{cases} \displaystyle {k_i \cdot \epsilon_j\over z_{ij}} &\quad i\neq j\,,\\
\displaystyle -\sum_{l=1,\,l\neq j}^n\hspace{-.5em}{k_l \cdot \epsilon_j\over z_{lj}} &\quad i=j\,.\end{cases}
\label{ABCmatrix}
\eea
The reduced Pfaffian of $\Psi_n$ is defined as ${\bf Pf}'\Psi_n\equiv{(-)^{a+b}\over z_{ab}}{\bf Pf}(\Psi_n)^{ab}_{ab}$,
where the notation $(\Psi_n)^{ab}_{ab}$ means the rows and columns $a$, $b$ in the matrix $\Psi_n$
have been deleted (with $1\leq a,b\leq n$). When all external particles are on-shell, it can be proved that the reduced Pfaffian defined in this way is independent of the choice of $a$ and $b$.
Analogous notation holds for ${\bf Pf}'A_n$.

The definition of $\Psi_n$ can be generalized to the $(2a+b)\times(2a+b)$
case $\Psi_{a,b:a}$  as
\bea
\Psi_{a,b:a}\equiv\left(
         \begin{array}{c|c}
           ~~A_{(a+b)\times (a+b)}~~ &  C_{(a+b)\times a} \\
           \hline
            -C^{\rm T}_{a\times (a+b)} & B_{a\times a} \\
         \end{array}
       \right)\,,~~~~\label{psi-aba}
\eea
where $A$ is a $(a+b)\times (a+b)$ matrix, $C$ is a $(a+b)\times a$ matrix, and $B$ is a $a\times a$ matrix.
The definitions of elements in $A$, $B$ and $C$ are the same as in \eref{ABCmatrix}. The reduced Pfaffian ${\bf Pf}'\Psi_{a,b:a}$
is defined in the same manner as ${\bf Pf}'\Psi_n$ and ${\bf Pf}'A_n$.

Originally, ${\cal I}_R$ for NLSM is given as $({\bf Pf}' A_n)^2$ in \cite{Cachazo:2014xea}. In this paper, the off-shell NLSM integrand with the odd number of external legs, which vanishes for the on-shell case, will be used. Since ${\bf Pf}' A_n$ can not be defined in this situation, we adopt the generalization of ${\cal I}_R$ for NLSM in \cite{Bjerrum-Bohr:2018jqe,Gomez:2019cik}. We first use
\bea
({\bf Pf} M)^2= {\bf det} M
\eea
to rewrite the NLSM integrand as
\bea
({\bf Pf}' A_n)^2\equiv{1\over z_{ab}^2}{\bf det}(A_n)^{ab}_{ab}\,.
\eea
However, ${\bf det}(A_n)^{ab}_{ab}$ also vanishes when the number of external legs is odd. A natural generalization is
\bea
{\bf det}' A_n\equiv{(-)^{a+c}\over z_{ab}z_{bc}}{\bf det}(A_n)^{ab}_{bc}\,,~~~~\label{IRNLSM}
\eea
where the matrix $(A_n)^{ab}_{bc}$ is obtained from $A_n$ by removing $a^{\rm th}$ and $b^{\rm th}$ rows, $b^{\rm th}$ and $c^{\rm th}$ columns. When the number of external legs is even, one can verify that ${(-)^{a+c}\over z_{ab}z_{bc}}{\bf det}(A_n)^{ab}_{bc}$ equals to ${1\over z^2_{ab}}{\bf det}(A_n)^{ab}_{ab}$. When the number of external legs is odd, ${1\over z^2_{ab}}{\bf det}(A_n)^{ab}_{ab}$ vanishes automatically, while ${(-)^{a+c}\over z_{ab}z_{bc}}{\bf det}(A_n)^{ab}_{bc}$ vanishes only when $k_i^2=0$ for all external momenta. The formula \eref{IRNLSM} is the definition of ${\cal I}_R$ for NLSM integrand in this paper.

For latter convenience, we will call the formulae introduced in this subsection the single-cover formulae.

%%%%%%%%%%%%%%%%%%%%%%%%%%%%
\subsection{Double-cover prescription}
\label{subsecDC}
%%%%%%%%%%%%%%%%%%%%%%%%%%%%

The double-cover prescription of CHY construction is given as a contour integral on $n$-punctured double-covered Rieman spheres \cite{Gomez:2016bmv,Cardona:2016bpi,Bjerrum-Bohr:2018lpz,Gomez:2018cqg,Bjerrum-Bohr:2018jqe,Gomez:2019cik}.
Restricted to the curves
$0=C_i\equiv y^2_i-\sigma^2_i+\Lambda^2$ for $i\in\{1,2,\cdots,n\}$, the pairs $(y_1,\sigma_1),\,(y_2,\sigma_2),\cdots,(y_n,\sigma_n)$ provide new
set of variables. Then, all ${1\over z_{ij}}$ in the single-covered version are replaced by
\bea
\tau_{ij}\equiv {(y\sigma)_i\over y_i}T_{ij}\,,
\eea
with
\bea
(y\sigma)_i\equiv y_i+\sigma_i\,,~~~~T_{ij}\equiv{1\over (y\sigma)_i-(y\sigma)_j}\,.
\eea
Especially, the scattering equations are turned to
\bea
0={\cal E}^\tau_i\equiv\sum_{j\in\{1,2,\cdots,n\}\setminus\{i\}}\,(2k_i\cdot k_j)\tau_{ij}\,.~~~~\label{SCeqDC}
\eea
Amplitudes in such framework are expressed as the contour integral
\bea
{\pmb A}_n=\int d\mu^\Lambda_n {{\cal I}^\tau_L(\sigma,y,k,\epsilon){\cal I}^\tau_R(\sigma,y,k,\W\epsilon)\over {\cal E}^\tau_m}\,,~~\label{DCamp}
\eea
where the measure $d\mu^\Lambda_n$ is defined through
\bea
d\mu^\Lambda_n\equiv {1\over {\rm vol}\,{\rm GL}(2,\mathbb{C})}{d\Lambda\over \Lambda}\Big(\prod_{i=1}^n\,{y_idy_id\sigma_i\over C_i}\Big){\Delta_{pqr}\over \prod_{j\neq p,q,r,m}{\cal E}^\tau_j}\,,
\eea
with $\Delta_{pqr}\equiv(\tau_{pq}\tau_{qr}\tau_{rp})^{-1}$.
Correspondingly, the contour is determined by poles $\Lambda=0$, $C_i=0$, as well as ${\cal E}^\tau_j=0$ for $j\neq p,q,r,m$.
To eliminate the gauge redundancy of ${\rm GL}(2,\mathbb{C})$ symmetry, one can use the Faddeev-Popov method to fix four
coordinates $\sigma_p$, $\sigma_q$, $\sigma_r$ and $\sigma_m$. This procedure yields the determinant
\bea
\Delta_{pqr|m}\equiv\sigma_p\Delta_{qrm}-\sigma_q\Delta_{rmp}+\sigma_r\Delta_{mpq}-\sigma_m\Delta_{pqr}\,,
\eea
and turns the measure to be
\bea
d\mu^\Lambda_n\equiv {1\over 2^2}{d\Lambda\over \Lambda}\Big(\prod_{i=1}^n\,{y_idy_i\over C_i}\Big)
\Big(\prod_{j\neq p,q,r,m}\,{d\sigma_j\over {\cal E}^\tau_j}\Big)\Delta_{pqr}\Delta_{pqr|m}\,.
\eea

Then we take a glance on ${\cal I}^\tau_L(\sigma,y,k,\epsilon)$ and ${\cal I}^\tau_R(\sigma,y,k,\W\epsilon)$, which are obtained from ${\cal I}_L(z,k,\epsilon)$ and ${\cal I}_R(z,k,\W\epsilon)$ in Table \ref{tab:theories} via the replacement ${1\over z_{ij}}\to\tau_{ij}$. Under the replacement, the Parke-Taylor factor becomes
\bea
PT^\tau_n(\sigma)\equiv\Big(\prod_{i=1}^n{(y\sigma)_i\over y_i}\Big)T_{\sigma_1\sigma_2}T_{\sigma_2\sigma_3}\cdots T_{\sigma_n\sigma_1}\,.
\eea
When the integrand contains Pfaffians, one can first replace all ${1\over z_{ij}}$ by $T_{ij}$, then times the obtained formula by the factor $\Big(\prod_{i=1}^n\,{(y\sigma)_i\over y_i}\Big)$.
Let us take ${\bf Pf}'\Psi_n$ as the example, the reduced Pfaffian with new variables $y_i$ and $\sigma_i$ is given by
\bea
({\bf Pf}'\Psi_n)^\tau\equiv\Big(\prod_{i=1}^n\,{(y\sigma)_i\over y_i}\Big)(-)^{a+b}T_{ab}{\bf Pf}(\Psi^\Lambda_n)^{ab}_{ab}\,,
\eea
where the matrix $\Psi^\Lambda_n$ is obtained from $\Psi_n$ via the replacement ${1\over z_{ij}}\to T_{ij}$.

\begin{figure}
  \centering
  \includegraphics[width=7cm]{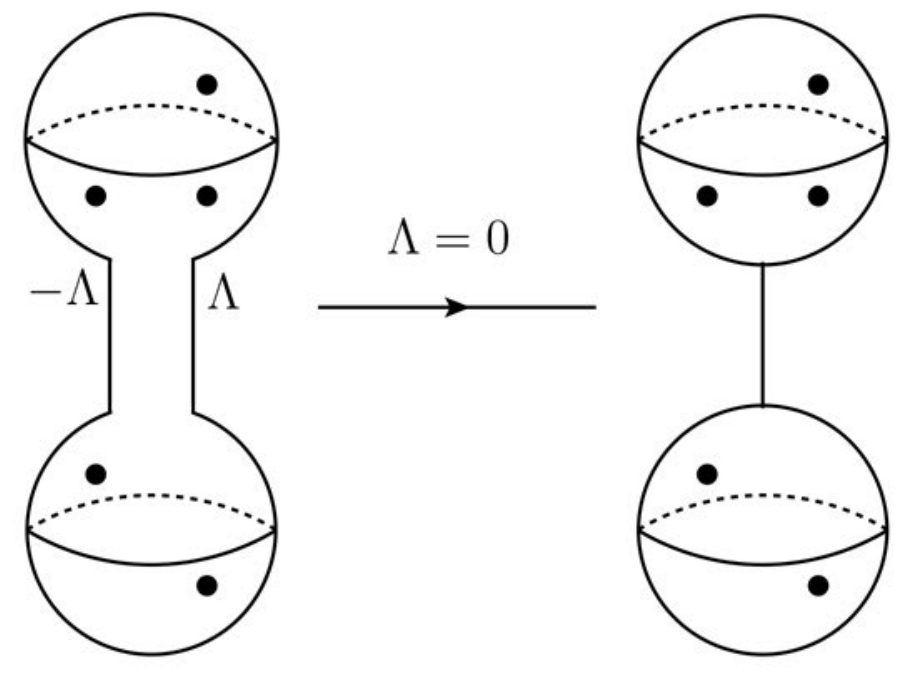} \\
  \caption{Factorization process}\label{factor}
\end{figure}

In the double-cover framework, the factorization for an amplitude is obtained naturally by integrating over variables $y_i$ and $\Lambda$.
After doing the integral over all $y_i$ encircle poles from solutions
\bea
C_i=0\Rightarrow y_i=\pm\sqrt{\sigma_i^2-\Lambda^2}\,,~~\forall\,i\,,
\eea
the obtained formula is the summation over $2^n$ possible configurations. In each configuration, coordinates $(y_i,\sigma_i)$ are separated to the so called upper and lower sheets, namely $(y_i=+\sqrt{\sigma_i^2-\Lambda^2},\sigma_i)$ and $(y_i=-\sqrt{\sigma_i^2-\Lambda^2},\sigma_i)$. This separation is the foundation of the factorization. After integrating $\Lambda$ encircles $\Lambda=0$, the double-cover formalism is factorized into two single-cover formulae on two separated sheets, attached by an off-shell propagator. Two separated sheets correspond to two sub-amplitudes ${\cal A}_L$ and ${\cal A}_R$, respectively.
The process of factorization can be represented diagrammatically as in Figure \ref{factor}.
For the integral over $\Lambda$ encloses $\Lambda=0$,
an important rule is as follows: in the $\Lambda\to 0$ limit, if a configuration has non-vanishing contribution, each sheet must contain two fixed punctures $\sigma_i$, i.e., two elements in the set $\{\sigma_p,\sigma_q,\sigma_r,\sigma_m\}$.

%%%%%%%%%%%%%%%%%%%%%%%
\subsection{Differential operators}
\label{differ-OP}
%%%%%%%%%%%%%%%%%%%%%

The differential operators proposed by Cheung, Shen and Wen transmute the tree-level amplitude of one theory to amplitudes of other theories \cite{Cheung:2017ems,Zhou:2018wvn,Bollmann:2018edb}. There are three kinds of basic operators:
\begin{itemize}
\item (1) Trace operator:
\bea
{\cal T}^\epsilon[i,j]\equiv {\partial\over\partial(\epsilon_i\cdot\epsilon_j)}\,,
\eea
where the superscript $\epsilon$ means the operator is defined through polarization vectors in the set $\{\epsilon_i\}$ rather than $\{\W\epsilon_i\}$.
\item (2) Insertion operator:
\bea
{\cal I}^\epsilon_{ikj}\equiv {\partial\over\partial(\epsilon_k\cdot k_i)}-{\partial\over\partial(\epsilon_k\cdot k_j)}\,.~~~~\label{defin-insertion}
\eea
When applying to physical amplitudes, the insertion operator ${\cal I}^\epsilon_{ikj}$ inserts the external leg $k$ between external legs $i$ and $j$ in the color-ordering $(\cdots,i,j,\cdots)$, generates $(\cdots,i,k,j,\cdots)$. One can also use the definition \eref{defin-insertion} to split ${\cal I}^\epsilon_{ikj}$ as
\bea
{\cal I}^\epsilon_{ikj}={\cal I}^\epsilon_{ika}+{\cal I}^\epsilon_{akj}\,.~~~~\label{split-Inser}
\eea
The operator ${\cal I}^\epsilon_{ika}$ at the RHS is interpreted as inserting the leg $k$ between $i$ and $a$,
while the operator ${\cal I}^\epsilon_{akj}$ is interpreted as inserting $k$ between $a$ and $j$.
\item (3) Longitudinal operator:
\bea
{\cal L}^\epsilon_i\equiv \sum_{j\neq i}\,k_i\cdot k_j{\partial\over\partial(\epsilon_i\cdot k_j)}\,.~~~~\label{defin-Long1}
\eea
\end{itemize}

Three combinatory operators, which are products of basic operators, are defined as:
\begin{itemize}
\item (1) For a length-$m$ ordered set $\{\alpha_1,\alpha_2,\cdots,\alpha_m\}$ of external legs, the trace operator
is given as
\bea
{\cal T}^\epsilon[\alpha_1,\alpha_2,\cdots,\alpha_m]\equiv {\cal T}^\epsilon[\alpha_1,\alpha_m]\cdot\prod_{i=2}^{m-1}\,{\cal I}^\epsilon_{\alpha_{i-1}\alpha_i\alpha_m}\,.
\eea
It creates the color-ordering $(\alpha_1,\alpha_m)$ through ${\cal T}^\epsilon[\alpha_1\alpha_m]$, and inserts other elements between these two legs to generate the color-ordering $(\alpha_1,\alpha_2,\cdots,\alpha_m)$.
Notice taht  we adopt the convention in \cite{Cheung:2017ems} for products of operators that ${\cal O}_1\cdot{\cal O}_2$ is understood as
\bea
\big({\cal O}_1\cdot{\cal O}_2\big)f(k,\epsilon)={\cal O}_2\big({\cal O}_1f(k,\epsilon)\big)\,,~~~~\label{order-OP}
\eea
where $f(k,\epsilon)$ is a function of momenta and polarization vectors.
The interpretation of insertion operators indicates that ${\cal T}^\epsilon[\alpha_1,\alpha_2,\cdots,\alpha_m]$ has various equivalent formulae
when applying to physical amplitudes, for example
\bea
& &{\cal T}^\epsilon[\alpha_1,\alpha_2,\cdots,\alpha_m]\equiv {\cal T}^\epsilon[\alpha_1,\alpha_2]\cdot\prod_{i=3}^{m}\,{\cal I}^\epsilon_{\alpha_{i-1}\alpha_i\alpha_1}\,,\nn
& &{\cal T}^\epsilon[\alpha_1,\alpha_2,\cdots,\alpha_m]\equiv {\cal T}^\epsilon[\alpha_1,\alpha_3]\cdot {\cal I}^\epsilon_{\alpha_{1}\alpha_2\alpha_3}\cdot\prod_{i=4}^{m}\,{\cal I}^\epsilon_{\alpha_{i-1}\alpha_i\alpha_1}\,,
\eea
and so on.
\item (2) For $n$-point amplitudes, the operator ${\cal L}^\epsilon$
is defined as \footnote{When number of external legs is even, there is another physically equivalent definition ${\cal L}^\epsilon\equiv\sum_{\rho\in{\rm pair}}\,\prod_{i_k,j_k\in\rho}\,{\cal L}^\epsilon_{i_kj_k}$, with ${\cal L}^\epsilon_{ij}\equiv -k_i\cdot k_j{\partial\over\partial(\epsilon_i\cdot \epsilon_j)}$. In this paper, we will apply the operator ${\cal L}^\epsilon$ to generate the off-shell NLSM integrands with the odd number of external legs, thus only the current definition works.}
\bea
{\cal L}^\epsilon\equiv\prod_i\,{\cal L}^\epsilon_i\,.~~~~\label{defin-Long2}
\eea
\item (3) For a length-$2m$ set, the operator ${\cal T}^\epsilon_{{\cal X}_{2m}}$ is defined as
\bea
{\cal T}^\epsilon_{{\cal X}_{2m}}\equiv\sum_{\rho\in{\rm pair}}\,\prod_{i_k,j_k\in\rho}\,\delta_{I_{i_k}I_{j_k}}{\cal T}^\epsilon[i_k,j_k]\,,
\eea
where $\delta_{I_{i_k}I_{j_k}}$ forbids the interaction between particles carry different flavors. For the special case $2m$ particles do not carry any flavor, the operator ${\cal T}^\epsilon_{X_{2m}}$ is defined by removing $\delta_{I_{i_k}I_{j_k}}$,
\bea
{\cal T}^\epsilon_{X_{2m}}\equiv\sum_{\rho\in{\rm pair}}\,\prod_{i_k,j_k\in\rho}\,{\cal T}^\epsilon[i_k,j_k]\,.
\eea
The explanation for the notation $\sum_{\rho\in{\rm pair}}\,\prod_{i_k,j_k\in\rho}$ is in order. Let $\Gamma$ be the set of all partitions of the set $\{a_1,a_2,\cdots, a_{2m}\}$ into pairs without regard to the order.
An element in $\Gamma$ can be written as
\bea
\rho=\{(i_1,j_1),(i_2,j_2),\cdots,(i_m,j_m)\}\,,
\eea
with conditions $i_i<i_2<\cdots<i_m$ and $i_t<j_t,\,\forall t$. Then, $\prod_{i_k,j_k\in\rho}$ stands for the product of ${\cal T}^\epsilon[i_kj_k]$
for all pairs $(i_k,j_k)$ in $\rho$, and $\sum_{\rho\in{\rm pair}}$ denotes the summation over all partitions.
\end{itemize}

The combinatory operators defined above link tree-level amplitudes of a wide range of theories together, by transmuting the GR amplitude to amplitudes of other theories, formally expressed as
\bea
{\pmb A}=\big({\cal O}^\epsilon\cdot{\cal O}^{\W\epsilon}\big){\pmb A}^{\epsilon,\W\epsilon}_{\rm GR}\,.~~~~\label{fund-uni-diff}
\eea
Operators ${\cal O}^\epsilon$ and ${\cal O}^{\W\epsilon}$ for different theories, which will be used in this paper, are listed in Table \ref{tab:unifying}.
\begin{table}[!h]
\begin{center}
\begin{tabular}{c|c|c}
Amplitude& ${\cal O}^\epsilon$  & ${\cal O}^{\W\epsilon}$ \\
\hline
${\pmb A}_{{\rm GR}}^{\epsilon,\W\epsilon}(\pmb{H}_n)$ & $\mathbb{I}$ & $\mathbb{I}$  \\
${\pmb A}_{{\rm YM}}^{\W\epsilon}(i_1,\cdots,i_n)$ &  ${\cal T}^{\epsilon}[i_1,\cdots, i_n]$ & $\mathbb{I}$ \\
${\pmb A}_{{\rm sYMS}}^{\W\epsilon}(\pmb{S}_{2m}||\pmb{G}_{n-2m};i_1,\cdots,i_n)$ &  ${\cal T}^{\epsilon}[i_1,\cdots, i_n]$ & ${\cal T}^{\W\epsilon}_{{\cal X}_{2m}}$ \\
${\pmb A}_{{\rm NLSM}}(i_1,\cdots,i_n)$ & ${\cal T}^{\epsilon}[i_1,\cdots, i_n]$ & ${\cal T}^{\W\epsilon}[a,b]\cdot {\cal L}^{\W\epsilon}$ \\
${\pmb A}_{{\rm BAS}}(i_1,\cdots,i_n;i'_1,\cdots,i'_n)$ &  ${\cal T}^{\epsilon}[i_1,\cdots, i_n]$ & ${\cal T}^{\W\epsilon}[i'_1,\cdots, i'_n]$ \\
\end{tabular}
\end{center}
\caption{\label{tab:unifying}Unifying relations for differential operators}
\end{table}
In this table, all amplitudes include $n$ external legs. The symbol $\mathbb{I}$ stands for the identical operator. Notations $\pmb{H}_a$, $\pmb{G}_a$ and $\pmb{S}_a$
denote un-ordered sets of gravitons, gluons and scalars respectively, where the subscript denotes the length of the set. In next sections these sets will be given explicitly by their elements, for example $\pmb{H}_n=\{1,\cdots,n\}$. Thus, in this paper ${\pmb A}(\{i_1,\cdots,i_n\})$ means $n$ legs are not color-ordered, while ${\pmb A}(i_1,\cdots,i_n)$ means legs are color-ordered.
The notation $;\cdots$ denotes the additional color-ordering among all external legs, such as $(i'_1,\cdots,i'_n)$ among all scalars and gluons in the sYMS example. In the notation ${\pmb A}_{{\rm sYMS}}^{\W\epsilon}(\pmb{S}_{2m}||\pmb{G}_{n-2m};i'_1,\cdots,i'_n)$, $||$ is used to separate sets of external scalars and gluons, and the polarization vectors $\W\epsilon_i$ are carried by gluons.

\begin{figure}
  \centering
  \includegraphics[width=5cm]{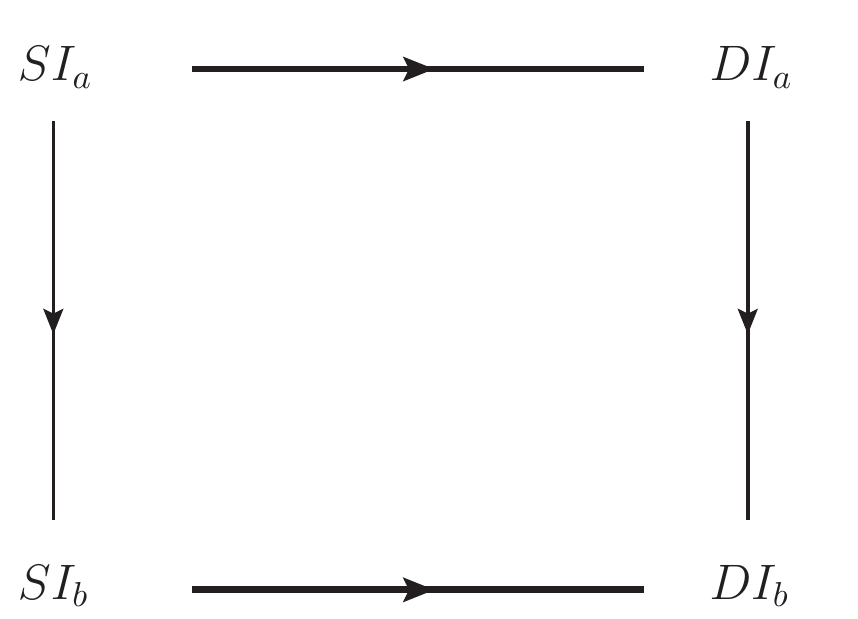} \\
  \caption{Transmuting integrands. $SI$ and $DI$ stand for single-cover and double-cover integrands respectively. Subscripts $a$ and $b$ refer to theory-$a$ and theory-$b$. The normal lines indicate the applying of differential operators, while the bold lines mean change the single-cover integrand to the double-cover integrand.}\label{transinte}
\end{figure}

The effect of differential operators can be understood at levels of both amplitudes and CHY integrands. As proved in \cite{Zhou:2018wvn,Bollmann:2018edb}, for on-shell amplitudes, the differential operators transmute the single-cover integrand of one theory to that
of another theory. In other words, relations in Table \ref{tab:unifying} hold for single-cover integrands. This is a way of understanding why these relations hold for on-shell amplitudes, due to the fact that the differential operators do not affect the single-cover measure therefore are commutable with CHY contour integral. Now we point out that for on-shell amplitudes these relations also hold for double-cover integrands. The technical details of proving this will be presented elsewhere \cite{Zhou}. Here we only discuss two general paths to understand this fact. One path to achieve the conclusion is that the differential operators will not affect the double-cover measure, thus relations for amplitudes must be satisfied by double-cover integrands. Another path to understand this conclusion is that the differential operators are commutable with the replacement ${1\over z_{ij}}\to\tau_{ij}$. To see this, we regard such replacement as two steps, the first one is ${1\over z_{ij}}\to T_{ij}$, and the second one is timing the factor $\Big(\prod_{i=1}^n\,{(y\sigma)_i\over y_i}\Big)$. The second step is obviously commutable with the differential operators since the factor $\Big(\prod_{i=1}^n\,{(y\sigma)_i\over y_i}\Big)$ is overall. For the first step, since $z_{ij}\equiv z_i-z_j$ and $T_{ij}\equiv{1\over (y\sigma)_i-(y\sigma)_j}$, $T_{ij}$ carries the same algebraical properties as ${1\over z_{ij}}$, such as antisymmetry. Thus replacing ${1\over z_{ij}}$ by $T_{ij}$ will not change the effect of differential operators. Consequently, the first step is also commutable with the differential operators. Thus, the differential operators transmute the numerator ${\cal I}^\tau_L(\sigma,y,k,\epsilon){\cal I}^\tau_R(\sigma,y,k,\W\epsilon)$ of the double-cover integrand in \eref{DCamp} and the single-cover integrand ${\cal I}_L(z,k,\epsilon){\cal I}_R(z,k,\W\epsilon)$ in the same manner. On the other hand, the denominate of the double-cover integrand, which is the scattering equation, will not be affected. Thus we arrive at the conclusion that relations in Table \ref{tab:unifying} hold for double-cover integrands. From the argument of the second path, we also know that the double-cover integrand of theory-$b$ can be generated from the single-cover integrand of theory-$a$ through two equivalent paths. One is to act the corresponding differential operator on the single-cover integrand of theory-$a$, to generate the single-cover integrand of theory-$b$, then change the single-cover integrand of theory-$b$ to the double-cover integrand of theory-$b$. Another one is to change the single-cover integrand of theory-$a$ to the double-cover integrand of theory-$a$ at first, then apply the differential operator on the double-cover integrand of theory-$a$, to generate the double-cover integrand of theory-$b$. Two paths are expressed diagrammatically in Figure \ref{transinte}.

In the next section, we will discuss if relations in Table \ref{tab:unifying} hold for off-shell amplitudes ${\cal A}_L$ and ${\cal A}_R$, which are the basic objects in the factorized formulae.

%%%%%%%%%%%%%%%%%%%%%%%%%%%%%%%%%%%%%%
\section{Applying differential operators to off-shell amplitudes}
\label{secfeasi}
%%%%%%%%%%%%%%%%%%%%%%%%%%%%%%%%%%%%%%%

As introduced in section \ref{secintro}, in the current work, the main idea is to act the differential operators on the factorized formula for one theory
\bea
{\pmb A}=\sum_{\rm channels}\,{\cal A}_L{1\over P^2}{\cal A}_R\,,
\eea
to generate the factorizations for other theories. To achieve the goal, we will factorize a differential operator as ${\cal O}\cong{\cal O}_L\cdot{\cal O}_R$, where $\cong$ means two operators are equivalent when acting on physical amplitudes. The operator ${\cal O}_L$ acts on ${\cal A}_L$ and annihilates ${\cal A}_R$, while ${\cal O}_R$ acts on ${\cal A}_R$ and annihilates ${\cal A}_L$. Such factorization for differential operators will be discussed in next sections. In this section, we explain that the relations in Table \ref{tab:unifying} hold for off-shell amplitudes ${\cal A}_L$ and ${\cal A}_R$, thus ${\cal O}_L$ and ${\cal O}_R$ transmute sub-amplitudes ${\cal A}_L$ and ${\cal A}_R$ to sub-amplitudes ${\cal A}'_L$ and ${\cal A}'_R$ of another theory.

%%%%%%%%%%%%%%%%%%%%%%%%%%%%%%
\subsection{Properties of off-shell single-cover amplitudes}
\label{property-OS}
%%%%%%%%%%%%%%%%%%%%%%%%%%%%%%%%

The sub-amplitudes ${\cal A}_L$ contains one off-shell external leg from the propagator $1\over P^2$, and so does ${\cal A}_R$.
In this subsection, we give the explicit definition of these off-shell amplitudes, and discuss the properties of them.

In this paper, the off-shell amplitudes ${\cal A}_L$ and ${\cal A}_R$ are defined in the single-cover version \eref{CHY}, formally expressed as
\bea
{\cal A}_n=\int {\big(\prod^n_{i=1,i\neq p,q,r}\,dz_i\big)|pqr|^2\over \prod_{i=1,i\neq p,q,r}^n\,{\cal E}_i(z)}\,{\cal I}_L(k,\epsilon,z){\cal I}_R(k,\W\epsilon,z)\,.~~~~\label{os-CHY}
\eea
For simplicity, the labels for removed scattering equations are chosen to be the same as labels for fixed punctures. In \eref{os-CHY}, the off-shell integrand has the same formula as the on-shell one. The difference is, for the off-shell particle $i$, the on-shell conditions $k_i^2=0$ and $\epsilon_i\cdot k_i=0$ will be violated. On the other hand, the momentum conservation law is still satisfied. The scattering equations are still written in the form \eref{SCeqS}, but $2k_i\cdot k_j$ can not be identified as Mandelstam variables $s_{ij}$ as in the on-shell case. For the on-shell case, there are $(n-3)$ independent scattering equations. However, for the off-shell case, only $(n-2)$ equations are independent, since the linear combination
$\sum_{j\neq i}\,z_i^2{\cal E}_i(z)$
only vanishes for the on-shell case. Thus, an important character of off-shell amplitudes is, $(n-3)$ poles in \eref{os-CHY} can not restrict all $n$ scattering equations to be satisfied.

One more significant character of off-shell amplitudes is the gauge dependence. The off-shell amplitudes depend on the fixed punctures $p,q,r$ in \eref{os-CHY}, and the choice of which rows and columns to be removed in the reduced matrices. The on-shell amplitudes are independent of all these choices. Notice that in this paper the choices of fixed punctures, and the removed rows and columns in the reduced matrices, are called the gauge choices.

Finally, if the momentum $k_r$ correspond to the fixed puncture $r$ is off-shell, the off-shell amplitude expressed by \eref{os-CHY} do not contain poles $(k_p+k_q+K)^2$, where $p,q$ are other two fixed punctures, $K$ is the combination of some external momenta which does not include $k_p,k_q,k_r$. The reason can be explained as follows. Kinematical variables in Pfaffians appear in the numerators therefore do not create propagators, thus the propagators only arise from coordinates $z_i$ restricted by scattering equations. In the expression \eref{os-CHY}, the scattering equations ${\cal E}_p$, ${\cal E}_q$ and ${\cal E}_r$ are removed, thus $k_p\cdot k_q$ can not appear in the remaining scattering equations directly. For the on-shell case, $k_p\cdot k_q$ can arise from other $k_i\cdot k_j$ through (remember that the scattering equations in this paper are defined through $k_i\cdot k_j$ rather than Mandelstm variables $s_{ij}$)
\bea
2k_p\cdot k_q=s_{pq}=s_{\overline{pq}}=\Big(\sum_{i\neq p,q}\,k_i\Big)^2=\sum_{i\neq p,q}\sum_{j\neq p,q,i}\,2k_i\cdot k_j\,,
\eea
where $\overline{pq}$ denotes all external legs except $p$ and $q$. When $k_r$ is off-shell, the above relation does not hold, thus $k_p\cdot k_q$
will not appear in the denominate. Thus, there is no way to create poles $(k_p+k_q+K)^2$. The above argument is not so strict, but the conclusion can be verified numerically. This observation is useful when discussing the relation between the factorized formula and the BCFW recursion in next sections.

%%%%%%%%%%%%%%%%%%%%%%%%%%%%%%%%
\subsection{Effect of differential operators}
\label{effect-OP}
%%%%%%%%%%%%%%%%%%%%%%%%%%%%%%%%%

With the understanding of off-shell amplitudes ${\cal A}_L$ and ${\cal A}_R$, now we explain the feasibility of relations in Table \ref{tab:unifying} for these off-shell amplitudes. As discussed in subsection \ref{differ-OP}, the differential operators are commutable with the single-cover contour integral, thus we only need to verify relations in Table \ref{tab:unifying} at the integrand level. More explicitly, we have
\bea
{\cal O}\,{\cal A}_n=\int {\big(\prod^n_{i=1,i\neq p,q,r}\,dz_i\big)|pqr|^2\over \prod_{i=1,i\neq p,q,r}^n\,{\cal E}_i(z)}\,{\cal O}\,\Big({\cal I}_L(k,\epsilon,z){\cal I}_R(k,\W\epsilon,z)\Big)\,.
\eea
If a differential operator transmutes the off-shell integrand of one theory to the off-shell integrand of another theory, it transmutes the off-shell amplitudes in the same manner.

In \cite{Zhou:2018wvn}, it has been proved that relations in Table \ref{tab:unifying} hold for on-shell single-cover integrands. To see if these relations can be generalized to the off-shell case, one can check the processes of proof in \cite{Zhou:2018wvn}. When applying ${\cal T}[\alpha_1,\cdots,\alpha_m]$ and ${\cal T}[{\cal X}]_{2m}$ to ${\bf Pf}'\Psi_n$, the manipulations in \cite{Zhou:2018wvn} are independent of the on-shell information about kinematical variables and scattering equations, thus are obviously correct for the off-shell case. Thus ${\cal T}[\alpha_1,\cdots,\alpha_m]$ and ${\cal T}[{\cal X}]_{2m}$ transmute the off-shell integrand of one theory to that of another theory.

When applying ${\cal T}[a,b]\cdot{\cal L}$ to ${\bf Pf}'\Psi_n$, the situation need to be analysed carefully,
since we have generalized the NLSM integrand to the case that the number of external legs is odd. The manipulation in \cite{Zhou:2018wvn}
gives
\bea
{\cal T}[a,b]\cdot{\cal L}\,{\bf Pf}'{\Psi}_n={\bf Pf}'{\widehat{A}}\,{\bf Pf}X_2\,.~~~~\label{act.LT2}
\eea
where
\bea
\widehat{ A}= \left(
         \begin{array}{c|c}
           ~~A_n~~ &  {\cal C}_{n\times(n-2)} \\
           \hline
          -{\cal C}^T_{(n-2)\times n} & 0 \\
         \end{array}
       \right)\,,~~~~~~~~X_2= \left(
         \begin{array}{cc}
           0 & {1\over z_{ab}} \\
          {1\over z_{ba}} & 0 \\
         \end{array}
       \right)\,,
\eea
with
\bea
{\cal C}_{ij} \equiv \begin{cases} \displaystyle {k_i \cdot k_j\over z_{ij}} &\quad i\neq j\,,\\
\displaystyle -\sum_{l=1,\,l\neq j}^n\hspace{-.5em}{k_l \cdot k_j\over z_{lj}} &\quad i=j\,,\end{cases}\,.
\eea
In \cite{Zhou:2018wvn}, the identities
\bea
\sum_{l=1,\,l\neq j}^n\hspace{-.5em}{k_l \cdot k_j\over z_{lj}}=0~~~~\label{iden1}
\eea
due to scattering equations indicate that ${\cal C}_{n\times(n-2)}=A_{n\times(n-2)}$ \footnote{This is the reason why we choose $k_i\cdot k_j$ rather than $s_{ij}$ to define scattering equations in this paper.}.
For the off-shell case in \eref{os-CHY}, three scattering equations ${\cal E}_i$ with $i=p,q,r$ are not satisfied.
However, if we choose $a$ and $b$ in ${\cal T}^\epsilon[a,b]$ belong to the set of fixed punctures (it is also the set of removed scattering equations) $\{p,q,r\}$ in \eref{os-CHY}, and choose the reduced matrix to be $(\widehat{A})^{ca}_{ca}$ with $c\in\{p,q,r\}$ and $c\neq a,b$, the three un-satisfied scattering equations will not appear in $({\cal C}_n)^{ca}_{ab}$, then the relation $({\cal C}_n)^{ca}_{ab}=(A_n)^{ca}_{ab}$ holds effectively.
Thus we arrive at
\bea
{\bf Pf}'{\widehat{A}}={(-)^{c+a}\over z_{ca}}{\bf Pf}\left(
         \begin{array}{c|c}
           ~~(A_n)^{ca}_{ca}~~ &  (A_n)^{ca}_{ab} \\
           \hline
          (A_n)^{ab}_{ca} & 0 \\
         \end{array}
       \right)\,.
\eea
For a $2m\times 2m$ skew symmetric matrix $M$, Pfaffian is defined as
\bea
{\bf Pf}M\equiv\sum_{\rho\in {\rm pair}}\,{\bf sgn}(\pi_\rho)\prod_{i_k,j_k\in \rho}M_{i_kj_k}\,,~~~~\label{defin-Pf}
\eea
where $\pi_\rho$ denotes the permutation $(1,2,\cdots, 2m)\to(i_1,j_1,i_2,j_2,\cdots, i_m,j_m)$. In the $(2n-4)\times(2n-4)$ matrix $\widehat{A}'= (\widehat{A})^{ca}_{ca}$, all elements $\widehat{A}'_{ij}$ with $i>n-2,j>n-2$ are zero. Thus, in the definition \eref{defin-Pf}, the non-zero contributions come from
the partitions that $i_k\in\{1,2,\cdots,n-2\}$ and $j_k\in\{n-1,n,\cdots,2n-4\}$, and we get
\bea
{\bf Pf} \widehat{A}'=(-)^{(n-3)(n-2)/2}\sum_{\pi\in S_{n-2}}\,{\bf sgn}(\pi)\prod_{k=1}^{n-2}\widehat{A}'_{kj_k}
=(-)^{(n-3)(n-2)/2}{\bf det}(A_n)^{ca}_{ab}\,.
\eea
Substituting it into \eref{act.LT2}, and using the definition of ${\bf det}'{A}_n$ in \eref{IRNLSM}, we see that
${\cal T}[a,b]\cdot{\cal L}$ transmutes ${\bf Pf}'{\Psi}_n$ to ${\bf det}'{A}_n$, up to an overall sign. We emphasize that when applying ${\cal T}[a,b]\cdot{\cal L}$ to
${\bf Pf}'{\Psi}_n$ in the off-shell case, one need to choose $a,b\in\{p,q,r\}$. Simultaneously,
$c$ in $(A_n)^{ca}_{ab}$ should also belong to $\{p,q,r\}$.

The above discussion leads to the conclusion that the differential operators transmute the off-shell single-cover integrand of one theory to off-shell single-cover integrands of other theories in the manner in Table \ref{tab:unifying}. Since the differential operators are commutable with the contour integral, we conclude that the relations in Table \ref{tab:unifying} hold for off-shell amplitudes ${\cal A}_L$ and ${\cal A}_R$. The argument can be expressed as
\bea
{\cal O}\,{\cal A}_n&=&\int {\big(\prod^n_{i=1,i\neq p,q,r}\,dz_i\big)|pqr|^2\over \prod_{i=1,i\neq p,q,r}^n\,{\cal E}_i(z)}\,{\cal O}\,\Big({\cal I}_L(k,\epsilon,z){\cal I}_R(k,\W\epsilon,z)\Big)\nn
&=&\int {\big(\prod^n_{i=1,i\neq p,q,r}\,dz_i\big)|pqr|^2\over \prod_{i=1,i\neq p,q,r}^n\,{\cal E}_i(z)}\,{\cal I}'_L(k,\epsilon,z){\cal I}'_R(k,\W\epsilon,z)\nn
&=&{\cal A}'_n\,.~~~~\label{operator-OS}
\eea

The off-shell amplitudes ${\cal A}'_n$ and ${\cal A}_n$ in \eref{operator-OS} depend on the gauge choices. As can be seen directly from \eref{operator-OS}, the gauge choice of fixed punctures for ${\cal A}'_n$ is the same as that for ${\cal A}_n$. This descendent relation arise from the commutability of the differential operators and the contour integral.

%%%%%%%%%%%%%%%%%%%%%%%%%%%%%%%%%%%%%%%%
\section{From GR to factorization for YM}
\label{secGRYM}
%%%%%%%%%%%%%%%%%%%%%%%%%%%%%%%%%%%%%%%%

In this section, by applying differential operators, we generate the factorization for the YM amplitude from the GR amplitude in the double-cover formula.
We first consider the simplest $4$-point GR amplitude ${\pmb A}^{\epsilon,\W\epsilon}_{\rm GR}(\{1,2,3,4\})$, with the fixed punctures $(p,q,r|m)=(1,2,3|4)$,
and the reduced matrices $(\Psi^\Lambda_4)^{13}_{13}$, $(\W\Psi^\Lambda_4)^{13}_{13}$. In the double-cover prescription, after integrating all $y_i$, $n$ coordinates $\sigma_i$ will be separated to upper and lower sheets. According to the rule introduced at the end of subsection \ref{subsecDC}, there are three separations which have non-vanishing contributions, $\{\{1,2\},\{3,4\}\}$, $\{\{1,4\},\{2,3\}\}$ and $\{\{1,3\},\{2,4\}\}$. As will be seen later, the first two separations correspond to channels which provide physical poles, while the third separation corresponds to a spurious pole. For two physical poles, the corresponding factorized GR terms can be obtained directly. Then, since the color-ordered YM amplitude ${\pmb A}^{\W\epsilon}_{\rm YM}(1,2,3,4)$ can be generated by ${\cal T}^\epsilon[1,2,3,4]{\pmb A}^{\epsilon,\W\epsilon}_{\rm GR}(\{1,2,3,4\})$, one can apply the trace operator ${\cal T}^\epsilon[1,2,3,4]$ to get the corresponding factorized terms contribute to ${\pmb A}^{\W\epsilon}_{\rm YM}(1,2,3,4)$. For the spurious pole, it is hard to find the factorized GR term. However, by using the trace operator, it is still straightforward to get the corresponding factorized YM term. Combining contributions from three channels together, the factorization for $4$-point YM amplitude in \cite{Bjerrum-Bohr:2018lpz} will be reproduced. Then we discuss the general case with arbitrary number of external legs. As will be shown, the factorized formula corresponds to two physical poles in the $4$-point case can be generalized to the general case directly, while the generalization of the formula corresponds to spurious pole is not clear. The difficulty arise from the evaluation of the factorized formula corresponds to spurious poles at the GR side. The relations between the factorized formula for GR and YM amplitudes, such as the factorization channels, the pole-structures, the gauge choices for off-shell sub-amplitudes, and the BCFW recursions, will be studied. The proper gauge choice in the double-cover prescription, which is consistent with the above relations, will also be discussed.

%%%%%%%%%%%%%%%%%%%%%%%%%%%%%%%%%%%%%%%%%%%
\subsection{$4$-point case: physical poles}
\label{4-point-YM-phy}
%%%%%%%%%%%%%%%%%%%%%%%%%%%%%%%%%%%%%%%%%%

In order to get the factorized formula for the color-ordered YM amplitude ${\pmb A}^{\W\epsilon}_{\rm YM}(1,2,3,4)$ via
\bea
{\pmb A}^{\W\epsilon}_{\rm YM}(1,2,3,4)={\cal T}^\epsilon[1,2,3,4]{\pmb A}^{\epsilon,\W\epsilon}_{\rm GR}(\{1,2,3,4\})\,,
\eea
we need to compute the expression of the GR amplitude ${\cal A}^{\epsilon,\W\epsilon}_{\pmb GR}(\{1,2,3,4\})$ in the double-cover prescription.
After integrating over all $y_i$, we have
\bea
{\pmb A}^{\epsilon,\W\epsilon}_{\rm GR}(\{1,2,3,4\})={1\over2^2}\sum_{{\rm C}}\,\int\,{d\Lambda\over\Lambda}\Delta_{123}\Delta_{123|4}{({\bf Pf}'\Psi_4)^\tau_{\rm C}({\bf Pf}'\W\Psi_4)^\tau_{\rm C}\over {\cal E}_4^\tau}\,,
\eea
where $\sum_{\rm C}$ means summing over all configurations, $({\bf Pf}'\Psi_4)^\tau_{\rm C}$ and $({\bf Pf}'\W\Psi_4)^\tau_{\rm C}$ represent that all $y_i$ in $({\bf Pf}'\Psi_4)^\tau$ and $({\bf Pf}'\W\Psi_4)^\tau$ take the values for the corresponding configuration.
For $2^4$ configurations, the rule introduced at the end of subsection \ref{subsecDC} indicates that there are three separations $\{\{1,2\},\{3,4\}\}$, $\{\{1,4\},\{2,3\}\}$ and $\{\{1,3\},\{2,4\}\}$ give non-vanishing contributions, and each separation corresponds to two mirrored configurations. In this subsection, we consider the first two separations. The third separation will be considered in the next subsection.

The factorized formula corresponds to the separation $\{\{1,2\},\{3,4\}\}$ can be obtained by integrating $\Lambda$. We can focus on the configuration that the sets of punctures $\{\sigma_1,\sigma_2\}$ and $\{\sigma_3,\sigma_4\}$ are respectably on two sheets as
\bea
& &(y_1=+\sqrt{\sigma_1^2-\Lambda^2}\,,~\sigma_1)\,,~~~~(y_2=+\sqrt{\sigma_2^2-\Lambda^2}\,,~\sigma_2)\,,\nn
& &(y_3=-\sqrt{\sigma_3^2-\Lambda^2}\,,~\sigma_3)\,,~~~~(y_4=-\sqrt{\sigma_4^2-\Lambda^2}\,,~\sigma_4)\,,
\eea
the mirrored configuration
\bea
& &(y_1=-\sqrt{\sigma_1^2-\Lambda^2}\,,~\sigma_1)\,,~~~~(y_2=-\sqrt{\sigma_2^2-\Lambda^2}\,,~\sigma_2)\,,\nn
& &(y_3=+\sqrt{\sigma_3^2-\Lambda^2}\,,~\sigma_3)\,,~~~~(y_4=+\sqrt{\sigma_4^2-\Lambda^2}\,,~\sigma_4)
\eea
can be treated similarly.
To do the contour integral over $\Lambda$ encircles $\Lambda=0$, one can expand the measure and the integrand around $\Lambda=0$, and pick up the $\Lambda^0$
term. Expanding ${\Delta_{123}\Delta_{123|4}\over {\cal E}^\tau_4}$ to the leading order of $\Lambda$ gives
\bea
{\Delta_{123}\Delta_{123|4}\over {\cal E}^\tau_4}\Big|^{1,2}_{3,4}&=&{2^5\over \Lambda^4}(\sigma_{12}\sigma_2\sigma_1)^2{1\over P_{34}^2}(\sigma_3\sigma_{34}\sigma_4)^2\nn
&=&{2^5\over \Lambda^4}(\sigma_{12}\sigma_{2P_{34}}\sigma_{P_{34}1})^2{1\over P_{34}^2}(\sigma_{P_{12}3}\sigma_{34}\sigma_{4P_{12}})^2\nn
&=&{2^5\over \Lambda^4}|12P_{34}|^2{1\over P_{34}^2}|P_{12}34|^2\,,~~~~\label{FP-det}
\eea
where $P_{ij\cdots k}\equiv k_i+k_j+\cdots+k_k$ and two new punctures are fixed as $\sigma_{P_{12}}=\sigma_{P_{34}}=0$.
Thus this part contributes $\Lambda^{-4}+{\cal O}(\Lambda^{-2})$. Throughout this paper, we use $P_{ij\cdots k}$ to denote both the combinatory momentum and the corresponding label of external leg.
On the other hand, expanding the reduced Pfaffian to the leading order yields
\bea
({\bf Pf}'\Psi_4)^\tau\Big|^{1,2}_{3,4}&=&\prod_{i=1}^4\,{(y\sigma)_i\over y_i}T_{13}{\bf Pf}(\Psi_4^\Lambda)^{13}_{13}\Big|^{1,2}_{3,4}\nn
&=&
-{\Lambda^2\over 2^2}\sum_{\epsilon^M}\,{1\over \sigma_{P_{34}1}\sigma_{P_{12}3}}{\bf Pf} \left(
         \begin{array}{cccc}
           0 & {k_2\cdot \epsilon^M_{34}\over \sigma_{2P_{34}}} & {k_2\cdot \epsilon_1\over \sigma_{21}} & C_{22} \\
          {\epsilon^M_{34}\cdot k_2\over \sigma_{P_{34}2}} & 0 & {\epsilon^M_{34}\cdot \epsilon_1\over \sigma_{P_{34}1}} & {\epsilon^M_{34}\cdot \epsilon_2\over \sigma_{P_{34}2}} \\
           {\epsilon_1\cdot k_2\over \sigma_{12}} &  {\epsilon_1\cdot \epsilon^M_{34}\over \sigma_{1P_{34}}} & 0  & {\epsilon_1\cdot \epsilon_2\over \sigma_{12}} \\
           -C_{22} &  {\epsilon_2\cdot \epsilon^M_{34}\over \sigma_{2P_{34}}}  & {\epsilon_2\cdot \epsilon_1\over \sigma_{21}} & 0 \\
         \end{array}
       \right)
      {\bf Pf} \left(
         \begin{array}{cccc}
           0 & {k_4\cdot \epsilon^M_{12}\over \sigma_{4P_{12}}} & {k_4\cdot \epsilon_3\over \sigma_{43}} & C_{44} \\
          {\epsilon^M_{12}\cdot k_4\over \sigma_{P_{12}4}} & 0 & {\epsilon^M_{12}\cdot \epsilon_3\over \sigma_{P_{12}3}} & {\epsilon^M_{12}\cdot \epsilon_4\over \sigma_{P_{12}4}} \\
           {\epsilon_3\cdot k_4\over \sigma_{34}} &  {\epsilon_3\cdot \epsilon^M_{12}\over \sigma_{3P_{12}}} & 0  & {\epsilon_3\cdot \epsilon_4\over \sigma_{34}} \\
           -C_{44} &  {\epsilon_4\cdot \epsilon^M_{12}\over \sigma_{4P_{12}}}  & {\epsilon_4\cdot \epsilon_3\over \sigma_{43}} & 0 \\
         \end{array}
       \right)\nn
&=&-{\Lambda^2\over 2^2}\sum_{\epsilon^M}\,{-1\over \sigma_{P_{34}1}}{\bf Pf}(\Psi_3)^{P_{34}1}_{P_{34}1}{-1\over\sigma_{P_{12}3}}{\bf Pf}(\Psi'_3)^{P_{12}3}_{P_{12}3}\nn
&=&-{\Lambda^2\over 2^2}\sum_{\epsilon^M}\,{\bf Pf}'\Psi_3{\bf Pf}'\Psi'_3\,,~~~~\label{exp-pf1}
\eea
where the polarization vectors of internal virtual particles are introduced as $\sum_{\epsilon^M} \epsilon^{M\mu}_i\epsilon^{M\nu}_j=\eta^{\mu\nu}$.
The matrices $\Psi_3$ and $\Psi'_3$ are for legs $\{1,2,P_{34}\}$ and $\{3,4,P_{12}\}$ respectively, with
\bea
C_{22}={\epsilon_2\cdot k_1\over \sigma_{21}}+{\epsilon_2\cdot P_{34}\over \sigma_{2P_{34}}}\,,~~~~C_{44}={\epsilon_4\cdot k_3\over \sigma_{43}}+{\epsilon_4\cdot P_{12}\over \sigma_{4P_{12}}}\,.
\eea
The GR integrand contains two reduced Pfaffians
$({\bf Pf}\Psi_4)^\tau$ and $({\bf Pf}\W\Psi_4)^\tau$. The leading order contribution in \eref{exp-pf1} is obviously correct for both two Pfaffians, with replacing $\epsilon^M$ by $\W\epsilon^{M'}$
for $\W\Psi_3$ and $\W\Psi'_3$ which arise from $\W\Psi_4$. Thus the Pfaffians give
$\Lambda^4+{\cal O}(\Lambda^6)$. Consequently, the desired $\Lambda^0$ term come from expanding all parts to the leading order. Integrating $\Lambda$ around $\Lambda=0$ gives rise to
the factorized formula
\bea
{\pmb A}^{\epsilon,\W\epsilon}_{\rm GR}(\{1,2,3,4\})\Big|^{1,2}_{3,4}
&=&{1\over 2}\sum_{\epsilon^M\W\epsilon^{M'}}\,{\cal A}^{\epsilon,\W\epsilon}_{\rm GR}(\{\underline{\hat{1}},\underline{2},\underline{\hat{P}}^{\epsilon^M,\W\epsilon^{M'}}_{34}\}){1\over P_{34}^2}{\cal A}^{\epsilon,\W\epsilon}_{\rm GR}(\{\underline{\hat{3}},\underline{4},\underline{\hat{P}}^{\epsilon^M,\W\epsilon^{M'}}_{12}\})\,,
\eea
where the expressions for two off-shell sub-amplitudes are in the single-cover forms
\bea
{\cal A}^{\epsilon,\W\epsilon}_{\rm GR}(\{\underline{\hat{1}},\underline{2},\underline{\hat{P}}^{\epsilon^M,\W\epsilon^{M'}}_{34}\})&=&|12P_{34}|^2{\bf Pf}'\Psi_3{\bf Pf}'{\W\Psi}_3\,,\nn
{\cal A}^{\epsilon,\W\epsilon}_{\rm GR}(\{\underline{\hat{3}},\underline{4},\underline{\hat{P}}^{\epsilon^M,\W\epsilon^{M'}}_{12}\})&=&
|P_{12}34|^2{\bf Pf}'\Psi'_3{\bf Pf}'{\W\Psi}'_3\,.
\eea
For these two sub-amplitudes, all coordinates are fully localized by gauge fixing, thus the contour integral does not appear. We have used $\underline{i}$ to denote the fixed punctures, $\hat{j}$ to denote the removed rows and columns in the reduced matrices, since the off-shell amplitudes depend on the choices of them.

Considering the mirrored configuration
\bea
& &(y_1=-\sqrt{\sigma_1^2-\Lambda^2}\,,~\sigma_1)\,,~~~~(y_2=-\sqrt{\sigma_2^2-\Lambda^2}\,,~\sigma_2)\,,\nn
& &(y_3=+\sqrt{\sigma_3^2-\Lambda^2}\,,~\sigma_3)\,,~~~~(y_4=+\sqrt{\sigma_4^2-\Lambda^2}\,,~\sigma_4)
\eea
gives the same result. Thus, summing over two configurations gives the factorized formula
\bea
{\pmb A}^{\epsilon,\W\epsilon}_{\rm GR}(\{1,2,3,4\})\Big|^{1,2}_{3,4}+{\pmb A}^{\epsilon,\W\epsilon}_{\rm GR}(\{1,2,3,4\})\Big|^{3,4}_{1,2}=\sum_{\epsilon^M\W\epsilon^{M'}}\,{\cal A}^{\epsilon,\W\epsilon}_{\rm GR}(\{\underline{\hat{1}},\underline{2},\underline{\hat{P}}^{\epsilon^M,\W\epsilon^{M'}}_{34}\}){1\over P_{12}^2}{\cal A}^{\epsilon,\W\epsilon}_{\rm GR}(\{\underline{\hat{3}},\underline{4},\underline{\hat{P}}^{\epsilon^M,\W\epsilon^{M'}}_{12}\})\,.~~\label{fct12-34}
\eea
The similar manipulation for the separation $\{\{1,4\},\{2,3\}\}$ gives
\bea
{\pmb A}^{\epsilon,\W\epsilon}_{\rm GR}(\{1,2,3,4\})\Big|^{1,4}_{2,3}+{\pmb A}^{\epsilon,\W\epsilon}_{\rm GR}(\{1,2,3,4\})\Big|^{2,3}_{1,4}=\sum_{\epsilon^M\W\epsilon^{M'}}\,{\cal A}^{\epsilon,\W\epsilon}_{\rm GR}(\{\underline{2},\underline{\hat{3}},\underline{\hat{P}}^{\epsilon^M,\W\epsilon^{M'}}_{14}\}){1\over P_{23}^2}{\cal A}^{\epsilon,\W\epsilon}_{\rm GR}(\{\underline{\hat{1}},\underline{4},\underline{\hat{P}}^{\epsilon^M,\W\epsilon^{M'}}_{23}\})\,.~~\label{fct14-23}
\eea
Combining them together, we obtain
\bea
{\pmb A}^{\epsilon,\W\epsilon}_{\rm GR}(\{1,2,3,4\})&=&\sum_{\epsilon^M\W\epsilon^{M'}}\,{\cal A}^{\epsilon,\W\epsilon}_{\rm GR}(\{\underline{\hat{1}},\underline{2},\underline{\hat{P}}^{\epsilon^M,\W\epsilon^{M'}}_{34}\}){1\over P_{12}^2}{\cal A}^{\epsilon,\W\epsilon}_{\rm GR}(\{\underline{\hat{3}},\underline{4},\underline{\hat{P}}^{\epsilon^M,\W\epsilon^{M'}}_{12}\})\nn
& &+\sum_{\epsilon^M\W\epsilon^{M'}}\,{\cal A}^{\epsilon,\W\epsilon}_{\rm GR}(\{\underline{2},\underline{\hat{3}},\underline{\hat{P}}^{\epsilon^M,\W\epsilon^{M'}}_{14}\}){1\over P_{23}^2}{\cal A}^{\epsilon,\W\epsilon}_{\rm GR}(\{\underline{\hat{1}},\underline{4},\underline{\hat{P}}^{\epsilon^M,\W\epsilon^{M'}}_{23}\})\nn
& &+\cdots\,,~~~~\label{fact-phy-1234}
\eea
where $\cdots$ denotes the term arise from the separation $\{\{1,3\},\{2,4\}\}$, which will be treated in the next subsection.
Both ${ P_{12}^2}$ and ${P_{23}^2}$ are physical poles, which will not be canceled by kinematical numerators.

The factorized GR terms for the first two channels have already been given in \eref{fact-phy-1234}. Now we apply the trace operator ${\cal T}^\epsilon[1,2,3,4]$ to the RHS of it. Since the color-ordered YM amplitude ${\pmb A}^{\W\epsilon}_{\rm YM}(1,2,3,4)$ can be generated from the $4$-point GR amplitude by applying ${\cal T}^\epsilon[1,2,3,4]$, we expect that applying the operator ${\cal T}^\epsilon[1,2,3,4]$ to the first two lines of \eref{fact-phy-1234} provides the corresponding YM terms. From the definition of the operator ${\cal T}^\epsilon[1,2,3,4]$, we know that this operator will not create or annihilate any propagator, thus one can expect that for terms survive under the action of the operator ${\cal T}^\epsilon[1,2,3,4]$, the physical poles for the GR amplitude are also physical poles for the YM amplitude, while the spurious poles for the GR amplitude are also spurious poles for the YM amplitude.

Let us consider the effect of applying the operator ${\cal T}^\epsilon[1,2,3,4]$ to the first line at the RHS of \eref{fact-phy-1234}. For simplicity, we choose the formula of ${\cal T}^\epsilon[1,2,3,4]$ among various equivalent choices to be
\bea
{\cal T}^\epsilon[1,2,3,4]={\cal T}^\epsilon[1,3]\cdot{\cal I}^\epsilon_{123}\cdot{\cal I}^\epsilon_{341}\,.
\eea
Now we explain that the above operator can be factorized as
\bea
{\cal T}^\epsilon[1,2,3,4]\cong \cancel{\sum}_{\epsilon^M}\,\cdot{\cal T}^\epsilon[1,2,P_{34}]\cdot{\cal T}^\epsilon[3,4,P_{12}]\,,~~~~\label{fact-trace-OP}
\eea
where $\cancel{\sum}_{\epsilon^M}$ means removing the summation over $\epsilon^M$, ${\cal T}^\epsilon[1,2,P_{34}]$ and ${\cal T}^\epsilon[3,4,P_{12}]$ are two trace operators which create color-orderings $(1,2,P_{34})$
and $(3,4,P_{12})$ respectively \footnote{Due to the convention in \eref{order-OP}, this operator is understood as removing the summation over $\epsilon^M$ first, then applying ${\cal T}^\epsilon[1,2,P_{34}]$ and ${\cal T}^\epsilon[3,4,P_{12}]$.}.
We first treat the operator ${\cal T}^\epsilon[1,3]\equiv{\partial\over\partial \epsilon_1\cdot\epsilon_3}$.
Since each polarization vector appear in each term of an amplitude once and only once, the effect of ${\cal T}^\epsilon[1,3]$
is just turning $\epsilon_1\cdot\epsilon_3$ to $1$ and annihilating terms do not contain $\epsilon_1\cdot\epsilon_3$. Due to the completeness relationship
$\sum_{\epsilon^M} \epsilon^{M\mu}_i\epsilon^{M\nu}_j=\eta^{\mu\nu}$, one can rewrite $\epsilon_1\cdot\epsilon_3$ as $\sum_{\epsilon^M} (\epsilon_1\cdot\epsilon^{M}_i)(\epsilon^{M}_j\epsilon_3)$. Thus, when applying to the first line of \eref{fact-phy-1234}, the operator ${\cal T}^\epsilon[1,3]$ removes the summation
over $\epsilon^M$, turns both $\epsilon_1\cdot\epsilon^{M}_{P_{34}}$ in ${\cal A}_L$
and $\epsilon^{M}_{P_{12}}\cdot\epsilon_3$ in ${\cal A}_R$ to $1$, and annihilate other terms do not include $\epsilon_1\cdot\epsilon^{M}_{P_{34}}$ and $\epsilon^{M}_{P_{12}}\cdot\epsilon_3$. Thus, although operators
\bea
{\cal T}^\epsilon[1,3]\equiv{\partial\over \partial\epsilon_1\cdot\epsilon_3}
\eea
and
\bea
\cancel{\sum}_{\epsilon^M}\,\cdot{\cal T}^\epsilon[1,P_{34}]\cdot{\cal T}^\epsilon[3,P_{12}]\equiv\cancel{\sum}_{\epsilon^M}\,\cdot{\partial\over \partial\epsilon_1\cdot\epsilon^{M}_{P_{34}}}\cdot{\partial\over \partial\epsilon_3\cdot\epsilon^{M}_{P_{12}}}~~~~\label{T[13]}
\eea
are not equivalent at the algebraic level, they are equivalent to each other when acting on the first line of \eref{fact-phy-1234}. The operator
${\cal T}^\epsilon[1,P_{34}]$ only acts on ${\cal A}_L$ and annihilates ${\cal A}_R$, while the operator ${\cal T}^\epsilon[3,P_{12}]$ only acts on ${\cal A}_R$ and annihilates ${\cal A}_L$.

Then we use the property \eref{split-Inser} to split ${\cal I}^\epsilon_{123}$ and ${\cal I}^\epsilon_{341}$ as follows
\bea
{\cal I}^\epsilon_{123}={\cal I}^\epsilon_{12P_{34}}+{\cal I}^\epsilon_{P_{34}23}\,,~~~~{\cal I}^\epsilon_{341}={\cal I}^\epsilon_{34P_{12}}+{\cal I}^\epsilon_{P_{12}41}\,.
\eea
Obviously, ${\cal I}^\epsilon_{12P_{34}}$ and ${\cal I}^\epsilon_{P_{34}23}$ annihilate ${\cal A}_R$ which does not contain $\epsilon_2$. Since $P_{34}=k_3+k_4$, when applying ${\cal I}^\epsilon_{P_{34}23}$ to ${\cal A}_L$, both ${\partial\over\partial \epsilon_2\cdot P_{34}}$ and ${\partial\over\epsilon_2\cdot \partial k_3}$ act on $(\epsilon_2\cdot P_{34})$, and give the same result. Thus ${\cal I}^\epsilon_{P_{34}23}$ also annihilates ${\cal A}_L$. Similar discussion holds for ${\cal I}^\epsilon_{34P_{12}}$ and ${\cal I}^\epsilon_{P_{12}41}$.
Thus,
only applying ${\cal I}^\epsilon_{12P_{34}}$ to ${\cal A}_L$ in the first line of \eref{fact-phy-1234} and ${\cal I}^\epsilon_{34P_{12}}$ to ${\cal A}_R$ gives the non-vanishing contribution. Notice that one can not use $\sum_{\epsilon^M} \epsilon^{M\mu}_{P_{34}}\epsilon^{M\nu}_{P_{12}}=\eta^{\mu\nu}$ to split the insertions operators because $\epsilon^{M\mu}_{P_{34}}$
and $\epsilon^{M\nu}_{P_{12}}$ are removed from ${\cal A}_L$ and ${\cal A}_R$ by the action of ${\cal T}^\epsilon[1,P_{34}]$ and ${\cal T}^\epsilon[3,P_{12}]$.
Consequently, when applying to the first line of \eref{fact-phy-1234}, the trace operator ${\cal T}^\epsilon[1,2,3,4]$ is equivalent to the operator
\bea
& &\cancel{\sum}_{\epsilon^M}\,\cdot{\cal T}^\epsilon[1,P_{34}]\cdot{\cal T}^\epsilon[3,P_{12}]\cdot{\cal I}^\epsilon_{12P_{34}}\cdot{\cal I}^\epsilon_{34P_{12}}\nn
&=&\cancel{\sum}_{\epsilon^M}\,\cdot\Big({\cal T}^\epsilon[1,P_{34}]\cdot{\cal I}^\epsilon_{12P_{34}}\Big)\cdot\Big({\cal T}^\epsilon[3,P_{12}]\cdot{\cal I}^\epsilon_{34P_{12}}\Big)\nn
&=&\cancel{\sum}_{\epsilon^M}\,\cdot{\cal T}^\epsilon[1,2,P_{34}]\cdot{\cal T}^\epsilon[3,4,P_{12}]\,,
\eea
where the commutability of ${\cal T}^\epsilon[3,P_{12}]$ and ${\cal I}^\epsilon_{12P_{34}}$ has been used.
Thus we achieve the factorized operator in \eref{fact-trace-OP}.

As discussed in subsection \ref{effect-OP}, relations in Table \ref{tab:unifying} hold for off-shell amplitudes ${\cal A}_L$ and ${\cal A}_R$. Thus, the operator ${\cal T}^\epsilon[1,2,P_{34}]$ transmutes ${\cal A}_L$ in the first line of \eref{fact-phy-1234} to the color-ordered off-shell YM amplitude ${\cal A}^{\W\epsilon}_{\rm YM}(1,2,P^{\W\epsilon^{M'}}_{34})$ and annihilates ${\cal A}_R$, while the operator ${\cal T}^\epsilon[3,4,P_{12}]$ transmutes ${\cal A}_R$ to ${\cal A}^{\W\epsilon}_{\rm YM}(3,4,P^{\W\epsilon^{M'}}_{12})$ and annihilates ${\cal A}_L$.
Using the factorized trace operator in \eref{fact-trace-OP}, we get the factorized form
\bea
& &{\cal T}^\epsilon[1,2,3,4]\,\Big(\sum_{\epsilon^M\W\epsilon^{M'}}\,{\cal A}^{\epsilon,\W\epsilon}_{\rm GR}(\{\underline{\hat{1}},\underline{2},\underline{\hat{P}}^{\epsilon^M,\W\epsilon^{M'}}_{34}\}){1\over P_{12}^2}{\cal A}^{\epsilon,\W\epsilon}_{\rm GR}(\{\underline{\hat{3}},\underline{4},\underline{\hat{P}}^{\epsilon^M,\W\epsilon^{M'}}_{12}\})\Big)\nn
&=&\sum_{\W\epsilon^{M'}}\,{\cal A}^{\W\epsilon}_{\rm YM}(\underline{\hat{1}},\underline{2},\underline{\hat{P}}^{\W\epsilon^{M'}}_{34}){1\over P_{12}^2}{\cal A}^{\W\epsilon}_{\rm YM}(\underline{\hat{3}},\underline{4},\underline{\hat{P}}^{\W\epsilon^{M'}}_{12})\,,~~\label{fct-YM-12-34}
\eea
which reproduces the first line of the result Eq.(42) in \cite{Bjerrum-Bohr:2018lpz}.

To understand the above process more clear, we notice that the trace operator ${\cal T}^\epsilon[1,2,3,4]$ defined via polarization vectors $\epsilon_i$ transmutes $\sum_{\epsilon^M}{\bf Pf}'\Psi_3{\bf Pf}'\Psi'_3$ to the Parke-Taylor factor $PT_3(1,2,P_{34})PT'_3(3,4,P_{12})$, and leaves $\sum_{\W\epsilon^{M'}}{\bf Pf}'\W\Psi_3{\bf Pf}'\W\Psi'_3$ un-altered. Then $PT_3(1,2,P_{34})PT'_3(3,4,P_{12})$ and $\sum_{\W\epsilon^{M'}}{\bf Pf}'\W\Psi_3{\bf Pf}'\W\Psi'_3$, together with the Faddeev-Popov determinants provided in \eref{FP-det}, give the integrands for ${\cal A}^{\W\epsilon}_{\rm YM}(1,2,P^{\W\epsilon^{M'}}_{34})$ and ${\cal A}^{\W\epsilon}_{\rm YM}(3,4,P^{\W\epsilon^{M'}}_{12})$.

In a similar way, one can find that applying ${\cal T}^\epsilon[1,2,3,4]$ to the second line of \eref{fact-phy-1234} gives
\bea
& &{\cal T}^\epsilon[1,2,3,4]\,\Big(\sum_{\epsilon^M\W\epsilon^{M'}}\,{\cal A}^{\epsilon,\W\epsilon}_{\rm GR}(\{\underline{2},\underline{\hat{3}},\underline{\hat{P}}^{\epsilon^M,\W\epsilon^{M'}}_{14}\}){1\over P_{23}^2}{\cal A}^{\epsilon,\W\epsilon}_{\rm GR}(\{\underline{\hat{1}},\underline{4},\underline{\hat{P}}^{\epsilon^M,\W\epsilon^{M'}}_{23}\})\Big)\nn
&=&\sum_{\W\epsilon^{M'}}\,{\cal A}^{\W\epsilon}_{\rm YM}(\underline{2},\underline{\hat{3}},\underline{\hat{P}}^{\W\epsilon^{M'}}_{14}){1\over P_{23}^2}{\cal A}^{\W\epsilon}_{\rm YM}(\underline{4},\underline{\hat{1}},\underline{\hat{P}}^{\W\epsilon^{M'}}_{23})\,,~~\label{fct-YM-14-23}
\eea
which reproduces the second line of Eq.(42) in \cite{Bjerrum-Bohr:2018lpz}.
Since the operator ${\cal T}^\epsilon[1,2,3,4]$ transmutes the GR amplitude ${\pmb A}^{\epsilon,\W\epsilon}_{\rm GR}(\{1,2,3,4\})$
to the color-ordered YM amplitude ${\pmb A}^{\W\epsilon}_{\rm YM}(1,2,3,4)$, putting \eref{fct-YM-12-34} and \eref{fct-YM-14-23} together gives rise to the factorized formula
\bea
{\pmb A}^{\W\epsilon}_{\rm YM}(1,2,3,4)&=&\sum_{\W\epsilon^{M'}}\,{\cal A}^{\W\epsilon}_{\rm YM}(\underline{\hat{1}},\underline{2},\underline{\hat{P}}^{\W\epsilon^{M'}}_{34}){1\over P_{12}^2}{\cal A}^{\W\epsilon}_{\rm YM}(\underline{\hat{3}},\underline{4},\underline{\hat{P}}^{\W\epsilon^{M'}}_{12})\nn
& &+\sum_{\W\epsilon^{M'}}\,{\cal A}^{\W\epsilon}_{\rm YM}(\underline{2},\underline{\hat{3}},\underline{\hat{P}}^{\W\epsilon^{M'}}_{14}){1\over P_{23}^2}{\cal A}^{\W\epsilon}_{\rm YM}(\underline{4},\underline{\hat{1}},\underline{\hat{P}}^{\W\epsilon^{M'}}_{23})\nn
& &+\cdots\,,
\eea
where the $\cdots$ part in the last line will be evaluated in the next subsection. One can verify that denominates $P_{12}^2$ and $P_{23}^2$ will not be canceled by kinematical numerators in ${\cal A}^{\W\epsilon}_{\rm YM}(\underline{\hat{1}},\underline{2},\underline{\hat{P}}^{\W\epsilon^{M'}}_{34}){\cal A}^{\W\epsilon}_{\rm YM}(\underline{\hat{3}},\underline{4},\underline{\hat{P}}^{\W\epsilon^{M'}}_{12})$ and ${\cal A}^{\W\epsilon}_{\rm YM}(\underline{2},\underline{\hat{3}},\underline{\hat{P}}^{\W\epsilon^{M'}}_{14}){\cal A}^{\W\epsilon}_{\rm YM}(\underline{4},\underline{\hat{1}},\underline{\hat{P}}^{\W\epsilon^{M'}}_{23})$, thus provide physical poles. As discussed before, these two physical poles are inherited from physical poles for GR amplitude in \eref{fact-phy-1234}, since the trace operator will not create or annihilate any pole.

%%%%%%%%%%%%%%%%%%%%%%%%%%%%%%%%%%%%%%%%
\subsection{$4$-point case: spurious pole}
\label{4-point-YM-spur}
%%%%%%%%%%%%%%%%%%%%%%%%%%%%%%%%%%%%%%%%

Then we consider the separation $\{\{1,3\},\{2,4\}\}$. We first focus on the configuration
\bea
& &(y_1=+\sqrt{\sigma_1^2-\Lambda^2}\,,~\sigma_1)\,,~~~~(y_3=+\sqrt{\sigma_3^2-\Lambda^2}\,,~\sigma_3)\,,\nn
& &(y_2=-\sqrt{\sigma_2^2-\Lambda^2}\,,~\sigma_2)\,,~~~~(y_4=-\sqrt{\sigma_4^2-\Lambda^2}\,,~\sigma_4)\,.
\eea
Similar as in the previous subsection, in order to do the contour integral over $\Lambda$ encircles $\Lambda=0$, we expand all elements around $\Lambda=0$. Expanding ${\Delta_{123}\Delta_{123|4}\over {\cal E}^\tau_4}$ to the leading order gives
\bea
{\Delta_{123}\Delta_{123|4}\over {\cal E}^\tau_4}\Big|^{1,3}_{2,4}&=&{2^5\over \Lambda^4}(\sigma_{13}\sigma_{3P_{24}}\sigma_{P_{24}1})^2{1\over P_{13}^2}(\sigma_{P_{13}2}\sigma_{24}\sigma_{4P_{13}})^2\nn
&=&{2^5\over \Lambda^4}|13P_{24}|^2{1\over P_{13}^2}|24P_{13}|^2\,,~~~~\label{ms-part}
\eea
where $\sigma_{P_{13}}=\sigma_{P_{24}}=0$. This part contributes $\Lambda^{-4}+{\cal O}(\Lambda^{-2})$ as before. However, expanding the reduced Pfaffian gives $\Lambda^0+{\cal O}(\Lambda^{2})$, thus the full $\Lambda^0$ term does not come from expanding all elements to the leading order. This fact makes the discussion of GR term for this channel to be extremely complicated.

However, we can still derive the corresponding factorized YM term by applying the trace operator. To do so, we use the observation that in the reduced Pfaffian
$({\bf Pf}'\Psi_4)^\tau$, the only effective term which will not vanish under the action of ${\cal T}^\epsilon[1,2,3,4]$ is
\bea
\Big(\prod_{i=1}^4\,{(y\sigma)_i\over y_i}\Big)T_{13}C_{22}C_{44}(T_{13}\epsilon_1\cdot\epsilon_3)\,.~~~~\label{contribute}
\eea
The reason is, in the reduced matrix $(\Psi^\Lambda_4)^{13}_{13}$, all quantities $\epsilon_2\cdot k_1$ and $\epsilon_2\cdot k_3$ are included in $C_{22}$, while all quantities
$\epsilon_4\cdot k_1$ and $\epsilon_4\cdot k_3$ are included in $C_{44}$. Other terms which do not include them will be annihilated by insertion operators ${\cal I}^\epsilon_{123}$ and ${\cal I}^\epsilon_{341}$. Notice that ${\cal T}^\epsilon[1,2,3,4]$ only acts on $({\bf Pf}'\Psi_4)^\tau$ since polarization vectors carried by $({\bf Pf}'\W\Psi_4)^\tau$ are $\W\epsilon_i$. Expanding $T_{13}$ and all ${(y\sigma)_i\over y_i}$ in \eref{contribute} to the leading order of $\Lambda$ gives
\bea
{\Lambda^4\over \sigma_2^2\sigma_4^2}{1\over 2^2\sigma_{13}^2}C_{22}C_{44}(\epsilon_1\cdot\epsilon_3)\,.
\eea
Then we expand $C_{22}$ and $C_{44}$ as
\bea
C_{22}&=&{1\over \Lambda^2}{2\sigma_2\sigma_4\over \sigma_{42}}\epsilon_2\cdot k_4
-{\epsilon_2\cdot k_1\over 2\sigma_1}-{\epsilon_2\cdot k_3\over 2\sigma_3}+\cdots\,,\\
C_{44}&=&{1\over \Lambda^2}{2\sigma_2\sigma_4\over \sigma_{24}}\epsilon_4\cdot k_2
-{\epsilon_4\cdot k_1\over 2\sigma_1}-{\epsilon_4\cdot k_3\over 2\sigma_3}+\cdots\,.
\eea
Obviously, the leading order terms of $C_{22}$ and $C_{44}$ will be annihilated by ${\cal I}^\epsilon_{123}$ and ${\cal I}^\epsilon_{341}$,
thus we need to pick up the next-to-leading order terms, and act the trace operator on
\bea
{\Lambda^4\over \sigma_2^2\sigma_4^2}{1\over 2^2\sigma_{13}^2}\Big({\epsilon_2\cdot k_1\over 2\sigma_1}+{\epsilon_2\cdot k_3\over 2\sigma_3}\Big)\Big({\epsilon_4\cdot k_1\over 2\sigma_1}+{\epsilon_4\cdot k_3\over 2\sigma_3}\Big)(\epsilon_1\cdot\epsilon_3)\,.~~~~\label{PT-pre}
\eea
Then we find the leading order term of ${\cal T}^\epsilon[1,2,3,4]({\bf Pf}'\Psi_4)^\tau$ is
\bea
{\cal T}^\epsilon[1,2,3,4]\,({\bf Pf}'\Psi_4)^\tau\Big|^{1,3}_{2,4}&=&{\Lambda^4\over 2^4}{1\over \sigma_1^2\sigma_2^2\sigma_3^2\sigma_4^2}={\Lambda^4\over 2^4}{1\over \sigma_{1P_{24}}^2\sigma_{3P_{24}}^2\sigma_{2P_{13}}^2\sigma_{4P_{13}}^2}\,.~~~~\label{PT-part}
\eea

Now we see that the leading order terms of ${\Delta_{123}\Delta_{123|4}\over {\cal E}^\tau_4}$, ${\cal T}^\epsilon[1,2,3,4]\,({\bf Pf}'\Psi_4)^\tau$ and  $({\bf Pf}'\W\Psi_4)^\tau$
contribute $\Lambda^{-4}$, $\Lambda^{4}$ and $\Lambda^0$, respectively. Combining all above leading order terms together gives rise to the desired $\Lambda^0$ term. Thus we can expand $({\bf Pf}'\W\Psi_4)^\tau$ to the leading order and factorize it as
\bea
& &\prod_{i=1}^n\,{(y\sigma)_i\over y_i}T_{13}{\bf Pf}(\W\Psi^\Lambda_4)^{13}_{13}\Big|^{1,3}_{2,4}\nn
&=&{2\sigma_{1P_{24}}\sigma_{3P_{24}}\sigma_{2P_{13}}\sigma_{4P_{13}}\over\sigma_{13}\sigma_{24}}
\sum_{\W\epsilon^L}\,{-1\over \sigma_{P_{24}1}}{\bf Pf}\left(
         \begin{array}{cccc}
           0 & {k_3\cdot \W\epsilon^L_{24}\over \sigma_{3P_{24}}} & {k_3\cdot \W\epsilon_1\over \sigma_{13}} & C_{33} \\
          {\W\epsilon^L_{24}\cdot k_3\over \sigma_{P_{24}2}} & 0 & {\W\epsilon^L_{24}\cdot \W\epsilon_1\over \sigma_{P_{24}1}} & {\W\epsilon^L_{24}\cdot \W\epsilon_3\over \sigma_{P_{24}3}} \\
           {\W\epsilon_1\cdot k_3\over \sigma_{13}} &  {\W\epsilon_1\cdot \W\epsilon^L_{24}\over \sigma_{1P_{24}}} & 0  & {\W\epsilon_1\cdot \W\epsilon_3\over \sigma_{13}} \\
           -C_{33} &  {\W\epsilon_3\cdot \W\epsilon^L_{24}\over \sigma_{3P_{24}}}  & {\W\epsilon_3\cdot \W\epsilon_1\over \sigma_{31}} & 0 \\
         \end{array}
       \right)
      {\bf Pf} \left(
         \begin{array}{cccc}
           0 & {k_4\cdot \W\epsilon^L_{13}\over \sigma_{4P_{13}}} & {k_4\cdot \W\epsilon_2\over \sigma_{42}} & C_{44} \\
          {\W\epsilon^L_{13}\cdot k_4\over \sigma_{P_{13}4}} & 0 & {\W\epsilon^L_{13}\cdot \W\epsilon_2\over \sigma_{P_{13}2}} & {\W\epsilon^L_{13}\cdot \W\epsilon_4\over \sigma_{P_{13}4}} \\
           {\W\epsilon_2\cdot k_4\over \sigma_{24}} &  {\W\epsilon_2\cdot \W\epsilon^L_{13}\over \sigma_{2P_{13}}} & 0  & {\W\epsilon_2\cdot \W\epsilon_4\over \sigma_{24}} \\
           -C_{44} &  {\W\epsilon_4\cdot \W\epsilon^L_{13}\over \sigma_{4P_{13}}}  & {\W\epsilon_4\cdot \W\epsilon_2\over \sigma_{42}} & 0 \\
         \end{array}
       \right)\nn
&=&{2\sigma_{1P_{24}}\sigma_{3P_{24}}\sigma_{2P_{13}}\sigma_{4P_{13}}\over\sigma_{13}\sigma_{24}}
\sum_{\W\epsilon^L}\,{\bf Pf}'\W\Psi_3{\bf Pf}'\W\Psi'_3\,,~~~~\label{pf-part}
\eea
where $\sum_{\W\epsilon^L}$ means summing over longitudinal degree of freedoms satisfy $\sum_{\W\epsilon^L}\W\epsilon^L_i\W\epsilon^L_j={P_i^\mu P_j^\nu\over P_i\cdot P_j}$.
The matrices $\W\Psi_3$ and $\W\Psi'_3$ are for legs $\{P_{24},1,3\}$ and $\{P_{13},2,4\}$, respectively. Elements $C_{33}$ and $C_{44}$ are given by
\bea
C_{33}={\W\epsilon_3\cdot P_{24}\over \sigma_{3P_{24}}}+{\W\epsilon_3\cdot k_1\over \sigma_{31}}\,,~~~~C_{44}={\W\epsilon_4\cdot P_{13}\over \sigma_{4P_{13}}}+{\W\epsilon_4\cdot k_2\over \sigma_{42}}\,.
\eea

Combining three parts \eref{ms-part}, \eref{PT-part} and \eref{pf-part} together, doing the integral over $\Lambda$, and summing over the
mirrored configurations, we arrive at the factorized formula
\bea
& &{\cal T}^\epsilon[1,2,3,4]\,\Big({\cal A}^{\epsilon,\W\epsilon}_{\rm GR}(\{1,2,3,4\})\Big|^{1,3}_{2,4}+{\cal A}^{\epsilon,\W\epsilon}_{\rm GR}(\{1,2,3,4\})\Big|^{2,4}_{1,3}\Big)\nn
&=&-2\sum_{\W\epsilon^L}\,\Big({|13P_{24}|^2{\bf Pf}'\W\Psi_3\over \sigma_{13}\sigma_{3P_{24}}\sigma_{P_{24}1}}\Big){1\over P_{13}^2}\Big({|24P_{13}|^2{\bf Pf}'\W\Psi'_3\over\sigma_{24}\sigma_{4P_{13}}\sigma_{P_{13}2}}\Big)\nn
&=&-2\sum_{\W\epsilon^L}\,{\cal A}^{\W\epsilon}_{\rm YM}(\underline{\hat{1}},\underline{3},\underline{\hat{P}}^{\W\epsilon^L}_{24}){1\over P_{13}^2}{\cal A}^{\W\epsilon}_{\rm YM}(\underline{\hat{2}},\underline{4},\underline{\hat{P}}^{\W\epsilon^L}_{13})\,,~~\label{fct-YM-13-24}
\eea
which reproduces the last line of the result in \cite{Bjerrum-Bohr:2018lpz}. The above formula \eref{fct-YM-13-24} together with \eref{fct-YM-12-34} and \eref{fct-YM-14-23} give the full factorized formula for the $4$-point YM amplitude ${\pmb A}^{\W\epsilon}_{\rm YM}(1,2,3,4)$ as
\bea
{\pmb A}^{\epsilon}_{\rm YM}(1,2,3,4)&=&\sum_{\epsilon^M}\,{\cal A}^{\epsilon}_{\rm YM}(\underline{\hat{1}},\underline{2},\underline{\hat{P}}^{\epsilon^{M}}_{34}){1\over P_{12}^2}{\cal A}^{\epsilon}_{\rm YM}(\underline{\hat{3}},\underline{4},\underline{\hat{P}}^{\epsilon^{M}}_{12})\nn
& &+\sum_{\epsilon^M}\,{\cal A}^{\epsilon}_{\rm YM}(\underline{2},\underline{\hat{3}},\underline{\hat{P}}^{\epsilon^{M}}_{14}){1\over P_{23}^2}{\cal A}^{\epsilon}_{\rm YM}(\underline{4},\underline{\hat{1}},\underline{\hat{P}}^{\epsilon^{M}}_{23})\nn
& &-2\sum_{\epsilon^L}\,{\cal A}^{\epsilon}_{\rm YM}(\underline{\hat{1}},\underline{3},\underline{\hat{P}}^{\epsilon^L}_{24}){1\over P_{13}^2}{\cal A}^{\epsilon}_{\rm YM}(\underline{\hat{2}},\underline{4},\underline{\hat{P}}^{\epsilon^L}_{13})\,.~~~~\label{fct-YM-4p}
\eea

Some remarks are in order. For the channel discussed in this subsection, the pole $P_{13}^2$ is a spurious pole for both GR and YM amplitudes, since the leading order term of
$({\bf Pf}'\W\Psi_4)^\tau$ includes $P_{13}^2$ which cancels the propagator. The trace operator will not create or eliminate any pole, thus transmutes the spurious pole for GR to the spurious pole for YM. At the YM side, this term without any propagator is interpreted by the $4$-point interaction vertex in \cite{Bjerrum-Bohr:2018lpz}. One can observe that the factorization for this part is not as natural as those in the previous subsection. For example, in the previous case, the Parke-Taylor factors come from only one piece of the integrand ${\cal I}^\tau_L(\sigma,y,k,\epsilon)$, the same as the situation for the single-cover prescription. But in the
current case, the Parke-Taylor factor in the final result comes from both two pieces \eref{PT-part} and $\eref{pf-part}$. For a term with physical pole, although the sub-amplitudes depend on the gauge choices, the factorization channel is uniquely determined. However, for a term with spurious pole, the factorization channel is not unique, as will be seen in the example in subsection \ref{subsecsYMS}.

%%%%%%%%%%%%%%%%%%%%%%%%%%%%%%%%%%%%%%%%
\subsection{General case}
\label{general-YM}
%%%%%%%%%%%%%%%%%%%%%%%%%%%%%%%%%%%%%%%%%

In this subsection, we demonstrate that the result corresponds to physical poles in subsection \ref{4-point-YM-phy} can be generalized to the general case with arbitrary number of external legs. The relations between the factorized terms of YM and GR amplitudes, and the proper gauge choice in the double-cover prescription consistent with these relations, will also be discussed by using the trace operator.

Similar as in the $4$-point case, the first step is to derive the GR terms in the double-cover prescription. We still choose the fixed punctures to be $(p,q,r|m)=(1,2,3|4)$, and the reduced matrices to be $(\Psi^\Lambda_n)^{13}_{13}$ and $(\W\Psi^\Lambda_n)^{13}_{13}$. Integrating all coordinates $y_i$ provides
\bea
{\pmb A}^{\epsilon,\W\epsilon}_{\rm GR}(\{1,\cdots,n\})={1\over2^2}\sum_{\rm C}\,\int\,{d\Lambda\over\Lambda}\Big(\prod_{i\neq1,2,3,4}\,{d\sigma_i\over{\cal E}_i^\tau}\Big)\Delta_{123}\Delta_{123|4}{({\bf Pf}'\Psi_n)^\tau_{\rm C}({\bf Pf}'\W\Psi_n)^\tau_{\rm C}\over {\cal E}_4^\tau}\,.
\eea
Based on the rule described at the end of subsection \ref{subsecDC}, the effective separations correspond to non-vanishing contributions are $\{\{1,2,{\pmb \alpha}_1\},\{3,4,{\pmb \beta}_1\}\}$, $\{\{1,4,{\pmb \alpha}_2\},\{2,3,{\pmb \beta}_2\}\}$, and $\{\{1,3,{\pmb \alpha}_3\},\{2,4,{\pmb \beta}_3\}\}$, where ${\pmb \alpha}_i$ and ${\pmb\beta}_i$ are sets of external legs satisfy ${\pmb\alpha}_i\cup{\pmb\beta}_i=\{5,6,\cdots,n\}$, ${\pmb\alpha}_i\cap{\pmb\beta}_i=\emptyset$. The first two types of separations correspond to physical poles, while the third type of separations correspond to spurious poles. Expanding all elements to the leading order of $\Lambda$, and integrating $\Lambda$ around the pole $\Lambda=0$, the factorized GR terms correspond to first two types of separations can be obtained as
\bea
{\pmb A}^{\epsilon,\W\epsilon}_{\rm GR}(\{1,\cdots,n\})&=&\sum_{\pmb{\alpha}_1}\,\sum_{\epsilon^M\W\epsilon^{M'}}\,{\cal A}^{\epsilon,\W\epsilon}_{\rm GR}(\{\underline{\hat{1}},\underline{2},{\pmb\alpha}_1,\underline{\hat{P}}_{34{\pmb\beta}_1}^{\epsilon^M,\W\epsilon^{M'}}\}){1\over P^2_{12{\pmb\alpha}_1}}
{\cal A}^{\epsilon,\W\epsilon}_{\rm GR}(\{\underline{\hat{3}},\underline{4},{\pmb\beta}_1,\underline{\hat{P}}_{12{\pmb\alpha}_1}^{\epsilon^M,\W\epsilon^{M'}}\})\nn
& &+\sum_{\pmb{\alpha}_2}\,
\sum_{\epsilon^M\W\epsilon^{M'}}\,{\cal A}^{\epsilon,\W\epsilon}_{\rm GR}(\{\underline{\hat{1}},\underline{4},{\pmb\alpha}_2,\underline{\hat{P}}_{23{\pmb\beta}_2}^{\epsilon^M,\W\epsilon^{M'}}\}){1\over P^2_{14{\pmb\alpha}_2}}
{\cal A}^{\epsilon,\W\epsilon}_{\rm GR}(\{\underline{2},\underline{\hat{3}},{\pmb\beta}_2,\underline{\hat{P}}_{14{\pmb\alpha}_2}^{\epsilon^M,\W\epsilon^{M'}}\})\nn
& &+\cdots\,.~~~~~~~~~~~~~\label{phy-pole-YM-gen}
\eea
The details of derivation can be seen in Appendix \ref{phypole}.

The $n$-point color-ordered YM amplitude ${\pmb A}^{\W\epsilon}_{\rm YM}(1,2,\cdots,n)$ can be generated from the $n$-point GR amplitude by acting the trace operator ${\cal T}^\epsilon[1,2,\cdots, n]$. We can apply the trace operator ${\cal T}^\epsilon[1,2,\cdots, n]$ to first two lines at the RHS of \eref{phy-pole-YM-gen}, to get the corresponding
factorized terms contribute to the YM amplitude ${\pmb A}^{\W\epsilon}_{\rm YM}(1,2,\cdots,n)$. Let us choose the formula of ${\cal T}^\epsilon[1,2,\cdots,n]$ to be
\bea
{\cal T}^\epsilon[1,2,\cdots, n]={\cal T}^\epsilon[1,3]\cdot{\cal I}^\epsilon_{123}\cdot\prod_{i=4}^n{\cal I}^\epsilon_{(i-1)i1}\,.
\eea
We first apply it to the first line of \eref{phy-pole-YM-gen}. Similar as in the $4$-point case, we factorize ${\cal T}^\epsilon[1,3]$ as
\bea
\cancel{\sum}_{\epsilon^M}\,\cdot{\cal T}^\epsilon[1,P_{34{\pmb\beta}_1}]\cdot{\cal T}^\epsilon[3,P_{12{\pmb\alpha}_1}]
\equiv\cancel{\sum}_{\epsilon^M}\,\cdot{\partial\over\partial\epsilon_1\cdot\epsilon^M_{P_{34{\pmb\beta}_1}}}\cdot
{\partial\over\partial\epsilon_3\cdot\epsilon^M_{P_{12{\pmb\alpha}_1}}}\,.
\eea
The operator ${\cal T}^\epsilon[1,P_{34{\pmb\beta}_1}]$ acts on ${\cal A}_L$ in the first line of \eref{phy-pole-YM-gen} and annihilates ${\cal A}_R$,
${\cal T}^\epsilon[3,P_{12{\pmb\alpha}_1}]$ acts on ${\cal A}_R$ and annihilates ${\cal A}_R$. Then we split insertion operators as
\bea
& &{\cal I}^\epsilon_{123}={\cal I}^\epsilon_{12P_{34{\pmb\beta}_1}}+{\cal I}^\epsilon_{P_{34{\pmb\beta}_1}23}\,,\nn
& &{\cal I}^\epsilon_{(i-1)i1}={\cal I}^\epsilon_{(i-1)iP_{12{\pmb\alpha}_1}}+{\cal I}^\epsilon_{P_{12{\pmb\alpha}_1}i1}\,,
~~~~{\rm for}~i\in\{4,{\pmb\beta}_1\}\,,\nn
& &{\cal I}^\epsilon_{(i-1)i1}={\cal I}^\epsilon_{(i-1)iP_{34{\pmb\beta}_1}}+{\cal I}^\epsilon_{P_{34{\pmb\beta}_1}i1}\,,
~~~~{\rm for}~i-1\in\{4,{\pmb\beta}_1\}\,,~i\in{\pmb\alpha}_1\,,\nn
& &{\cal I}^\epsilon_{(i-1)i1}={\cal I}^\epsilon_{(i-1)i1}\,,
~~~~~~~~~~~~~~~~~~~~~~{\rm for}~i-1\in{\pmb\alpha}_1\,,~i\in{\pmb\alpha}_1\,.
\eea
In the $4$-point case, we have explained that both ${\cal A}_L$ and ${\cal A}_R$ will be annihilated by ${\cal I}^\epsilon_{P_{34}23}$. Using the similar argument, one can conclude that ${\cal I}^\epsilon_{P_{34{\pmb\beta}_1}23}$, ${\cal I}^\epsilon_{P_{12{\pmb\alpha}_1}i1}$ and ${\cal I}^\epsilon_{(i-1)iP_{34{\pmb\beta}_1}}$ annihilate both ${\cal A}_L$ and ${\cal A}_R$ in the first line of \eref{phy-pole-YM-gen},
thus the effective operator can be extracted as in the following factorized formula
\bea
{\cal T}^\epsilon[1,2,\cdots, n]\cong \cancel{\sum}_{\epsilon^M}\,\cdot{\cal T}^\epsilon_L\cdot{\cal T}^\epsilon_R\,,
\eea
where
\bea
& &{\cal T}^\epsilon_L={\cal T}^\epsilon[1,P_{34{\pmb\beta}_1}]\cdot{\cal I}^\epsilon_{12P_{34{\pmb\beta}_1}}\cdot\Big(\prod_{\substack{i-1\in\{4,{\pmb\beta}_1\}\\i\in{\pmb\alpha}_1}}\,{\cal I}^\epsilon_{P_{34{\pmb\beta}_1}i1}\Big)\cdot\Big(\prod_{\substack{i-1\in{\pmb\alpha}_1\\i\in{\pmb\alpha}_1}}\,{\cal I}^\epsilon_{(i-1)i1}\Big)\,,\nn
& &
{\cal T}^\epsilon_R={\cal T}^\epsilon[3,P_{12{\pmb\alpha}_1}]\cdot\prod_{i\in\{4,{\pmb\beta}_1\}}\,{\cal I}^\epsilon_{(i-1)iP_{12{\pmb\alpha}_1}}\,.
\eea
Obviously, ${\cal T}^\epsilon_L$ annihilates ${\cal A}_R$ and ${\cal T}^\epsilon_R$ annihilates ${\cal A}_L$. But it is still possible that ${\cal A}_R$ will be annihilated by ${\cal T}^\epsilon_R$. If the operator ${\cal I}^\epsilon_{(i-1)iP_{12{\pmb\alpha}_1}}$ gives non-vanishing contribution when acting on ${\cal A}_R$, not only the external leg $i$, but also the leg $(i-1)$, should be included in ${\cal A}_R$.
For the same reason, if the leg $(i-1)$ is included in ${\cal A}_R$, the leg $(i-2)$ should also be included in ${\cal A}_R$.
This recursive pattern indicates that the set ${\pmb\beta}_1$ should be $\{5,6,\cdots,j\}$. Subsequently, the set ${\pmb\alpha}_1$ should be $\{j+1,j+2,\cdots,n\}$.
For other ${\pmb\alpha}_1$ and ${\pmb\beta}_1$, the corresponding terms in the first line of \eref{phy-pole-YM-gen} will be annihilated by ${\cal T}^\epsilon[1,2,\cdots,n]$.
Thus ${\cal T}^\epsilon_L$ and ${\cal T}^\epsilon_R$ can be identified as trace operators
\bea
& &{\cal T}^\epsilon_L={\cal T}^\epsilon[1,2,P_{34{\pmb\beta}_1},j+1,j+2,\cdots,n]\,,\nn
& &{\cal T}^\epsilon_R={\cal T}^\epsilon[3,4,\cdots,j,P_{12{\pmb\alpha}_1}]\,.
\eea
Applying these two trace operators we get
\bea
& &{\cal T}^\epsilon_L\,{\cal A}_L={\cal A}^{\W\epsilon}_{\rm YM}(j+1,\cdots,\underline{\hat{1}},\underline{2},\underline{\hat{P}}^{\W\epsilon^{M'}}_{3:j})\,,\nn
& &{\cal T}^\epsilon_R\,{\cal A}_R={\cal A}^{\W\epsilon}_{\rm YM}(\underline{\hat{3}},\underline{4},\cdots,j,\underline{\hat{P}}^{\W\epsilon^{M'}}_{j+1:2})\,,
\eea
therefore
\bea
& &{\cal T}^\epsilon[1,2,\cdots,n]\Big(\sum_{\pmb{\alpha}_1}\,\sum_{\epsilon^M\W\epsilon^{M'}}\,{\cal A}^{\epsilon,\W\epsilon}_{\rm GR}(\{j+1,\cdots,\underline{\hat{1}},\underline{\hat{2}},\underline{\hat{P}}^{\W\epsilon^{M'}}_{3:j}\}){1\over P^2_{j+1:2}}{\cal A}^{\epsilon,\W\epsilon}_{\rm GR}(\{\underline{\hat{3}},\underline{4},\cdots,j,\underline{\hat{P}}^{\W\epsilon^{M'}}_{j+1:2}\})\Big)\nn
&=&\sum_{\pmb{\alpha}_1}\,\sum_{\W\epsilon^{M'}}\,{\cal A}^{\W\epsilon}_{\rm YM}(j+1,\cdots,\underline{\hat{1}},\underline{\hat{2}},\underline{\hat{P}}^{\W\epsilon^{M'}}_{3:j}){1\over P^2_{j+1:2}}{\cal A}^{\W\epsilon}_{\rm YM}(\underline{\hat{3}},\underline{4},\cdots,j,\underline{\hat{P}}^{\W\epsilon^{M'}}_{j+1:2})\,,
\eea
where $P_{a:b}$ denotes $\sum_{i=a}^bk_i$.
For non-vanishing terms under the action of the trace operator ${\cal T}^\epsilon[1,2,\cdots,n]$,  summing over proper separations $\{\{1,2,{\pmb\alpha}_1\},\{3,4,{\pmb\beta}_1\}\}$ equivalents to summing over $j$. Thus the factorized YM terms for the first type of channels are given as
\bea
\sum_{j=4}^n\sum_{\W\epsilon^{M'}}\,{\cal A}^{\W\epsilon}_{\rm YM}(j+1,\cdots,\underline{\hat{1}},\underline{2},\underline{\hat{P}}^{\W\epsilon^{M'}}_{3:j}){1\over P^2_{j+1:2}}{\cal A}^{\W\epsilon}_{\rm YM}(\underline{\hat{3}},\underline{4},\cdots,j,\underline{\hat{P}}^{\W\epsilon^{M'}}_{j+1:2})\,.~~~~\label{fact-YM-channel1}
\eea

Then we apply  ${\cal T}^\epsilon[1,2,\cdots,n]$ to the second line at the RHS of \eref{phy-pole-YM-gen}. Similar as before, we factorize ${\cal T}^\epsilon[1,3]$ as
\bea
\cancel{\sum}_{\epsilon^M}\,\cdot{\cal T}^\epsilon[1,P_{23{\pmb\beta}_2}]\cdot{\cal T}^\epsilon[3,P_{14{\pmb\alpha}_2}]
\cong\cancel{\sum}_{\epsilon^M}\,\cdot{\partial\over\partial\epsilon_1\cdot\epsilon^M_{P_{23{\pmb\beta}_2}}}\cdot{\partial\over\partial\epsilon_3\cdot\epsilon^M_{P_{14{\pmb\alpha}_2}}}\,,
\eea
and split insertion operators as
\bea
& &{\cal I}^\epsilon_{123}={\cal I}^\epsilon_{12P_{14{\pmb\alpha}_2}}+{\cal I}^\epsilon_{P_{14{\pmb\alpha}_2}23}\,,\nn
& &{\cal I}^\epsilon_{(i-1)i1}={\cal I}^\epsilon_{(i-1)iP_{14{\pmb\alpha}_2}}+{\cal I}^\epsilon_{P_{14{\pmb\alpha}_2}i1}\,,
~~~~{\rm for}~i\in{\pmb\beta}_2\,,\nn
& &{\cal I}^\epsilon_{(i-1)i1}={\cal I}^\epsilon_{(i-1)iP_{23{\pmb\beta}_2}}+{\cal I}^\epsilon_{P_{23{\pmb\beta}_2}i1}\,,
~~~~{\rm for}~i-1\in\{3,{\pmb\beta}_2\}\,,~i\in\{4,{\pmb\alpha}_2\}\,,\nn
& &{\cal I}^\epsilon_{(i-1)i1}={\cal I}^\epsilon_{(i-1)i1}\,,
~~~~~~~~~~~~~~~~~~~~~~{\rm for}~i-1\in{\pmb\alpha}_2\,,~i\in{\pmb\alpha}_2\,.
\eea
The effective operators can be extracted as
\bea
{\cal T}^\epsilon[1,2,\cdots, n]\cong \cancel{\sum}_{\epsilon^M}\,\cdot{\cal T}'^\epsilon_L\cdot{\cal T}'^\epsilon_R\,,
\eea
where
\bea
& &{\cal T}'^\epsilon_L={\cal T}^\epsilon[1,P_{23{\pmb\beta}_2}]\cdot\Big(\prod_{\substack{i-1\in\{3,{\pmb\beta}_2\}\\i\in\{4,{\pmb\alpha}_2\}}}\,{\cal I}^\epsilon_{P_{23{\pmb\beta}_2}i1}\Big)\cdot\Big(\prod_{\substack{i-1\in{\pmb\alpha}_2\\i\in{\pmb\alpha}_2}}\,{\cal I}^\epsilon_{(i-1)i1}\Big)\,,\nn
& &
{\cal T}'^\epsilon_R={\cal T}^\epsilon[3,P_{14{\pmb\alpha}_1}]\cdot{\cal I}^\epsilon_{P_{14{\pmb\alpha}_2}23}\cdot\prod_{i\in{\pmb\beta}_2}\,{\cal I}^\epsilon_{(i-1)iP_{14{\pmb\alpha}_2}}\,.
\eea
If the operator ${\cal I}^\epsilon_{(i-1)iP_{14{\pmb\alpha}_2}}$ does not annihilate ${\cal A}_R$, not only the leg $i$ but also $(i-1)$ should be included in ${\cal A}_R$. The recursive pattern can not be satisfied since the leg $4$ is included in ${\cal A}_L$. Thus we conclude that the non-vanishing contribution requires ${\pmb\beta}_2=\emptyset$. Thus two trace operators are identified as
\bea
& &{\cal T}'^\epsilon_L={\cal T}^\epsilon[1,P_{23{\pmb\beta}_2},4,\cdots,n]\,,\nn
& &
{\cal T}'^\epsilon_R={\cal T}^\epsilon[2,3,P_{12{\pmb\alpha}_2}]\,.
\eea
Using these two operators, we find that the factorized YM terms for the second type of separations are given as follows
\bea
& &{\cal T}^\epsilon[1,2,\cdots,n]\Big(\sum_{\pmb{\alpha}_2}\,
\sum_{\epsilon^M\W\epsilon^{M'}}\,{\cal A}^{\epsilon,\W\epsilon}_{\rm GR}(\{\underline{\hat{1}},\underline{4},{\pmb\alpha}_2,\underline{\hat{P}}_{23{\pmb\beta}_2}^{\epsilon^M,\W\epsilon^{M'}}\}){1\over P^2_{14{\pmb\alpha}_2}}
{\cal A}^{\epsilon,\W\epsilon}_{\rm GR}(\{\underline{2},\underline{\hat{3}},{\pmb\beta}_2,\underline{\hat{P}}_{14{\pmb\alpha}_2}^{\epsilon^M,\W\epsilon^{M'}}\})\Big)\nn
&=&\sum_{\W\epsilon^{M'}}\,{\cal A}^{\W\epsilon}_{\rm YM}(\underline{4},\cdots,n,\underline{\hat{1}},\underline{\hat{P}}^{\W\epsilon^{M'}}_{23}){1\over P^2_{23}}{\cal A}^{\W\epsilon}_{\rm YM}(\underline{2},\underline{\hat{3}},\underline{\hat{P}}^{\W\epsilon^{M'}}_{4:1})\,.~~~~\label{fact-YM-channel2}
\eea

Combining results in \eref{fact-YM-channel1} and \eref{fact-YM-channel2} together, we find that applying the trace operator ${\cal T}^\epsilon[1,2,\cdots,n]$ to the first two lines at the RHS of \eref{phy-pole-YM-gen} gives the factorized formula
\bea
{\cal A}^{\epsilon}_{\rm YM}(1,\cdots,n)&=&\sum_{i=4}^n\sum_{\epsilon^{M}}\,{\cal A}^{\epsilon}_{\rm YM}(i+1,\cdots,\underline{\hat{1}},\underline{2},\underline{\hat{P}}^{\epsilon^{M}}_{3:i}){1\over P^2_{i+1:2}}{\cal A}^{\epsilon}_{\rm YM}(\underline{\hat{3}},\underline{4},\cdots,i,\underline{\hat{P}}^{\epsilon^{M}}_{i+1:2})\nn
& &+\sum_{\epsilon^{M}}\,{\cal A}^{\epsilon}_{\rm YM}(\underline{4},\cdots,n,\underline{\hat{1}},\underline{\hat{P}}^{\epsilon^{M}}_{23}){1\over P^2_{23}}{\cal A}^{\epsilon}_{\rm YM}(\underline{2},\underline{\hat{3}},\underline{\hat{P}}^{\epsilon^{M}}_{4:1})\nn
& &+\cdots\,,~~~~\label{phy-pole-YM-gen2}
\eea
which reproduces the first two lines in the result in \cite{Bjerrum-Bohr:2018lpz}.

We can compare the factorized formula for the YM amplitude in \eref{phy-pole-YM-gen2} with the factorized formula for the GR amplitude in \eref{phy-pole-YM-gen}, and understand the relationship between them by using the trace operator ${\cal T}^\epsilon[1,2,\cdots,n]$. One can see that the set of factorization channels in \eref{phy-pole-YM-gen2} is a subset of factorization channels in \eref{phy-pole-YM-gen},
because the trace operator ${\cal T}^\epsilon[1,2,\cdots,n]$ selects these channels by annihilating other channels which are not compatible with the color-ordering $(1,2,\cdots,n)$. Then, the resulted channels are fully determined by the color-ordering, as required by the definition of the color-ordered amplitude. For instance, the channel corresponds to the combinatory momentum $P_{235}=k_2+k_3+k_5$, which appears in the second line of \eref{phy-pole-YM-gen}, is not permitted by the color-ordering $(1,2,3,4\cdots,n)$. As discussed before, the trace operator annihilates all terms in the second line of \eref{phy-pole-YM-gen} except ${\pmb\beta}_2=\emptyset$. Thus the trace operator eliminates the $P_{235}$-channel and other channels which are not compatible with the color-ordering, only leaves the compatible $P_{23}$-channel.

The denominates $P_{i+1:2}^2$ and $P_{23}^2$, which provide physical poles for GR amplitudes in \eref{phy-pole-YM-gen}, also serve as physical poles for YM amplitudes in \eref{phy-pole-YM-gen2}, as can be verified directly. The reason has been explained in previous subsections, the trace operator will not create or annihilate any pole, thus transmutes physical poles to physical poles.

The gauge choices for sub-amplitudes ${\cal A}_L$ and ${\cal A}_R$ in \eref{phy-pole-YM-gen2} are the same as that for the corresponding sub-amplitudes in \eref{phy-pole-YM-gen}. For example, the gauge choice for ${\cal A}_L$ in the second line of \eref{phy-pole-YM-gen2} is fixing $\sigma_4$, $\sigma_1$, $\sigma_{P_{23}}$, removing $1^{\rm th}$ and $P_{23}^{\rm th}$ rows and columns in the reduced matrix. For the corresponding part with ${\pmb\beta}_2=\emptyset$ in the second line of \eref{phy-pole-YM-gen}, the gauge choice is totally the same. As discussed in subsection \ref{effect-OP}, for the descendent relation of fixed punctures, the underlying reason is that the trace operator is commutable with the CHY contour integral, thus only acts on the integrand, leaves the measure un-affected. For the removed rows and columns, the descendent relation can be understood as, the trace operator defined via polarization vectors $\epsilon_i$ affect only one of two reduced Pfaffians in the GR integrand which carries $\epsilon_i$, thus another un-altered reduced Pfaffian carries the choice of removed rows and columns to the YM integrand.

It is necessary to notice that the factorized formula for an amplitude obtained by the double-cover method depend on the fixed punctures $(p,q,r|m)$ in \eref{DCamp}, and the removed rows and columns in the reduced matrices, i.e., the factorized formula is not unique. Thus we can not expect the relations mentioned above exist for general factorized formulae for the GR and YM amplitudes. Instead, we need to figure out the proper gauge choices which lead to the factorized formula appear in these relations. The gauge choice in the double-cover prescription for the GR amplitude, was chosen as $(1,2,3|4)$, $(\Psi^\Lambda_n)^{13}_{13}$ and $(\W\Psi^\Lambda_n)^{13}_{13}$ at the beginning. The proper gauge choice which we need to seek is for the YM amplitude. Since the above relations are indicated by the trace operator, it is equivalent to ask: when acting the trace operator to the double-cover integral \eref{DCamp} for GR, what gauge choice for the YM amplitude should be created? This generated gauge choice is consistent with the above relations. The solution is $(1,2,3|4)$ and $(\W\Psi^\Lambda_n)^{13}_{13}$, which was used in \cite{Bjerrum-Bohr:2018lpz} to derive the result in \eref{phy-pole-YM-gen2}. The reason can be explained as follows. As discussed in subsection \ref{differ-OP}, the differential operators will not affect the measure part and the denominate of the integrand. Thus we conclude that under the action of differential operators, the choice of $(p,q,r|m)$ will be carried from GR to YM.
Since the trace operator only acts on one reduced Pfaffian, another reduced Pfaffian will not be changed. Thus the choice of removed rows and columns for another reduced Pfaffian will be transmitted from the GR amplitude to the YM amplitude.
Thus, the relations discussed above exist for the factorized YM amplitude obtained by the double-cover prescription \eref{DCamp} with the gauge choice $(1,2,3|4)$ and $(\W\Psi^\Lambda_n)^{13}_{13}$, which is inherited from the gauge choice for the GR amplitude in the double-cover prescription.
In general, for any gauge choice for the GR amplitude in the double-cover prescription, if the gauge choice for the YM amplitude in the double-cover prescription is inherited from the GR amplitude, the relations discussed above are always correct.

The factorized formula corresponds to physical poles obtained in \eref{phy-pole-YM-gen2} arise from two types of separations, $\{\{1,2,{\pmb\alpha}_1\},\{3,4,{\pmb\beta}_1\}\}$ and $\{\{1,4,{\pmb\alpha}_2\},\{2,3,{\pmb\beta}_2\}\}$. Before ending this subsection, we give a brief discussion about the remaining third type separations $\{\{1,3,{\pmb\alpha}_3\},\{2,4,{\pmb\beta}_3\}\}$. It is easy to generalize results in \eref{ms-part} and \eref{PT-part} for the $4$-point example
to the current general case. But the generalization of the result in \eref{pf-part} is really hard. However, we can still discuss some general properties for the factorized formula corresponds to the current separation. First, one can still conclude that the propagator arise from the scattering equation ${\cal E}^\tau_4$ will be canceled by the leading order term of
$({\bf Pf}'\W\Psi_n)^\tau$, therefore provides a spurious pole for both GR and YM amplitudes. Secondly, the separation indicates that coordinates $\sigma_1,\sigma_3$ on one sheet while $\sigma_2,\sigma_4$ on another sheet, thus we know that legs $1,3$ are included in one sub-amplitude, legs $2,4$ are included in another one.
In \cite{Bjerrum-Bohr:2018lpz}, the corresponding YM terms are conjectured as
\bea
-2\sum_{i=4}^n\sum_{\W\epsilon^L}\,{\cal A}^{\W\epsilon}_{\rm YM}(i+1,\cdots,\underline{\hat{1}},\underline{3},\underline{\hat{P}}^{\W\epsilon^{L}}_{2(4:i)}){1\over P^2_{(i+1:1)3}}{\cal A}^{\W\epsilon}_{\rm YM}(\underline{\hat{2}},\underline{4},\cdots,i,\underline{\hat{P}}^{\W\epsilon^{L}}_{(i+1:1)3})\,,
\eea
but the proof is still lacking.

In summary, the factorization for the general color-ordered YM amplitude is given as
\bea
{\cal A}^{\epsilon}_{\rm YM}(1,\cdots,n)&=&\sum_{\epsilon^M}\,{\cal A}^{\epsilon}_{\rm YM}(\underline{4},\cdots,n,\underline{\hat{1}},\underline{\hat{P}}^{\epsilon^{M}}_{23}){1\over P^2_{23}}{\cal A}^{\epsilon}_{\rm YM}(\underline{2},\underline{\hat{3}},\underline{\hat{P}}^{\epsilon^{M}}_{4:1})\nn
& &+\sum_{i=4}^n\sum_{\epsilon^M}\,{\cal A}^{\epsilon}_{\rm YM}(i+1,\cdots,\underline{\hat{1}},\underline{2},\underline{\hat{P}}^{\epsilon^{M}}_{3:i}){1\over P^2_{i+1:2}}{\cal A}^{\epsilon}_{\rm YM}(\underline{\hat{3}},\underline{4},\cdots,i,\underline{\hat{P}}^{\epsilon^{M}}_{i+1:2})\nn
& &-2\sum_{i=4}^n\sum_{\epsilon^L}\,{\cal A}^{\epsilon}_{\rm YM}(i+1,\cdots,\underline{\hat{1}},\underline{3},\underline{\hat{P}}^{\epsilon^{L}}_{2(4:i)}){1\over P^2_{(i+1:1)3}}{\cal A}^{\epsilon}_{\rm YM}(\underline{\hat{2}},\underline{4},\cdots,i,\underline{\hat{P}}^{\epsilon^{L}}_{(i+1:1)3})\,.~~~~\label{fct-YM}
\eea
The first two lines can be derived by our method illustrated in this paper, while the third line is a conjecture.

%%%%%%%%%%%%%%%%%%%%%%%%%%%%
\subsection{BCFW recursion}
\label{BCFW-YM}
%%%%%%%%%%%%%%%%%%%%%%%%%%%%%

In this subsection, we discuss the relationship between the BCFW recursion relation \cite{Britto:2004ap,Britto:2005fq,Feng:2011np} for the color-ordered YM amplitude, and the factorized formula in \eref{fct-YM}, in a manner different from that in \cite{Bjerrum-Bohr:2018lpz}. Our manner allows us to relate the recursive part and the boundary term to corresponding terms in \eref{fct-YM} without the conjectured explicit formula in the last line of \eref{fct-YM}. In other words, in our discussion, the ignorance of terms correspond to spurious poles will not affect the recognizing of the recursive part and the boundary term. In this subsection, we only focus on the BCFW recursion for YM amplitudes, thus will not discuss the effect of the trace operator. In the next section, we will show that the differential operators link the BCFW recursion for the YM amplitude to the BCFW recursions for NLSM and BAS amplitudes.

We first analyse the pole-structures of the recursive part and the boundary term. Suppose the BCFW deformation is chosen as
\bea
k_i(z)=k_i+zq\,,~~~~~~k_j(z)=k_j-zq\,,~~~~\label{BCFW-shift}
\eea
with the on-shell condition $q^2=k_i\cdot q=k_j\cdot q=0$. This deformation divides all physical poles $P_t^2$ of a tree amplitude into two categories. Physical poles in the first category depend on $z$, therefore are detectable by the BCFW deformation in \eref{BCFW-shift}. Physical poles in the second category are independent of $z$ thus are un-detectable. We denote the first set by ${\cal D}$ and the second set by ${\cal U}$. As a rational function of $z$, an amplitude under the deformation \eref{BCFW-shift} can be decomposed as
\bea
{\pmb A}(z)={N(z)\over \prod\,P_t^2(z)}=-\sum_{P_t^2\in{\cal D}}\,{{\pmb A}_L(z_t){\pmb A}_R(z_t)\over P_t^2(z)}+C_0+\sum_i\,C_iz^i\,,~~~~\label{decomp-amp}
\eea
where $z_t$ denotes the special value of $z$ satisfies the on-shell condition $P_t^2(z_t)=0$. The physical amplitude is evaluated at $z=0$, which can be expressed as
\bea
{\pmb A}(0)=-\sum_{P_t^2\in{\cal D}}\,{{\pmb A}_L(z_t){\pmb A}_R(z_t)\over P_t^2(z)}\Big|_{z=0}+C_0\,.
\eea
Considering the contour integral
\bea
\oint\,{{\pmb A}_L(z_t){\pmb A}_R(z_t)\over zP_t^2(z)}\,,
\eea
where the contour encircles poles $z=0$ and $P_t^2(z_t)=0$, one can get
\bea
-{{\pmb A}_L(z_t){\pmb A}_R(z_t)\over P_t^2(z)}\Big|_{z=0}={{\pmb A}_L(z_t){\pmb A}_R(z_t)\over P_t^2}\,,
\eea
Thus the full physical amplitude can be expressed by the BCFW recursion relation
\bea
{\pmb A}(0)=\sum_{P_t^2\in{\cal D}}\,{{\pmb A}_L(z_t){\pmb A}_R(z_t)\over P_t^2}+C_0\,.~~~~\label{recursion}
\eea
In the above expression, the first part at the RHS can be evaluated recursively, thus is called the recursive part. The second part $C_0$ is called the boundary term. The above definition of the recursive part and the boundary term is obviously equivalent to the standard definition in \cite{Britto:2004ap,Britto:2005fq,Feng:2011np}.
The above discussion shows that detectable poles in ${\cal D}$ can only be contained in the recursive part, while the boundary term only contains un-detectable poles in ${\cal U}$.

With the understanding of the pole-structures, we now discuss how terms in the factorized formula \eref{fct-YM} contribute to the recursive part and the boundary term. The recursive part and the boundary term depend on the choice of BCFW deformation. We will show that under the special deformation
\bea
k_2(z)=k_2+zq,~~~~~~k_3(z)=k_3-zq\,,~~~~\label{special-shift}
\eea
there is an elegant correspondence between terms in the factorized formula \eref{fct-YM} and terms in the BCFW recursion relation \eref{recursion}.

Under the deformation in \eref{special-shift}, the detectable physical poles of the color-ordered YM amplitude ${\pmb A}^\epsilon_{\rm YM}(1,\cdots,n)$, which will depend on $z$, are $P_a^2=(\sum_{l=3}^ak_l)^2$ with $4\leq a\leq n$ (or equivalently $P_b^2=(\sum_{l=b}^2k_l)^2$ with $5\leq b\leq 1$).
All these detectable physical poles require legs $3$ and $4$ to be included in one sub-amplitude, legs $1$ and $2$ to be included in another sub-amplitude. This observation indicates that in the first line of \eref{fct-YM}, not only the denominate $P_{23}^2$, but also poles contained in sub-amplitudes ${\cal A}_L$ and ${\cal A}_R$
are un-detectable. Thus the first line of \eref{fct-YM} contributes to the boundary term. On the other hand, all terms in the second line of \eref{fct-YM} contain detectable poles thus contribute to the recursive part. For terms with spurious poles, the general discussion about this part in subsection \ref{general-YM} shows that legs $1,3$ belong to one sub-amplitude while legs $2,4$ belong to another one. This configuration excludes all detectable physical poles. Thus this part can be recognized as the boundary contribution without the explicit formula.

Furthermore, as explained at the end of subsection \ref{property-OS}, for an off-shell amplitude ${\cal A}^{\epsilon}_{\rm YM}(\underline{\hat{3}},\underline{4},\cdots,i,\underline{\hat{P}}^{\epsilon^{M}}_{i+1:2})$ in the second line of \eref{fct-YM}, it does not contain any pole $(k_3+k_4+K)^2$. Similarly, an off-shell amplitude ${\cal A}^{\epsilon}_{\rm YM}(i+1,\cdots,\underline{\hat{1}},\underline{2},\underline{\hat{P}}^{\epsilon^{M}}_{3:i})$ does not contain any pole $(k_1+k_2+K')^2$. This observation indicates that one term
\bea
{\cal A}^{\epsilon}_{\rm YM}(i+1,\cdots,\underline{\hat{1}},\underline{2},\underline{\hat{P}}^{\epsilon^{M}}_{3:i}){1\over P^2_{i+1:2}}{\cal A}^{\epsilon}_{\rm YM}(\underline{\hat{3}},\underline{4},\cdots,i,\underline{\hat{P}}^{\epsilon^{M}}_{i+1:2})
\eea
in the second line of \eref{fct-YM} contains only one detectable pole $P^2_{i+1:2}$, thus only contributes to the residue at $P^2_{i+1:2}(z_{i+1:2})=0$.
Consequently, there is a one to one map from terms in the second line of \eref{fct-YM} to recursive terms
\bea
{\pmb A}^{\epsilon}_{\rm YM}(i+1,\cdots,1,2(z_{i+1:2}),P^{\epsilon^{M}}_{3:i}(z_{i+1:2})){1\over P^2_{i+1:2}}{\pmb A}^{\epsilon}_{\rm YM}(3(z_{i+1:2}),4,\cdots,i,P^{\epsilon^{M}}_{i+1:2}(z_{i+1:2}))
\eea
in the BCFW recursion relation \eref{recursion}.

Thus we arrive at the conclusion that under the BCFW deformation in \eref{special-shift}, the first and third lines correspond to the boundary term in the BCFW recursion, and the terms in the second line have the one to one correspondence to terms in the recursive part. Since the discussion is independent of the conjectured formula in the last line of \eref{fct-YM}, our conclusion is strict.

At the end of this subsection, we emphasize again that the recursive part and the boundary term depend on the choice of deformation. For example, if we deform $k_1$ and $k_3$, physical poles $(\sum_{l=c}^1 k_l)^2$ with $4\leq c\leq n$ become detectable. Sub-amplitudes in all three lines in \eref{fct-YM} can contain some of these poles therefore can contribute to the recursive part. Furthermore, under the current deformation, each term at the RHS of \eref{fct-YM} can contain more than one detectable physical pole, thus contributes to more than one term in the recursive part. Then the relationship between terms in \eref{fct-YM} and terms in the BCFW recursion becomes complicated. Thus, the elegant correspondence found under the deformation in \eref{special-shift} does not hold for general deformations.

%%%%%%%%%%%%%%%%%%%%%%%%%%%%%%%%%%%%%%%
\section{From factorization for YM to other theories}
\label{secYMother}
%%%%%%%%%%%%%%%%%%%%%%%%%%%%%%%%%%%%%%%

In this section, we apply differential operators to the factorized formula for the YM amplitude, to generate the factorizations for other theories.
We will consider amplitudes of three theories, sYMS, NLSM, BAS, which are generated from the YM amplitude via three types of operators ${\cal T}_{{\cal X}_{2m}}$, ${\cal L}\cdot{\cal T}[a,b]$, ${\cal T}[i_1,\cdots,i_n]$, respectively. In the factorized formula for the YM amplitude given in \eref{fct-YM}, the first two lines for physical poles are strict, while the conjectured third line for spurious poles is only strict for the $4$-point case. For the sYMS case, the third line of \eref{fct-YM} will contribute, thus we only consider $4$-point examples. For NLSM and BAS cases, the third line of \eref{fct-YM} will not contribute, and we will give the general results. The effects of differential operators will be discussed through four angles: factorization channels, pole-structures, gauge choices, as well as BCFW recursions. The proper gauge choices for sYMS, NLSM and BAS amplitudes in the double-cover prescription will also be discussed.

%%%%%%%%%%%%%%%%%%%%%%%%%%%%%%%%%%%%%%%%%%%
\subsection{Factorization for sYMS amplitude}
\label{subsecsYMS}
%%%%%%%%%%%%%%%%%%%%%%%%%%%%%%%%%%%%%%%%%%%%

From Table \ref{tab:unifying}, one can see the color-ordered sYMS  amplitude can be generated from the YM amplitude as
\bea
& &{\pmb A}^\epsilon_{\rm sYMS}({\pmb S}_{2m}||{\pmb G}_{n-2m};1,\cdots,n)={\cal T}^\epsilon_{{\cal X}_{2m}}\,
{\pmb A}^\epsilon_{\rm YM}(1,\cdots,n)\,.
\eea
In this subsection, we apply operators ${\cal T}^\epsilon_{{\cal X}_{2m}}$ to the factorized $4$-point YM
amplitude, to generate the factorizations for the $4$-point sYMS amplitudes. We will consider three $4$-point examples. Notice that the explicit formula of the operator ${\cal T}^\epsilon_{{\cal X}_{2m}}$ depend on the number and the flavors of scalar particles. Thus although all three examples are $4$-point amplitudes, the operators ${\cal T}^\epsilon_{{\cal X}_{2m}}$ in three examples are different to each other.

Our first example is the color-ordered $4$-point sYMS amplitude ${\pmb A}^\epsilon_{\rm sYMS}(1_g,2_g,3_s,4_s)$, which contains two gluons $1_g,2_g$ and two scalar particles $3_s,4_s$. This amplitude can be obtained by acting ${\cal T}^\epsilon_{{\cal X}_2}={\cal T}^\epsilon[3,4]$ on the YM amplitude ${\pmb A}^\epsilon_{\rm YM}(1,2,3,4)$. To get the factorized formula for ${\pmb A}^\epsilon_{\rm sYMS}(1_g,2_g,3_s,4_s)$, we apply the operator ${\cal T}^\epsilon[3,4]$ to the RHS of \eref{fct-YM-4p}. For the first line, ${\cal T}^\epsilon[3,4]$
only acts on ${\cal A}_R$, and gives
\bea
\sum_{\epsilon^M}\,{\cal A}^\epsilon_{\rm YM}(\underline{\hat{1}},\underline{2},\underline{\hat{P}}^{\epsilon^M}_{34}){1\over P_{12}^2}{\cal A}^\epsilon_{\rm sYMS}(\underline{\hat{3}}_s,\underline{4}_s,(\underline{\hat{P}}_{12})_g^{\epsilon^M})\,.
\eea
For the second line of \eref{fct-YM-4p}, we factorize ${\cal T}^\epsilon[3,4]$ as $\cancel{\sum}_{\epsilon^M}\cdot{\cal T}^\epsilon[3,P_{14}]\cdot{\cal T}^\epsilon[4,P_{23}]$. The operator ${\cal T}^\epsilon[3,P_{14}]$ only acts on ${\cal A}_L$, ${\cal T}^\epsilon[4,P_{23}]$ only acts on ${\cal A}_R$. Thus we get
\bea
{\cal A}^\epsilon_{\rm sYMS}(\underline{2}_g,\underline{\hat{3}}_s,(\underline{\hat{P}}_{14})_s){1\over P_{23}^2}
{\cal A}^\epsilon_{\rm sYMS}(\underline{4}_s,\underline{\hat{1}}_g,(\underline{\hat{P}}_{23})_s)\,.
\eea
The last line of \eref{fct-YM-4p} will be annihilated by ${\cal T}^\epsilon[3,4]$, since it can not be factorized into operators act on ${\cal A}_L$ and ${\cal A}_R$ respectively, and $\epsilon_3$ can not contract with $\epsilon_4$ across the propagator ${1\over P_{13}^2}$. Combining the results for three lines together, we obtain
\bea
{\pmb A}^\epsilon_{\rm sYMS}(1_g,2_g,3_s,4_s)&=&\sum_{\epsilon^M}\,{\cal A}^\epsilon_{\rm YM}(\underline{\hat{1}},\underline{2},\underline{\hat{P}}^{\epsilon^M}_{34}){1\over P_{12}^2}{\cal A}^\epsilon_{\rm sYMS}(\underline{\hat{3}}_s,\underline{4}_s,(\underline{\hat{P}}_{12})_g^{\epsilon^M})\nn
& &+{\cal A}^\epsilon_{\rm sYMS}(\underline{2}_g,\underline{\hat{3}}_s,(\underline{\hat{P}}_{14})_s){1\over P_{23}^2}
{\cal A}^\epsilon_{\rm sYMS}(\underline{4}_s,\underline{\hat{1}}_g,(\underline{\hat{P}}_{23})_s)\,,~~~~\label{fct-sYMS-1}
\eea
which reproduces Eq.(9.9) in \cite{Gomez:2018cqg}.

The second example is the color-ordered sYMS amplitude ${\pmb A}_{\rm sYMS}(1^{I_1}_s,2^{I_2}_s,3^{I_1}_s,4^{I_2}_s)$, which includes four scalar particles, two scalar particles $1$ and $3$ carry the
flavor $I_1$, two scalar particles $2$ and $4$ carry the flavor $I_2$. This amplitude can be generated by acting ${\cal T}^\epsilon_{{\cal X}_4}={\cal T}^\epsilon[1,3]\cdot{\cal T}^\epsilon[2,4]$ on the YM amplitude ${\pmb A}^\epsilon_{\rm YM}(1,2,3,4)$. The operator ${\cal T}^\epsilon[1,3]{\cal T}^\epsilon[2,4]$ annihilates the first line of \eref{fct-YM-4p}, since if we factorize ${\cal T}^\epsilon[1,3]$ as $\cancel{\sum}_{\epsilon^M}\cdot{\cal T}^\epsilon[3,P_{14}]\cdot
{\cal T}^\epsilon[1,P_{23}]$ and apply it at the first step, then $\epsilon^M_{P_{14}}$ and $\epsilon^M_{P_{23}}$ will be removed from ${\cal A}_L$
and ${\cal A}_R$. The remaining object will be annihilated by ${\cal T}^\epsilon[2,4]$, since the similar factorization for the operator ${\cal T}^\epsilon[2,4]$ can not work without $\epsilon^M_{P_{14}}$ and $\epsilon^M_{P_{23}}$, and $\epsilon_2$ can not contract with $\epsilon_4$ across the propagator ${1\over P_{12}^2}$. For the same reason, the second line of \eref{fct-YM-4p} will also be annihilated. The non-vanishing contribution arises from the third line, which is given by
\bea
{\pmb A}_{\rm sYMS}(1^{I_1}_s,2^{I_2}_s,3^{I_1}_s,4^{I_2}_s)=-2\sum_{\epsilon^L}\,{\cal A}^{\epsilon}_{\rm sYMS}(\underline{\hat{1}}_s,\underline{3}_s,(\underline{\hat{P}}_{24})^{\epsilon^L}_g){1\over P_{13}^2}
{\cal A}^{\epsilon}_{\rm sYMS}(\underline{\hat{2}}_s,\underline{4}_s,(\underline{\hat{P}}_{13})^{\epsilon^L}_g)\,.~~~\label{syms-resul1}
\eea
In the literature \cite{Gomez:2018cqg}, the corresponding result is given as
\bea
{\pmb A}_{\rm sYMS}(1^{I_1}_s,2^{I_2}_s,3^{I_1}_s,4^{I_2}_s)=2\sum_{\epsilon^L}\,{\cal A}^{\epsilon}_{\rm sYMS}(\underline{\hat{1}}_s,\underline{4}_s,(\underline{\hat{P}}_{23})^{\epsilon^L}_g){1\over P_{14}^2}
{\cal A}^{\epsilon}_{\rm sYMS}(\underline{\hat{2}}_s,\underline{3}_s,(\underline{\hat{P}}_{14})^{\epsilon^L}_g)\,.~~~\label{syms-resul2}
\eea
Two formulae \eref{syms-resul1} and \eref{syms-resul2} contain different spurious poles. As can be verified straightforwardly, both of them give ${1\over2}$, thus two formulae are equal to each other. This is an example of that for the term with spurious pole the factorization channel is not unique.

The third example is the sYMS amplitude ${\pmb A}^\epsilon_{\rm sYMS}(1^{I_1}_s,2^{I_1}_s,3^{I_2}_s,4^{I_2}_s)$, which can be generated from ${\pmb A}^\epsilon_{\rm YM}(1,2,3,4)$ by applying ${\cal T}'^\epsilon_{{\cal X}_4}={\cal T}^\epsilon[1,2]\cdot{\cal T}^\epsilon[3,4]$. This operator annihilates the second and third lines in \eref{fct-YM-4p}. When acting on the first line, it gives
\bea
{\pmb A}_{\rm sYMS}(1^{I_1}_s,2^{I_1}_s,3^{I_2}_s,4^{I_2}_s)=\sum_{\epsilon^M}\,{\cal A}^\epsilon_{\rm sYMS}(\underline{\hat{1}}_s,\underline{2}_s,(\underline{\hat{P}}_{34})^{\epsilon^M}_g){1\over P_{12}^2}{\cal A}^\epsilon_{\rm sYMS}(\underline{\hat{3}}_s,\underline{4}_s,(\underline{\hat{P}}_{12})^{\epsilon^M}_g)\,.~~~~\label{syms-resul3}
\eea
In \cite{Gomez:2018cqg}, the corresponding expression is
\bea
{\pmb A}_{\rm sYMS}(1^{I_1}_s,2^{I_1}_s,3^{I_2}_s,4^{I_2}_s)={1\over2}+\sum_{\epsilon^M}\,{\cal A}^\epsilon_{\rm sYMS}(\underline{\hat{1}}_s,\underline{2}_s,(\underline{\hat{P}}_{34})^{\epsilon^M}_g){1\over P_{12}^2}{\cal A}^\epsilon_{\rm sYMS}(\underline{3}_s,\underline{\hat{4}}_s,(\underline{\hat{P}}_{12})^{\epsilon^M}_g)\,,~~~~\label{syms-resul4}
\eea
where ${1\over2}$ comes from the term with spurious pole similar as in \eref{syms-resul1} and \eref{syms-resul2}. One can check that
both \eref{syms-resul3} and \eref{syms-resul4} give $-{s_{13}\over 2s_{12}}$.

Let us give a brief discussion about how the operators ${\cal T}^\epsilon_{{\cal X}_{2m}}$ determine the properties of the factorization for sYMS amplitudes. The operator selects factorization channels from channels for the YM amplitude, by annihilating some lines in \eref{fct-YM-4p}. The definition of the operator ${\cal T}^\epsilon_{{\cal X}_{2m}}$ indicates that this operator will not create or annihilate any pole, therefore transmutes physical poles to physical poles, and transmutes spurious poles to spurious poles. Thus, as can be verified, poles $P_{12}^2$ and $P_{23}^2$ in \eref{fct-sYMS-1} will not be canceled by the numerators of sub-amplitudes therefore are physical poles, the same as in the factorized YM amplitude \eref{fct-YM-4p}. Similarly, $P_{13}^2$ in \eref{syms-resul1} is a spurious pole, $P_{12}^2$ in \eref{syms-resul3} is a physical pole, both of them are inherited from the factorized YM amplitude \eref{fct-YM-4p}. In three examples, the choices of fixed punctures for sub-amplitudes ${\cal A}_L$ and ${\cal A}_R$ at the sYMS side, are the same as those at the YM side in \eref{fct-YM-4p}, since the operator ${\cal T}^\epsilon_{{\cal X}_{2m}}$ and the CHY contour integral are commutable. The choices of removed rows and columns are also the same as in \eref{fct-YM-4p}, since the operator ${\cal T}^\epsilon_{{\cal X}_{2m}}$ will not affect the reduced sub-matrix $(A_4)^{ab}_{ab}$.

In sections \ref{secintro} and
\ref{general-YM}, we have pointed out that the factorization arise from the double-cover prescription \eref{DCamp} depend on the gauge choice. As discussed in subsection \ref{general-YM}, the gauge choice for the YM amplitude in the double-cover prescription is fixed as $(1,2,3|4)$, $(\Psi^\Lambda_4)^{13}_{13}$. Now we discuss the proper gauge choice in \eref{DCamp} for the sYMS amplitudes, which is consistent with the relations mentioned above. For the $4$-point examples, four fixed punctures have only one choice. Since the differential operator will not affect the denominate of the integrand, the choice of $(p,q,r|m)$ for the sYMS amplitudes is $(1,2,3|4)$, which is the same as that for the YM amplitude.
The removed rows and columns in the reduced matrix, which lead to our results \eref{fct-sYMS-1}, \eref{syms-resul1} and \eref{syms-resul3}, are $1$ and $3$. This choice is inherited from the choice $(\Psi^\Lambda_4)^{13}_{13}$ for the YM amplitude, since ${\cal T}^\epsilon_{{\cal X}_{2m}}$ will not affect the reduced sub-matrix $(A^\Lambda_4)^{13}_{13}$. The gauge choice for the YM amplitude is inherited from the GR amplitude.
Thus, we see that the gauge choice in the double-cover prescription has been transmitted from the GR amplitude to amplitudes of other theories via the differential operators.
In \cite{Gomez:2018cqg}, the second and third examples are obtained by $(A^\Lambda_4)^{14}_{14}$ with the gauge different from our choice, thus the formulae \eref{syms-resul2} and \eref{syms-resul4} are different from our results.

%%%%%%%%%%%%%%%%%%%%%%%%%%%%%%%%%%%%%%%%
\subsection{Factorization for NLSM amplitude}
\label{subsec-fct-NLSM}
%%%%%%%%%%%%%%%%%%%%%%%%%%%%%%%%%%%%%%%%%%%

Now we turn to the factorization for the color-ordered NLSM amplitude ${\pmb A}_{\rm NLSM}(1,\cdots,n)$, which can be generated from the YM amplitude ${\pmb A}^\epsilon_{\rm YM}(1,\cdots,n)$ via the operator ${\cal T}^\epsilon[a,b]\cdot {\cal L}^\epsilon$. To get the factorization for ${\pmb A}_{\rm NLSM}(1,\cdots,n)$, let us choose $a=1,b=3$, and apply the operator ${\cal T}^\epsilon[1,3]\cdot {\cal L}^\epsilon$ to the RHS of \eref{fct-YM}.

For the first line at the RHS of \eref{fct-YM}, we can factorize ${\cal T}^\epsilon[1,3]$ as $\cancel{\sum}_{\epsilon^M}\cdot{\cal T}^\epsilon[1,P_{23}]\cdot{\cal T}^\epsilon[3,P_{4:1}]$, where ${\cal T}^\epsilon[1,P_{23}]$ only acts  on ${\cal A}_L$, ${\cal T}^\epsilon[3,P_{4:1}]$
only acts on ${\cal A}_R$. On the other hand, the longitudinal operator ${\cal L}^\epsilon_i$ acts on ${\cal A}_L$ if the leg $i$ is contained in ${\cal A}_L$, and acts on ${\cal A}_R$ if the leg $i$ is contained in ${\cal A}_R$. Thus one can re-group the operators as
\bea
{\cal T}^\epsilon[1,3]\cdot {\cal L}^\epsilon\cong\cancel{\sum}_{\epsilon^M}\cdot\Big({\cal T}^\epsilon[1,P_{23}]\cdot {\cal L}^\epsilon_L\Big)\cdot\Big({\cal T}^\epsilon[3,P_{4:1}]\cdot {\cal L}^\epsilon_R\Big)\,,
\eea
where
\bea
& &{\cal L}^\epsilon_{L}=\prod_{i\in\{4,\cdots,1\}}\,{\cal L}^\epsilon_i\,,\nn
& &{\cal L}^\epsilon_{R}=\prod_{i\in\{2,3\}}\,{\cal L}^\epsilon_i\,.
\eea
The operator ${\cal T}^\epsilon[1,P_{23}]\cdot {\cal L}^\epsilon_L$ only acts on ${\cal A}_L$, while ${\cal T}^\epsilon[3,P_{4:1}]\cdot {\cal L}^\epsilon_R$
only acts on ${\cal A}_R$. Based on the discussion in subsection \ref{effect-OP}, we know that ${\cal T}^\epsilon[1,P_{23}]\cdot {\cal L}^\epsilon_L$ transmutes ${\cal A}_L$ to the off-shell NLSM amplitude $
{\cal A}_{\rm NLSM}(\underline{\bar{4}},\cdots,n,\underline{\bar{\hat{1}}},\underline{\hat{P}}_{23})$, where we use $\bar{i}$ and $\hat{j}$
to denote removed rows and columns in the reduced matrix $(A_n)^{i_1i_2}_{j_1j_2}$, respectively. Similarly, ${\cal T}^\epsilon[3,P_{4:1}]\cdot {\cal L}^\epsilon_R$ transmutes ${\cal A}_R$ to the off-shell NLSM amplitude $
{\cal A}_{\rm NLSM}(\underline{\bar{2}},\underline{\bar{\hat{3}}},\underline{\hat{P}}_{4:1})$. Thus the first line of \eref{fct-YM} is transmuted to
\bea
{\cal A}_{\rm NLSM}(\underline{\bar{4}},\cdots,n,\underline{\bar{\hat{1}}},\underline{\hat{P}}_{23}){1\over P_{23}^2}{\cal A}_{\rm NLSM}(\underline{\bar{2}},\underline{\bar{\hat{3}}},\underline{\hat{P}}_{4:1})\,.~~~~\label{NLSM-lin1}
\eea
Similar manipulation for the second line gives
\bea
\sum_{i=4}^n\,{\cal A}_{\rm NLSM}(i+1\cdots\underline{\bar{\hat{1}}},\underline{\bar{2}},\underline{\hat{P}}_{3:i}){1\over P_{i+1:2}^2}{\cal A}_{\rm NLSM}(\underline{\bar{\hat{3}}},\underline{\bar{4}},\cdots,i,\underline{\hat{P}}_{i+1:2})\,.~~~~\label{NLSM-lin2}
\eea

In the third line of \eref{fct-YM}, the expression is a conjecture rather than a strict result. However, it is easy to conclude that this part will be annihilated by ${\cal T}^\epsilon[1,3]\cdot{\cal L}^\epsilon$. The argument is independent of the explicit formula for the YM terms. It is clear that both legs $1$ and $3$ are included in ${\cal A}_L$ since the corresponding two punctures are on the same sheet, as discussed in subsection \ref{4-point-YM-spur}. Thus ${\cal T}^\epsilon[1,3]$ only acts on ${\cal A}_L$. Then, suppose the number of external legs of ${\cal A}_R$ is $m$, one can see there are $(m-1)$ longitudinal operators ${\cal L}^\epsilon_i$ act on ${\cal A}_R$, due to the definition of ${\cal L}^\epsilon$. The mass dimension of $m$-point YM amplitude is $(4-m)$,
the propagators contribute $(6-2m)$, thus the numerator contributes $(m-2)$. Thus, if all $(m-1)$ polarization vectors $\epsilon_i$ are contracted with momenta, the correct mass dimension will be violated. This fact indicates that ${\cal A}_R$ will be annihilated by $(m-1)$ longitudinal operators.

Consequently, the factorization for NLSM amplitudes is obtained by combining \eref{NLSM-lin1} and \eref{NLSM-lin2}, which is given as
\bea
{\pmb A}_{\rm NLSM}(1,\cdots,n)&=&{\cal A}_{\rm NLSM}(\underline{\bar{4}},\cdots,n,\underline{\bar{\hat{1}}},\underline{\hat{P}}_{23}){1\over P_{23}^2}{\cal A}_{\rm NLSM}(\underline{\bar{2}},\underline{\bar{\hat{3}}},\underline{\hat{P}}_{4:1})\nn
& &+\sum_{i=1}^n\,{\cal A}_{\rm NLSM}(i+1\cdots\underline{\bar{\hat{1}}},\underline{\bar{2}},\underline{\hat{P}}_{3:i}){1\over P_{i+1:2}^2}{\cal A}_{\rm NLSM}(\underline{\bar{\hat{3}}},\underline{\bar{4}},\cdots,i,\underline{\hat{P}}_{i+1:2})\,.~~~~\label{fct-NLSM}
\eea
This expression reproduces the result in \cite{Bjerrum-Bohr:2018jqe}.

The operator ${\cal T}^\epsilon[1,3]\cdot{\cal L}^\epsilon$ selects the factorization channels from \eref{fct-YM} by eliminating channels correspond to spurious poles for the YM amplitude, gives rise to channels in \eref{fct-NLSM}. The definition of longitudinal operator ${\cal L}^\epsilon_i$ carries the quantities $k_i\cdot k_j$, it indicates the possibility
that some propagators will be canceled after applying the operator ${\cal T}^\epsilon[1,3]\cdot{\cal L}^\epsilon$. Thus, the operator ${\cal T}^\epsilon[1,3]\cdot{\cal L}^\epsilon$ can transmute physical poles to both physical poles or spurious poles. Actually, as analysed in \cite{Bjerrum-Bohr:2018jqe}, when the numbers of external legs for ${\cal A}_L$ and ${\cal A}_R$ are even, the corresponding pole is a physical pole. When the numbers of external legs for ${\cal A}_L$ and ${\cal A}_R$ are odd, the corresponding pole is a spurious pole \footnote{The number of external legs for a physical NLSM amplitude is even, thus one will not encounter the situation one of ${\cal A}_L$ and ${\cal A}_R$ contains even number of external legs while another one contains odd number of external legs.}. In sub-amplitudes ${\cal A}_L$ and ${\cal A}_R$ in \eref{fct-NLSM}, the choices of fixed punctures are inherited from that in \eref{fct-YM},
since the operator ${\cal T}^\epsilon[1,3]\cdot{\cal L}^\epsilon$ is commutable with the CHY contour integral. As discussed in subsection \ref{effect-OP}, the choices of removed rows and columns in the reduced matrices in ${\cal A}_L$ and ${\cal A}_R$ in \eref{fct-NLSM} are determined by both \eref{fct-YM}, and the choice of $a,b$ for the operator ${\cal T}^\epsilon[a,b]$.

The proper gauge choice for the NLSM amplitude in the double-cover prescription \eref{DCamp}, which is consistent with the above relations, is $(1,2,3|4)$, as well as removing $1^{\rm th}$ and $3^{\rm th}$ rows and columns in the reduced matrix. The reason is similar as those for YM and sYMS cases mentioned in subsections \ref{general-YM} and \ref{subsecsYMS}. Thus the proper gauge choice for the NLSM amplitude in the double-cover prescription is also inherited from the GR amplitude in the double-cover prescription.

%%%%%%%%%%%%%%%%%%%%%%%%%%%%%%%%%%%%%%%%%%%%
\subsection{Factorization for BAS amplitude}
\label{subsec-fct-BAS}
%%%%%%%%%%%%%%%%%%%%%%%%%%%%%%%%%%%%%%%%%%%%%%%

Then we consider the factorization for the double color-ordered BAS amplitude ${\pmb A}_{\rm BAS}(1,\cdots,n;i_1,\cdots,i_n)$, which can be generated from the YM amplitude ${\pmb A}^\epsilon_{\rm YM}(1,\cdots,n)$
by applying the trace operator ${\cal T}^\epsilon[i_1,i_2,\cdots,i_n]$. Let us choose the formula of trace operator as
\bea
{\cal T}^\epsilon[i_1,i_2,\cdots,i_n]&=&{\cal T}^\epsilon[1,i_2,\cdots,i_{k-1},3,i_{k+1},\cdots,i_n]\nn
&=&{\cal T}^\epsilon[1,3]\cdot\Big(\prod_{j=2}^{k-1}\,{\cal I}_{i_{j-1}i_j3}\Big)\cdot\Big(\prod_{j=k+1}^n\,{\cal I}_{i_{j-1}i_j1}\Big)\,,
\eea
with $i_1=1,i_{k}=3$. This choice can always be achieved due to the cyclic symmetry of color-ordering.
Now we apply this operator to the RHS of \eref{fct-YM}, to get the factorized formula for the BAS amplitude.

For the first line of \eref{fct-YM}, we again factorize ${\cal T}^\epsilon[1,3]$ as $\cancel{\sum}_{\epsilon^M}\cdot{\cal T}^\epsilon[1,P_{23}]\cdot{\cal T}^\epsilon[3,P_{4:1}]$, where ${\cal T}^\epsilon[1,P_{23}]$ only acts  on ${\cal A}_L$, ${\cal T}^\epsilon[3,P_{4:1}]$
only acts on ${\cal A}_R$. Further more, we split insertions operators as
\bea
& &{\cal I}^\epsilon_{i_{j-1}i_j3}={\cal I}^\epsilon_{i_{j-1}i_jP_{23}}+{\cal I}^\epsilon_{P_{23}i_j3}\,,~~~~~{\rm for}~i_j\in\{4,\cdots,n\}\nn
& &{\cal I}^\epsilon_{i_{j-1}i_j3}={\cal I}^\epsilon_{i_{j-1}i_jP_{4:1}}+{\cal I}^\epsilon_{P_{4:1}i_j3}\,,~~~~{\rm for}~i_j=2\,,~i_{j-1}\in\{4,\cdots,1\}\nn
& &{\cal I}^\epsilon_{i_{j-1}i_j1}={\cal I}^\epsilon_{i_{j-1}i_jP_{4:1}}+{\cal I}^\epsilon_{P_{4:1}i_j1}\,,
~~~~{\rm for}~i_j=2\,,~i_{j-1}=3\,,\nn
& &{\cal I}^\epsilon_{i_{j-1}i_j1}={\cal I}^\epsilon_{i_{j-1}i_jP_{23}}+{\cal I}^\epsilon_{P_{23}i_j1}\,,
~~~~~{\rm for}~i_{j-1}\in\{2,3\}\,,~i_j\in\{4,\cdots,n\}\,,\nn
& &{\cal I}^\epsilon_{i_{j-1}i_j1}={\cal I}^\epsilon_{i_{j-1}i_j1}\,,
~~~~~~~~~~~~~~~~~~~~{\rm for}~i_{j-1}\in\{4,\cdots,n\}\,,~i_j\in\{4,\cdots,n\}\,.
\eea
If the leg $2$ is at the LHS of the leg $3$ in the color-ordering $(1,i_2,\cdots,i_n)$, suppose $2$ is $i_l$, $l$ can only be $l=k-1$ since otherwise
the operator ${\cal I}^\epsilon_{2i_{l+1}3}$ will annihilate both ${\cal A}_L$ and ${\cal A}_R$. Then the effective operator can be given as
\bea
{\cal T}^\epsilon[i_1,i_2,\cdots,i_n]\cong\cancel{\sum}_{\epsilon^M}\cdot{\cal T}^\epsilon_L\cdot{\cal T}^\epsilon_R\,,
\eea
where
\bea
& &{\cal T}^\epsilon_L={\cal T}^\epsilon[1,P_{23}]\cdot\Big(\prod_{j=2}^{k-2}\,{\cal I}^\epsilon_{i_{j-1}i_jP_{23}}\Big)\cdot
{\cal I}^\epsilon_{P_{23}i_{k+1}1}\cdot\Big(\prod_{j=k+2}^n\,{\cal I}^\epsilon_{i_{j-1}i_j1}\Big)\,,\\
& &{\cal T}^\epsilon_R={\cal T}^\epsilon[3,P_{4:1}]\cdot{\cal I}^\epsilon_{P_{4:1}23}\,.
\eea
These two operators can be identified as trace operators
\bea
& &{\cal T}^\epsilon_L={\cal T}^\epsilon[1,i_2,\cdots,i_{k-2},P_{23},i_{k+1},\cdots,i_n]={\cal T}^\epsilon[\pi_4,\cdots,\pi_1,P_{23}]\,,\nn
& &{\cal T}^\epsilon_R={\cal T}^\epsilon[2,3,P_{4:1}]={\cal T}^\epsilon[\pi'_2,\pi'_3,P_{4:1}]\,,
\eea
where $\pi$ is a permutations of $(4,\cdots,1)$, and $\pi'$ is a permutation of $(2,3)$.
It requires that the color-ordering $(i_1,\cdots,i_n)$ is equivalent to $(\pi_4,\cdots,\pi_1,\pi'_2,\pi'_3)$. Otherwise, the first line at the RHS of \eref{fct-YM} will be annihilated by ${\cal T}^\epsilon[i_1,\cdots,i_n]$.
If the leg $2$ is at the RHS of the leg $3$ in the color-ordering $(1,i_2,\cdots,i_n)$, suppose $2$ is $i_l$, $l$ has only one choice $l=k+1$ since otherwise ${\cal I}^\epsilon_{i_{l-1}21}$ will annihilate both ${\cal A}_L$ and ${\cal A}_R$. Thus we have the following two effective operators
\bea
& &{\cal T}^\epsilon_L={\cal T}^\epsilon[1,P_{23}]\cdot\Big(\prod_{j=2}^{k-1}\,{\cal I}^\epsilon_{i_{j-1}i_jP_{23}}\Big)\cdot
{\cal I}^\epsilon_{P_{23}i_{k+2}1}\cdot\Big(\prod_{j=k+3}^n\,{\cal I}^\epsilon_{i_{j-1}i_j1}\Big)\,,\\
& &{\cal T}^\epsilon_R={\cal T}^\epsilon[3,P_{4:1}]\cdot{\cal I}^\epsilon_{32P_{4:1}}\,,
\eea
which can also be identified as
\bea
& &{\cal T}^\epsilon_L={\cal T}^\epsilon[1,i_2,\cdots,i_{k-1},P_{23},i_{k+2},\cdots,i_n]={\cal T}^\epsilon[\pi_4,\cdots,\pi_1,P_{23}]\,,\nn
& &{\cal T}^\epsilon_R={\cal T}^\epsilon[3,2,P_{4:1}]={\cal T}^\epsilon[\pi'_2,\pi'_3,P_{4:1}]\,.
\eea
Thus, the trace operator ${\cal T}^\epsilon[i_1,\cdots,i_n]$ transmutes the first line of \eref{fct-YM} to
\bea
\theta(\pi\pi'){\cal A}_{\rm BAS}(\underline{4},\cdots,n,\underline{1},\underline{P}_{23};\pi_4,\cdots,\pi_1,P_{23}){1\over P_{23}^2}
{\cal A}_{\rm BAS}(\underline{2},\underline{3},\underline{P}_{4:1};\pi'_2,\pi'_3,P_{4:1})\,,
\eea
where $\theta(\pi\pi')=1$ if the color-ordering $(i_1,\cdots,i_n)$ is equivalent to $(\pi_4,\cdots,\pi_n,\pi'_2,\pi'_3)$, and vanishes otherwise.

For the second line of \eref{fct-YM}, similar manipulation gives
\bea
\sum_{i=4}^n\,\theta(\pi\pi'){\cal A}_{\rm BAS}(i+1,\cdots,\underline{1},\underline{2},\underline{P}_{3:i};\pi_{i+1},\cdots,\pi_2,P_{3:i}){1\over P_{i+1:2}^2}
{\cal A}_{\rm BAS}(\underline{3},\underline{4},\cdots,i,\underline{P}_{i+1:2};\pi'_3,\cdots,\pi'_i,P_{i+1:2})\,.
\eea
For the third line, the analysing of mass dimension shows that ${\cal A}_R$ will be annihilated by insertion operators. The argument is similar as that in the previous subsection for the NLSM case.

Thus, the factorized formula for the BAS amplitude is given by
\bea
& &{\pmb A}_{\rm BAS}(1,\cdots,n;i_1\,\cdots,i_n)\nn
&=&\theta(\pi\pi'){\cal A}_{\rm BAS}(\underline{4},\cdots,n,\underline{1},\underline{P}_{23};\pi_4,\cdots,\pi_1,P_{23}){1\over P_{23}^2}
{\cal A}_{\rm BAS}(\underline{2},\underline{3},\underline{P}_{4:1};\pi'_2,\pi'_3,P_{4:1})\nn
& &+\sum_{i=4}^n\,\theta(\pi\pi'){\cal A}_{\rm BAS}(i+1,\cdots,\underline{1},\underline{2},\underline{P}_{3:i};\pi_{i+1},\cdots,\pi_2,P_{3:i}){1\over P_{i+1:2}^2}
{\cal A}_{\rm BAS}(\underline{3},\underline{4},\cdots,i,\underline{P}_{i+1:2};\pi'_3,\cdots,\pi'_i,P_{i+1:2})\,.~~~~\label{fct-BAS}
\eea

\begin{figure}
  \centering
  \includegraphics[width=6cm]{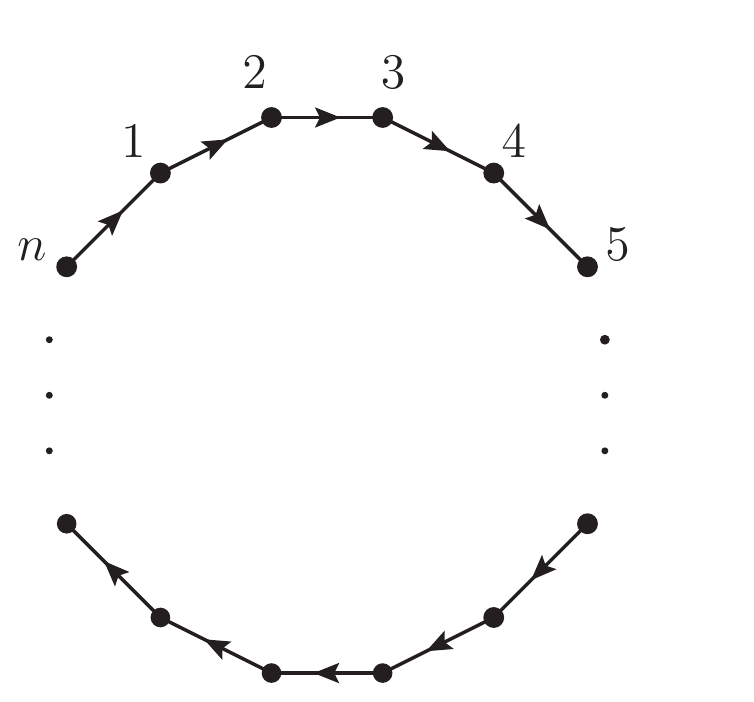} \\
  \caption{$PT^\tau_n(1,\cdots,n)$}\label{PT}
\end{figure}

Since the factorization for the BAS amplitude has not been given in the literature, we now derive it from the standard double-copy method, to check our result \eref{fct-BAS}. Since the purpose is the verification, some details will be omitted. The BAS integrand includes two Parke-Taylor factors $PT^\tau_n(1,\cdots,n)$ and $PT^\tau_n(i_1,\cdots,i_n)$, without any Pfaffian. Thus our gauge choice can be made as $(p,q,r|m)=(1,2,3|4)$. One can express $PT^\tau_n(1,\cdots,n)$ diagrammatically as in Figure \ref{PT}. The factor $PT^\tau_n(i_1,\cdots,i_n)$ can be expressed similarly. Based on the rule discussed at the end of subsection \ref{subsecDC}, there are three types of allowed cuts, the first one separates $\{1,2\}$ and $\{3,4\}$, the second one separates $\{1,4\}$ and $\{2,3\}$, the third one separates $\{1,3\}$ and $\{2,4\}$. We will consider them in turn.

\begin{figure}
  \centering
  \includegraphics[width=6cm]{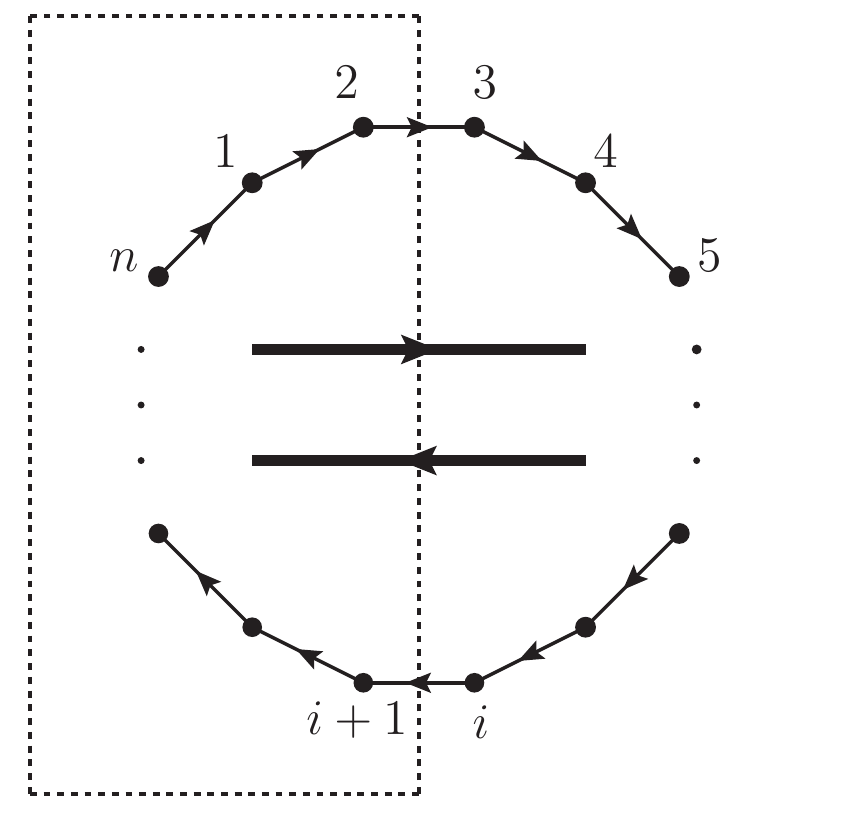} \\
  \caption{Cut-$1$, two bold lines are from $PT^\tau_n(i_1,\cdots,i_n)$}\label{cut1}
\end{figure}

The first type of cuts intersects lines $\tau_{23}$, $\tau_{i_ki_{k+1}}$ in $PT^\tau_n(1,\cdots,n)$, as well as lines in $PT^\tau_n(i_1,\cdots,i_n)$, as can be seen in Figure \ref{cut1}. The $\Lambda$-theorem in \cite{Gomez:2016bmv} indicates that the cut which gives non-vanishing contribution in the $\Lambda\to 0$ limit intersects up to $4$ lines. It means only two lines in $PT^\tau_n(i_1,\cdots,i_n)$ can be intersected, as shown in Figure \ref{cut1}.
Suppose these two lines are $\tau_{i_ai_{a'}}$ and $\tau_{i_{b'}i_b}$, with $i_a,i_b\in\{i+1,\cdots,2\}$ and $i_{a'},i_{b'}\in\{3,\cdots,i\}$, we have
\bea
& &\{i_b,i_{b+1},\cdots,i_{a-1},i_a\}=\{i+1,\cdots,2\}\,,\nn
& &\{i_{a'},i_{a'+1},\cdots,i_{b'-1},i_{b'}\}=\{3,\cdots,i\}\,.
\eea
Thus, the non-vanishing contribution corresponds to the Parke-Taylor factor
\bea
PT^\tau_n(i_1,\cdots,i_n)=PT^\tau_n(\pi_{i+1},\cdots,\pi_2,\pi'_3,\cdots,\pi'_{i})\,.
\eea

With the understanding of $PT^\tau_n(i_1,\cdots,i_n)$, we expand the measure and the integrand to the leading order of $\Lambda$.
The measure part contributes \eref{ms-part}, which has been evaluated previously. The integrand part gives
\bea
& &PT^\tau_n(1,\cdots,n)\Big|^{i+1,\cdots,2}_{3,\cdots,i}={\Lambda^2\over2^2}{1\over \sigma_{P_{3:i}(i+1)}\sigma_{(i+1)(i+2)}\cdots\sigma_{12}\sigma_{2P_{3:i}}}{1\over\sigma_{P_{i+1:2}3}\sigma_{34}\cdots\sigma_{(i-1)i}
\sigma_{iP_{i+1:2}}}\,,\nn
& &PT^\tau_n(i_1,\cdots,i_n)\Big|^{i+1,\cdots,2}_{3,\cdots,i}={\Lambda^2\over2^2}{1\over \sigma_{P_{3:i}\pi_{i+1}}\sigma_{\pi_{i+1}\pi_{i+2}}\cdots\sigma_{\pi_1\pi_2}\sigma_{\pi_2P_{3:i}}}{1\over\sigma_{P_{i+1:2}\pi'_3}
\sigma_{\pi'_3\pi'_4}\cdots\sigma_{\pi'_{i-1}\pi'_i}
\sigma_{\pi'_iP_{i+1:2}}}\,,~~~~\label{PT-BAS}
\eea
where the possible $-$ signs in $PT^\tau_n(1,\cdots,n)\Big|^{i+1,\cdots,2}_{3,\cdots,i}$ and $PT^\tau_n(i_1,\cdots,i_n)\Big|^{i+1,\cdots,2}_{3,\cdots,i}$ have been neglected since they will cancel each other.
Combining \eref{ms-part} and \eref{PT-BAS} together and integrating over $\Lambda$ gives
\bea
{\theta(\pi\pi')\over2}{\cal A}_{\rm BAS}(i+1,\cdots,\underline{1},\underline{2},\underline{P}_{3:i};\pi_{i+1},\cdots,\pi_2,P_{3:i}){1\over P_{i+1:2}^2}
{\cal A}_{\rm BAS}(\underline{3},\underline{4},\cdots,i\underline{P}_{i+1:2};\pi'_3,\cdots,\pi'_i,P_{i+1:2})\,.
\eea
Summing over the mirrored configurations and $i$, we get the second line of \eref{fct-BAS}.

\begin{figure}
  \centering
  \includegraphics[width=6cm]{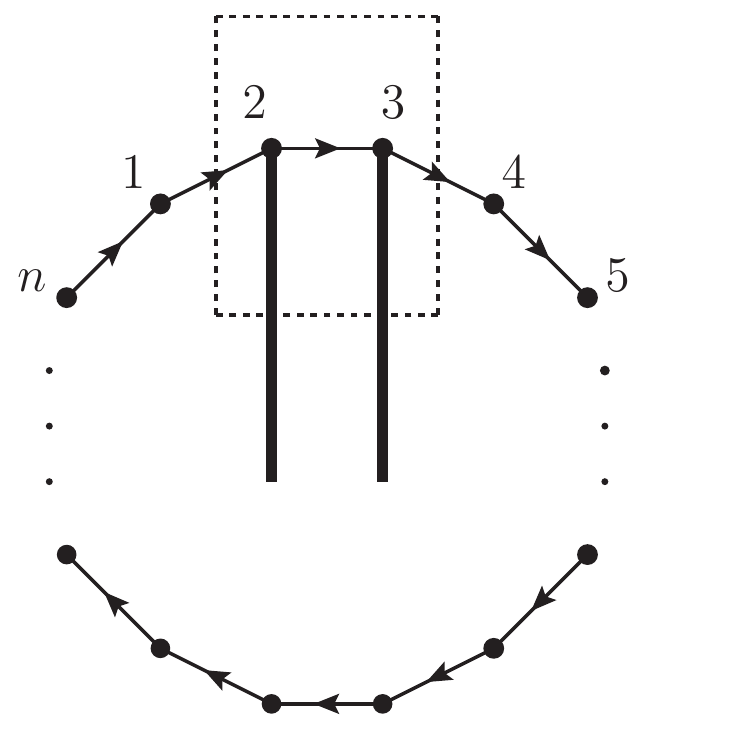} \\
  \caption{Cut-$2$, two bold lines are from $PT^\tau_n(i_1,\cdots,i_n)$}\label{cut2}
\end{figure}

For the second type of cuts, the only allowed cut which gives non-vanishing contribution is given in Figure \ref{cut2}.
The similar manipulation reproduces the first line of \eref{fct-BAS}.
For the third type of cuts, all configurations vanish in the $\Lambda\to 0$ limit, due to the $\Lambda$-Theorem.
Thus our result \eref{fct-BAS} is the correct factorized formula for the BAS amplitude in the double-cover framework.

The effect of trace operator for the current case is similar as linking the GR amplitude and the color-ordered YM amplitude in the previous section.
The factorization channels in \eref{fct-BAS} is selected from channels in \eref{fct-YM} by the trace operator. In \eref{fct-BAS}, all poles are physical poles, the same as in the first and second lines of \eref{fct-YM}, since the trace operator transmutes physical poles to physical poles. Terms with spurious poles in the third line in \eref{fct-YM} are annihilated by the trace operator, thus the factorized BAS amplitude does not include any term with spurious pole. This fact is quite natural since it is obvious that the BAS amplitude does not contain any kinematical numerator which can cancel some propagators. The gauge choices of fixed punctures for ${\cal A}_L$ and ${\cal A}_R$ in\eref{fct-BAS} are inherited from the YM terms in \eref{fct-YM}, since the trace operator is commutable with the CHY contour integral.

The proper gauge choice for the BAS amplitude in the double-cover prescription \eref{DCamp}, consistent with the relations mentioned above, is $(1,2,3|4)$, which has been used when calculating the factorized formula by the standard double-cover method. This choice is also inherited from the GR amplitude in the double-cover prescription, as can be discussed similarly as in the previous cases.

%%%%%%%%%%%%%%%%%%%%%%%%%%
\subsection{BCFW recursion}
\label{BCFW-NLSM-BAS}
%%%%%%%%%%%%%%%%%%%%%%%%%%

In this subsection, we consider the relationship between the factorized formula for amplitudes and the BCFW recursion. Since for sYMS amplitudes we have not obtained the general factorized formula valid for any configuration of external legs, in this subsection we only discuss NLSM and BAS amplitudes. We will focus on how the differential operators link the BCFW recursion for YM amplitudes to those for NLSM and BAS amplitudes.

To see the descendent relation, we choose the special BCFW deformation
\bea
k_2(z)=k_2+zq\,,~~~~~~k_3(z)=k_3-zq\,,~~~~\label{BCFW-shift'}
\eea
the same as in subsection \ref{BCFW-YM}.
We first consider the BAS amplitude. Since the factorization channels for the BAS amplitude is selected from channels for the YM amplitude by the trace operator, the set of detectable physical poles for the BAS amplitude is a subset of detectable poles for the YM amplitude, i.e., a subset of $\{P_a^2=(\sum_{l=3}^a k_l)^2\}$ with $4\leq a\leq n$. Under the deformation in \eref{BCFW-shift'}, the pole $P_{23}^2$ in the first line of \eref{fct-BAS} is independent of $z$ thus is un-detectable, the same as for the YM case.
It is easy to see that all sub-amplitudes ${\cal A}_L$ and ${\cal A}_R$ in the first line of \eref{fct-BAS} do not contain any detectable pole in the set $\{P_a^2=(\sum_{l=3}^a k_l)^2\}$. Since the set of detectable poles for the BAS amplitude is a subset of $\{P_a^2=(\sum_{l=3}^a k_l)^2\}$, we conclude that the first line of \eref{fct-BAS} does not contain any detectable pole.
Thus, this part contributes to the boundary term. For the second line of \eref{fct-BAS}, poles $P_{i+1:2}^2$ are detectable. Thus terms in this line contribute to the recursive part. Further more, as pointed out at the end of subsection \ref{property-OS}, an off-shell amplitude ${\cal A}_{\rm BAS}(i+1,\cdots,\underline{1},\underline{2},\underline{P}_{3:i};\pi_{i+1},\cdots,\pi_2,P_{3:i})$ does not contain any pole $(k_1+k_2+K)^2$, and an off-shell amplitude ${\cal A}_{\rm BAS}(\underline{3},\underline{4},\cdots,i,\underline{P}_{i+1:2};\pi'_3,\cdots,\pi'_i,P_{i+1:2})$ does not contain any pole $(k_3+k_4+K)^2$. Thus we can also conclude that
${\cal A}_L$ and ${\cal A}_R$ in terms in the second line do not contain any detectable pole.
In other words, a term
\bea
{\cal A}_{\rm BAS}(i+1,\cdots,\underline{1},\underline{2},\underline{P}_{3:i};\pi_{i+1},\cdots,\pi_2,P_{3:i}){1\over P_{i+1:2}^2}
{\cal A}_{\rm BAS}(\underline{3},\underline{4},\cdots,i,\underline{P}_{i+1:2};\pi'_3,\cdots,\pi'_i,P_{i+1:2})
\eea
in the second line of \eref{fct-BAS} only contributes to the residue at $P_{i+1:2}^2(z_{i+1:2})=0$, which is given as
\bea
{\pmb A}_L(z_{i+1:2}){1\over P_{i+1:2}^2}{\pmb A}_R(z_{i+1:2})\,,
\eea
with
\bea
{\pmb A}_L(z_{i+1:2})&=&{\pmb A}_{\rm BAS}(i+1,\cdots,1,2(z_{i+1:2}),P_{3:i}(z_{i+1:2});\pi_{i+1},\cdots,\pi_2,P_{3:i})\,,\nn
{\pmb A}_R(z_{i+1:2})&=&{\pmb A}_{\rm BAS}(3(z_{i+1:2}),4,\cdots,i,P_{i+1:2}(z_{i+1:2});\pi'_3,\cdots,\pi'_i,P_{i+1:2})\,.
\eea
Thus, there is a one to one correspondence from terms in the second line of \eref{fct-BAS} to terms in the recursive part, similar as in the YM case.
Now we can see that since the trace operator will not create or annihilate any pole, it links recursive terms of the YM amplitude to recursive terms of the BAS amplitude, and boundary contributions of the YM amplitude to boundary contributions of the BAS amplitude.

Then we turn to the NLSM amplitude. Similarly, the set of detectable physical poles is a subset of $\{P_a^2=(\sum_{l=3}^a k_l)^2\}$. For terms with physical poles in \eref{fct-NLSM}, we can conclude that there is a one to one map from these terms to terms in the recursive part. For terms with spurious poles in \eref{fct-NLSM}, since both ${\cal A}_L$ and ${\cal A}_R$ do not contain any detectable physical pole, they can be recognized as the boundary contributions. We have seen that, since the operator ${\cal T}^\epsilon[a,b]\cdot{\cal L}^\epsilon$ can transmute physical poles to both physical poles or spurious poles, when a physical pole is transmuted to a physical pole, this operator links a recursive term to a recursive term. When a physical pole is transmuted to a spurious pole, this operator links a recursive term to a boundary term.

%%%%%%%%%%%%%%%%%%%%%%%%%%%%%%%%%%%%%%
\section{Summary and discussion}
\label{secconclu}
%%%%%%%%%%%%%%%%%

In this paper, we have demonstrated that the factorizations for amplitudes of various theories in the double-cover framework can be generated from the GR amplitude in the double-cover prescription, by applying proper differential operators. Using this method, we first derived the factorized formula for the YM amplitude by applying the trace operator to the GR amplitude. Then, by applying three types of operators to the factorized YM amplitude, we have derived the factorized formula for sYMS, NLSM and BAS amplitudes. The factorized formulae for YM, sYMS and NLSM amplitudes are coincide with the results in the literature, while the factorized formula for the BAS amplitude is verified by the standard double-cover method.
The effects of all three types of differential operators proposed in \cite{Cheung:2017ems} have been covered.

Our method can explain some non-trivial relationships among factorized formulae for amplitudes of different theories. Suppose the amplitude ${\pmb A}'$ of theory-$b$ can be generated from the amplitude ${\pmb A}$ of theory-$a$ via the operator ${\cal O}$, the factorization channels for ${\pmb A}'$ are selected from the channels for ${\pmb A}$ by ${\cal O}$. The pole-structure of the factorized ${\pmb A}'$ is also arise from the pole-structure of the factorized ${\pmb A}$ via ${\cal O}$. The gauge choices of fixed punctures for off-shell sub-amplitudes in the factorized ${\pmb A}'$ are inherited from those in the factorized ${\pmb A}$,
since the differential operator is commutable with the CHY contour integral. Similar descendent relation also exist for the choices of removed rows and columns in the reduced matrices. The differential operators also relate terms in the BCFW recursion for theory-$a$ to those for theory-$b$ by relating pole-structures. Notice that the factorized formula for an amplitude depend on the fixed punctures $(p,q,r|m)$, as well as the removed rows and columns in the reduced matrices, in the double-cover integral \eref{DCamp}, thus is not unique. The proper gauge choices for YM, sYMS, NLSM and BAS amplitudes, which is consistent with the above relations, is nothing but the gauge choice in the GR amplitude in the double-cover prescription. This gauge choice has been transmitted from the GR to other theories by the differential operator.

The results in this paper also provides further understanding for differential operators. From the relation ${\pmb A}'={\cal O}{\pmb A}$, one can not conclude that the operator ${\cal O}$ transmutes the factorization for ${\pmb A}$ to the factorization for ${\pmb A}'$ directly, since logically it is possible that applying the operator to the the factorization for ${\pmb A}$ gives a formula which is equivalent but totally different to the factorization for ${\pmb A}'$. However, our calculation excludes this possibility. It is a quite non-trivial phenomenon, which implies that the relations among amplitudes of different theories not only unified the full expressions of amplitudes, but also
link the inner structures of amplitudes together.

Since the full factorized formula for the GR amplitude is hard to be obtained, for now we can not derive factorizations for all theories in the unified web in \cite{Cheung:2017ems}. How to calculate the full factorized GR amplitude and fill this gap is a potential future direction. In \cite{Feng:2019cbe,Zhou:2019mbe} it has been proved that all amplitudes in the unified web can be expanded to BAS amplitudes, and the coefficients can be computed via systematic rules. Thus, maybe another possible path is to generate factorizations for amplitudes from the factorized BAS amplitudes.

%%%%%%%%%%%%%%%%%%%%%%%%%%%%%%%%%%%%%%%%%%%%%%%%%%
\section*{Acknowledgments}

The author is indebted to Prof. Bo Feng for helpful discussions and valuable comments on the original manuscript, and to Prof. H. Gomez for answering the queries about the double-cover prescription. This
work is supported by Chinese NSF funding under
contracts No.11805163, as well as NSF of Jiangsu Province under Grant No.BK20180897.

%%%%%%%%%%%%%%%%%%%
\appendix
%%%%%%%%%%%%%%%%%%%

%%%%%%%%%%%%%%%%%%%%%%%%%%%%%%%%%%%%%%%%%%%%
\section{Factorized formula for GR: physical poles}
\label{phypole}
%%%%%%%%%%%%%%%%%%%%%%%%%%%%%

In this section, we provide details of deriving \eref{phy-pole-YM-gen}.

We begin with the separation $\{\{1,2,{\pmb \alpha}_1\},\{3,4,{\pmb \beta}_1\}\}$ by considering the configuration
\bea
& &(y_1=+\sqrt{\sigma_1^2-\Lambda^2}\,,~\sigma_1)\,,~~~~(y_2=+\sqrt{\sigma_2^2-\Lambda^2}\,,~\sigma_2)\,,~~~~
(y_{\alpha_l}=+\sqrt{\sigma_{\alpha_l}^2-\Lambda^2}\,,~\sigma_{\alpha_l})\,,\nn
& &(y_3=-\sqrt{\sigma_3^2-\Lambda^2}\,,~\sigma_3)\,,~~~~(y_4=-\sqrt{\sigma_4^2-\Lambda^2}\,,~\sigma_4)\,,~~~~
(y_{\beta_m}=-\sqrt{\sigma_{\beta_m}^2-\Lambda^2}\,,~\sigma_{\beta_m})\,,
\eea
where $\alpha_l$ and $\beta_m$ are elements in sets ${\pmb\alpha}_1$ and ${\pmb\beta}_1$, respectively.
The first step is to generalize \eref{FP-det}
to the current general case.
The expansions of $\Delta_{123}$ and $\Delta_{123|4}$ to the leading order of $\Lambda$ have not been changed, thus we only need to treat
the scattering equation ${\cal E}^\tau_4$. Expanding ${\cal E}^\tau_i$ with $i\in\{3,4,{\pmb\beta_1}\}$ to the leading order of $\Lambda$ gives
\bea
{\cal E}^\tau_i\Big|^{1,2,{\pmb\alpha}_1}_{3,4,{\pmb\beta}_1}&=&-2\sum_{j\in\{3,4,{\pmb\beta}_1\}\setminus\{i\}}\,{\sigma_j\over \sigma_i}{k_i\cdot k_j\over \sigma_{ji}}\,.~~~~\label{seq-gen}
\eea
Then one can obtain the relation
\bea
\sum_{i\in\{3,4,{\pmb\beta}_1\}}\,\sigma_i{\cal E}^\tau_i\Big|^{1,2,{\pmb\alpha}_1}_{3,4,{\pmb\beta}_1}&=&-2\sum_{i,j\in\{3,4,{\pmb\beta}_1\}}\,\sigma_j{k_i\cdot k_j\over \sigma_{ji}}=-P_{34{\pmb\beta}_1}^2\,,~~~~\label{seq-gen-rel1}
\eea
where the identity ${\sigma_i\over\sigma_{ij}}+{\sigma_j\over\sigma_{ji}}=1$ has been used. To continue, notice that scattering equations ${\cal E}^\tau_a$ with $a\neq 1,2,3,4$ are still poles for the contour integral, i.e., all these equations are satisfied. Thus expanding scattering equations ${\cal E}^\tau_a$ with $a\in{\pmb\beta}_1$ to the leading order provides
\bea
{\cal E}^\tau_a\Big|^{1,2,{\pmb\alpha}_1}_{3,4,{\pmb\beta}_1}=0\,,~~~~\forall\,a\in{\pmb\beta}_1\,.~~~~\label{constra}
\eea
Substituting \eref{constra} into \eref{seq-gen-rel1}, we arrive at an equation for ${\cal E}^\tau_4\Big|^{1,2,{\pmb\alpha}_1}_{3,4,{\pmb\beta}_1}$ and ${\cal E}^\tau_3\Big|^{1,2,{\pmb\alpha}_1}_{3,4,{\pmb\beta}_1}$ as
\bea
\sigma_3{\cal E}^\tau_3\Big|^{1,2,{\pmb\alpha}_1}_{3,4,{\pmb\beta}_1}+\sigma_4{\cal E}^\tau_4\Big|^{1,2,{\pmb\alpha}_1}_{3,4,{\pmb\beta}_1}
=-P^2_{34{\pmb\beta}_1}\,.~~~~\label{eq1}
\eea
To solve ${\cal E}^\tau_4\Big|^{1,2,{\pmb\alpha}_1}_{3,4,{\pmb\beta}_1}$, we need another equation which can be found via the observation
\bea
\sum_{i\in\{3,4,{\pmb\beta}_1\}}\,\sigma_i^2{\cal E}^\tau_i\Big|^{1,2,{\pmb\alpha}_1}_{3,4,{\pmb\beta}_1}
&=&-2\sum_{i,j\in\{3,4,{\pmb\beta}_1\}}\,{\sigma_i\sigma_j\over\sigma_{ji}}k_i\cdot k_j=0\,,
\eea
due to the anti-symmetry of $\sigma_{ij}$. Thus \eref{constra} indicates
\bea
\sigma_3^2{\cal E}^\tau_3\Big|^{1,2,{\pmb\alpha}_1}_{3,4,{\pmb\beta}_1}+\sigma_4^2{\cal E}^\tau_4\Big|^{1,2,{\pmb\alpha}_1}_{3,4,{\pmb\beta}_1}
=0\,.~~~~\label{eq2}
\eea
Solving equations \eref{eq1} and \eref{eq2} gives
\bea
{\cal E}^\tau_4\Big|^{1,2,{\pmb\alpha}_1}_{3,4,{\pmb\beta}_1}={\sigma_3\over \sigma_4\sigma_{43}}P^2_{34{\pmb\beta}_1}\,,~~~~
{\cal E}^\tau_3\Big|^{1,2,{\pmb\alpha}_1}_{3,4,{\pmb\beta}_1}={\sigma_4\over \sigma_3\sigma_{34}}P^2_{34{\pmb\beta}_1}\,.
\eea
The only difference between the general case ${\cal E}^\tau_4\Big|^{1,2,{\pmb\alpha}_1}_{3,4,{\pmb\beta}_1}$ and the special case ${\cal E}^\tau_4\Big|^{1,2}_{3,4}$
is replacing $P^2_{34}$ by $P^2_{34{\pmb\beta}_1}$.
Consequently, for the general case we have
\bea
{\Delta_{123}\Delta_{123|4}\over S^\tau_4}\Big|^{1,2}_{3,4}={2^5\over \Lambda^4}|12P_{34{\pmb\beta}_1}|^2{1\over P^2_{34{\pmb\beta}_1}}|P_{12{\pmb\alpha}_1}34|^2\,,~~~~\label{FP-det-gen}
\eea
where punctures $\sigma_{P_{34{\pmb\beta}_1}}$ and $\sigma_{P_{12{\pmb\alpha}_1}}$ are fixed at $\sigma_{P_{34{\pmb\beta}_1}}=\sigma_{P_{12{\pmb\alpha}_1}}=0$.
Thus this part is factorized into two determinants and one propagator.

Then we turn to the reduced Pfaffian $({\bf Pf}'\Psi^\Lambda_n)^\tau$.
One can exchange the order of rows and columns in $(\Psi^\Lambda_n)^{13}_{13}$ to get
\bea
{\bf Pf}(\Psi^\Lambda_n)^{13}_{13}={\bf Pf}\left(
         \begin{array}{c|c}
           (\Psi^\Lambda)^{\{1,2,{\pmb\alpha}_1\}}_{\{1,2,{\pmb\alpha}_1\}} & (\Psi^\Lambda)^{\{1,2,{\pmb\alpha}_1\}}_{\{3,4,{\pmb\beta}_1\}} \\
           \\
           \hline\\
         (\Psi^\Lambda)^{\{3,4,{\pmb\beta}_1\}}_{\{1,2,{\pmb\alpha}_1\}} & (\Psi^\Lambda)^{\{3,4,{\pmb\beta}_1\}}_{\{3,4,{\pmb\beta}_1\}} \\
         \end{array}
       \right)^{13}_{13}\,,
\eea
where the block $(\Psi^\Lambda)^{\{1,2,{\pmb\alpha}_1\}}_{\{3,4,{\pmb\beta}_1\}}$ contains elements with the row-indexes for $k_i$ and $\epsilon_i$
belong to the set $\{1,2,{\pmb\alpha}_1\}$, while the column-indexes belong to $\{3,4,{\pmb\beta}_1\}$. Analogous notations hold for other three blocks.
The possible $-$ sign has been omitted, since it will be canceled by the totally same sign from another reduced Pfaffian in the integrand.
When expanding to the leading order of $\Lambda$, we have
\bea
(\Psi^\Lambda)^{\{1,2,{\pmb\alpha}_1\}}_{\{1,2,{\pmb\alpha}_1\}}\Big|^{1,2,{\pmb\alpha}_1}_{3,4,{\pmb\beta}_1}= \left(\begin{array}{c|c}
           (A')^{\{1,2,{\pmb\alpha}_1\}}_{\{1,2,{\pmb\alpha}_1\}} & (C')^{\{1,2,{\pmb\alpha}_1\}}_{\{1,2,{\pmb\alpha}_1\}} \\
           \\
           \hline\\
         (-C'^T)^{\{1,2,{\pmb\alpha}_1\}}_{\{1,2,{\pmb\alpha}_1\}} & (B')^{\{1,2,{\pmb\alpha}_1\}}_{\{1,2,{\pmb\alpha}_1\}} \\
         \end{array}
       \right)\,,
\eea
\bea
(\Psi^\Lambda)^{\{1,2,{\pmb\alpha}_1\}}_{\{3,4,{\pmb\beta}_1\}}\Big|^{1,2,{\pmb\alpha}_1}_{3,4,{\pmb\beta}_1}= \left(\begin{array}{c|c}
           (A'')^{\{1,2,{\pmb\alpha}_1\}}_{\{3,4,{\pmb\beta}_1\}} & (C'')^{\{1,2,{\pmb\alpha}_1\}}_{\{3,4,{\pmb\beta}_1\}} \\
           \\
           \hline\\
         (-C''^T)^{\{1,2,{\pmb\alpha}_1\}}_{\{3,4,{\pmb\beta}_1\}} & (B'')^{\{1,2,{\pmb\alpha}_1\}}_{\{3,4,{\pmb\beta}_1\}} \\
         \end{array}
       \right)\,,
\eea
\bea
(\Psi^\Lambda)^{\{3,4,{\pmb\beta}_1\}}_{\{3,4,{\pmb\beta}_1\}}\Big|^{1,2,{\pmb\alpha}_1}_{3,4,{\pmb\beta}_1}= \left(\begin{array}{c|c}
           (A''')^{\{3,4,{\pmb\beta}_1\}}_{\{3,4,{\pmb\beta}_1\}} & (C''')^{\{3,4,{\pmb\beta}_1\}}_{\{3,4,{\pmb\beta}_1\}} \\
           \\
           \hline\\
         (-C'''^T)^{\{3,4,{\pmb\beta}_1\}}_{\{3,4,{\pmb\beta}_1\}} & (B''')^{\{3,4,{\pmb\beta}_1\}}_{\{3,4,{\pmb\beta}_1\}} \\
         \end{array}
       \right)\,.
\eea
In the first block, elements $A'_{ij}$, $B'_{ij}$ and $C'_{ij}$ are elements in \eref{ABCmatrix} times the factor ${1\over2}$, except
\bea
C_{jj}&=&-\sum_{l\in\{1,2,{\pmb\alpha}_1\},\,l\neq j}\hspace{-.4em}{k_l \cdot \epsilon_j\over 2\sigma_{lj}}+\sum_{l\in\{3,4,{\pmb\beta}_1\}}{k_l \cdot \epsilon_j\over 2\sigma_{j}}\nn
&=&-\sum_{l\in\{1,2,{\pmb\alpha}_1\},\,l\neq j}\hspace{-.4em}{k_l \cdot \epsilon_j\over 2\sigma_{lj}}-\sum_{l\in\{1,2,{\pmb\alpha}_1\},\,l\neq j}\hspace{-.4em}{k_l \cdot \epsilon_j\over 2\sigma_{j}}\nn
&=&-\sum_{l\in\{1,2,{\pmb\alpha}_1\},\,l\neq j}\hspace{-.4em}{\sigma_l\over2\sigma_j}{k_l\cdot\epsilon_j\over\sigma_{lj}}\,.
\eea
For other blocks, we have
\bea
& &A''_{ij} = {k_{i}\cdot k_j\over 2\sigma_i}\,,
~~~~ B''_{ij} = {\epsilon_i\cdot\epsilon_j\over 2\sigma_i}\,,~~~~
C''_{ij} = {k_i \cdot \epsilon_j\over 2\sigma_i}\,,
\eea
and
\bea
& &A'''_{ij} = \begin{cases} \displaystyle -{2\sigma_i\sigma_j\over\Lambda^2}{k_{i}\cdot k_j\over \sigma_{ij}} & i\neq j\,,\\
\displaystyle  ~~~ 0 & i=j\,,\end{cases} \qquad\qquad\qquad\qquad B'''_{ij} = \begin{cases} \displaystyle -{2\sigma_i\sigma_j\over\Lambda^2}{\epsilon_i\cdot\epsilon_j\over \sigma_{ij}} & i\neq j\,,\\
\displaystyle ~~~ 0 & i=j\,,\end{cases} \nn
& &C'''_{ij} = \begin{cases} \displaystyle -{2\sigma_i\sigma_j\over\Lambda^2}{k_i \cdot \epsilon_j\over \sigma_{ij}} &\quad i\neq j\,,\\
\displaystyle \sum_{l\in\{3,4,{\pmb\beta}_1\},\,l\neq j}\hspace{-.5em}{2\sigma_l\sigma_j\over\Lambda^2}{k_l \cdot \epsilon_j\over \sigma_{lj}} &\quad i=j\,.\end{cases}
\eea
For later convenience, we re-write $C'''_{jj}$ as
\bea
C'''_{jj}&=&\sum_{l\in\{3,4,{\pmb\beta}_1,P_{12{\pmb\alpha}_1}\},\,l\neq j}\hspace{-.5em}{2\sigma_l\sigma_j\over\Lambda^2}{k_l \cdot \epsilon_j\over \sigma_{lj}}\nn
&=&{2\sigma_j\over \Lambda^2}\sum_{l\in\{3,4,{\pmb\beta}_1,P_{12{\pmb\alpha}_1}\},\,l\neq j}\hspace{-.5em}\Big(1-{\sigma_j\over\sigma_{jl}}\Big)
k_l\cdot\epsilon_j\nn
&=&-{2\sigma_j^2\over \Lambda^2}\sum_{l\in\{3,4,{\pmb\beta}_1,P_{12{\pmb\alpha}_1}\},\,l\neq j}\hspace{-.5em}-{k_l\cdot\epsilon_j\over\sigma_{lj}}\,,
\eea
where the momentum conservation has been used in the last step.
Similar calculation gives
\bea
C'_{jj}=\sum_{l\in\{1,2,{\pmb\alpha}_1,P_{34{\pmb\beta}_1}\},\,l\neq j}\hspace{-.5em}-{k_l\cdot\epsilon_j\over2\sigma_{lj}}\,.
\eea
From above expansions, one can observe that each element in the block $(\Psi^\Lambda)^{\{3,4,{\pmb\beta}_1\}}_{\{3,4,{\pmb\beta}_1\}}\Big|^{1,2,{\pmb\alpha}_1}_{3,4,{\pmb\beta}_1}$ contributes $\Lambda^{-2}$, while all elements in other blocks contribute $\Lambda^0$. Thus, the leading order term of ${\bf Pf}(\Psi^\Lambda_n)^{13}_{13}$ comes from the corresponding terms in the definition of Pfaffian \eref{defin-Pf} which contain as much indexes for the block $(\Psi^\Lambda)^{\{3,4,{\pmb\beta}_1\}}_{\{3,4,{\pmb\beta}_1\}}$ as possible. Since rows and columns $1$ and $3$ were removed in the reduced matrix $(\Psi^\Lambda_n)^{13}_{13}$, the block
$(\Psi^\Lambda)^{\{3,4,{\pmb\beta}_1\}}_{\{3,4,{\pmb\beta}_1\}}$ is a $(2|{\pmb\beta}_1|+3)\times(2|{\pmb\beta}_1|+3)$ matrix, and contributes up to $(2|{\pmb\beta}_1|+2)$ indexes to the Pfaffian \eref{defin-Pf}. The notation $|{\pmb a}|$
is used to denote the length of the set ${\pmb a}$. Thus we find the leading order term is
\bea
{\bf Pf}(\Psi^\Lambda_n)^{13}_{13}\Big|^{1,2,{\pmb\alpha}_1}_{3,4,{\pmb\beta}_1}=
\sum_{i_a,j_a}\sum_{\rho_1\rho_2}\,\Big(\prod_{k_b,l_b\in \rho_1}\bar{\Psi}^\Lambda_{k_bl_b}\Big)\bar{\Psi}^\Lambda_{i_aj_a}\Big(\prod_{m_c,n_c\in \rho_2}\bar{\Psi}^\Lambda_{m_cn_c}\Big)\,,~~~~\label{gener-leading}
\eea
where $\bar{\Psi}^\Lambda_{i_aj_a}$ is an element of $(\Psi^\Lambda)^{\{1,2,{\pmb\alpha}_1\}}_{\{3,4,{\pmb\beta}_1\}}\Big|^{1,2,{\pmb\alpha}_1}_{3,4,{\pmb\beta}_1}$, $\bar{\Psi}^\Lambda_{k_bl_b}$
are elements of $(\Psi^\Lambda)^{\{1,2,{\pmb\alpha}_1\}}_{\{1,2,{\pmb\alpha}_1\}}\Big|^{1,2,{\pmb\alpha}_1}_{3,4,{\pmb\beta}_1}$, and $\bar{\Psi}^\Lambda_{m_cn_c}$ are elements of $(\Psi^\Lambda)^{\{3,4,{\pmb\beta}_1\}}_{\{3,4,{\pmb\beta}_1\}}\Big|^{1,2,{\pmb\alpha}_1}_{3,4,{\pmb\beta}_1}$.
Again, the possible $-$ sign is neglected for the same reason.
To factorize \eref{gener-leading}, we observe that the element $\bar{\Psi}^\Lambda_{i_aj_a}$ is always given as
${V_{i_a}\cdot V_{j_a}\over \sigma_{i_a}}$, where
\bea
V_i= \begin{cases} \displaystyle k_i & i<n\,,\\
\displaystyle  \epsilon_{i-n} & i>n\,,\end{cases} \qquad\qquad\qquad\qquad \sigma_i = \begin{cases} \displaystyle \sigma_i & i<n\,,\\
\displaystyle \sigma_{i-n} & i>n\,.\end{cases}
\eea
Thus one can factorize it as
\bea
\bar{\Psi}^\Lambda_{i_aj_a}=\sigma_{j_a}\sum_{\epsilon^M}\,{V_{i_a}\cdot \epsilon^M_{34{\pmb\beta}_1}\over \sigma_{i_a}}{\epsilon^M_{12{\pmb\alpha}_1}\cdot V_{j_a}\over \sigma_{j_a}}\,,
\eea
with the completeness relationship $\sum_{\epsilon^M}\epsilon^{M\mu}_i\epsilon^{M\nu}_j=\eta^{\mu\nu}$.
Thus we obtain
\bea
{\bf Pf}(\Psi^\Lambda_n)^{13}_{13}\Big|^{1,2,{\pmb\alpha}_1}_{3,4,{\pmb\beta}_1}&=&
\sum_{i_a,j_a}\sum_{\rho_1\rho_2}\sum_{\epsilon^M}\,\sigma_{ja}\Big({V_{i_a}\cdot \epsilon^M_{34{\pmb\beta}_1}\over \sigma_{i_a}}\prod_{k_b,l_b\in \rho_1}\bar{\Psi}^\Lambda_{k_bl_b}\Big)\nn
& &\Big({\epsilon^M_{12{\pmb\alpha}_1}\cdot V_{j_a}\over \sigma_{j_a}}\prod_{m_c,n_c\in \rho_2}\bar{\Psi}^\Lambda_{m_cn_c}\Big)\nn
&=&{2^{|{\pmb\beta}_1|-|{\pmb\alpha}_1|-1}\sigma_3\prod_{i\in\{4,{\pmb\beta}_1\}}\sigma_i^2\over \Lambda^{2|{\pmb\beta}_1|+2}}\sum_{\epsilon^M}\,{\bf Pf}(\Psi_{3+|{\pmb\alpha}_1|})^{P_{34{\pmb\beta}_1}1}_{P_{34{\pmb\beta}_1}1}{\bf Pf}(\Psi'_{3+|{\pmb\beta}_1|})^{P_{12{\pmb\alpha}_1}3}_{P_{12{\pmb\alpha}_1}3}\,,
\eea
where $\Psi_{3+|{\pmb\alpha}_1|}$ and $\Psi'_{3+|{\pmb\beta}_1|}$ are the single-cover matrices for punctures $\{1,2,{\pmb\alpha}_1,P_{34{\pmb\beta}_1}\}$ and $\{3,4,{\pmb\beta}_1,P_{12{\pmb\alpha}_1}\}$, respectively. Using this result, we get
\bea
({\bf Pf}'\Psi^\Lambda_n)^\tau&=&\prod_{i=1}^n\,{(y\sigma)_i\over y_i}T_{13}{\bf Pf}(\Psi^\Lambda_n)^{13}_{13}\nn
&=&{\Lambda^2\over 2^2}\sum_{\epsilon^M}\,{1\over \sigma_{P_{34{\pmb\beta}_1}1}}{1\over \sigma_{P_{12{\pmb\alpha}_1}}}{\bf Pf}(\Psi'_{3+|{\pmb\alpha}_1|})^{P_{34{\pmb\beta}_1}1}_{P_{34{\pmb\beta}_1}1}{\bf Pf}(\Psi'_{3+|{\pmb\beta}_1|})^{P_{12{\pmb\alpha}_1}3}_{P_{12{\pmb\alpha}_1}3}\nn
&=&{\Lambda^2\over 2^2}\sum_{\epsilon^M}\,{\bf Pf}'\Psi_{3+|{\pmb\alpha}_1|}{\bf Pf}'\Psi'_{3+|{\pmb\beta}_1|}\,.~~~~\label{pf-gen-1}
\eea
For another reduced Pfaffian $({\bf Pf}'\W\Psi^\Lambda_n)^\tau$, we have the same factorized formula
\bea
({\bf Pf}'\W\Psi^\Lambda_n)^\tau={\Lambda^2\over 2^2}\sum_{\W\epsilon^{M'}}\,{\bf Pf}'\W\Psi_{3+|{\pmb\alpha}_1|}{\bf Pf}'\W\Psi'_{3+|{\pmb\beta}_1|}\,.~~~~\label{pf-gen-2}
\eea

Until now, two basic objects in the $4$-point case have been generalized to the case with arbitrary number of external legs. However, in the current general case, there are $|{\pmb\alpha}_1|+|{\pmb\beta}_1|$ scattering equations which provide poles encircled by the contour. Now we discuss the behavior of them under the $\Lambda\to 0$ limit. For equations ${\cal E}_i$ with $i\in{\pmb\beta}_1$, whose leading order contribution are shown in \eref{seq-gen}, we can rewrite them as
\bea
{\cal E}_i\Big|^{1,2,{\pmb\alpha}_1}_{3,4,{\pmb\beta}_1}&=&2\sum_{\substack{j\in\{3,4,{\pmb\beta}_1,P_{12{\pmb\alpha}_1}\}\\j\neq i}}\,
{\sigma_{jP_{12{\pmb\alpha}_1}}\over \sigma_{iP_{12{\pmb\alpha}_1}}\sigma_{ij}}k_i\cdot k_j\nn
&=&2\sum_{\substack{j\in\{3,4,{\pmb\beta}_1,P_{12{\pmb\alpha}_1}\}\\j\neq i}}\,
\Big({1\over\sigma_{ij}}-{1\over \sigma_{iP_{12{\pmb\alpha}_1}}}\Big)k_i\cdot k_j\nn
&=&\sum_{\substack{j\in\{3,4,{\pmb\beta}_1,P_{12{\pmb\alpha}_1}\}\\j\neq i}}\,
{2k_i\cdot k_j\over\sigma_{ij}}\,,~~~~\label{eq-1type}
\eea
which are the single-cover scattering equations for punctures in the set $\{3,4,{\pmb\beta}_1,P_{12{\pmb\alpha}_1}\}$.
For equations ${\cal E}_i$ with $i\in{\pmb\alpha}_1$, there leading order contributions are given by
\bea
{\cal E}_i\Big|^{1,2,{\pmb\alpha}_1}_{3,4,{\pmb\beta}_1}&=&2\Big(\sum_{\substack{j\in\{1,2,{\pmb\alpha}_1\}\\j\neq i}}\,{k_i\cdot k_j\over\sigma_{ij}}
+\sum_{j\in\{3,4,{\pmb\beta}_1\}}\,{k_i\cdot k_j\over \sigma_i}\Big)\nn
&=&2\sum_{\substack{j\in\{1,2,{\pmb\alpha}_1\}\\j\neq i}}\,\Big({1\over\sigma_{ij}}
-{1\over\sigma_i}\Big)k_i\cdot k_j\nn
&=&2\sum_{\substack{j\in\{1,2,{\pmb\alpha}_1\}\\j\neq i}}\,{\sigma_j\over\sigma_i\sigma_{ij}}k_i\cdot k_j\,.~~~~\label{seq-gen-1}
\eea
The next treatment is totally the same as in \eref{eq-1type}, and gives
\bea
{\cal E}_i\Big|^{1,2,{\pmb\alpha}_1}_{3,4,{\pmb\beta}_1}=\sum_{\substack{j\in\{1,2,{\pmb\alpha}_1,P_{34{\pmb\beta}_1}\}\\j\neq i}}\,
{2k_i\cdot k_j\over\sigma_{ij}}\,,~~~~\label{seq-gen-2}
\eea
which are the single-cover scattering equations for punctures in $\{1,2,{\pmb\alpha}_1,P_{34{\pmb\beta}_1}\}$.

Combining \eref{FP-det-gen}, \eref{pf-gen-1}, \eref{pf-gen-2}, \eref{seq-gen-1} and \eref{seq-gen-2} together, we finally arrive at the factorized expression
\bea
{\cal A}^{\epsilon,\W\epsilon}_{\rm GR}(\{1,2,\cdots,n\})\Big|^{1,2,{\pmb\alpha}_1}_{3,4,{\pmb\beta}_1}
&=&{1\over2}\sum_{\epsilon^M\W\epsilon^{M'}}\,\Big(\int\Big(\prod_{i\in{\pmb\alpha_1}}\,{d\sigma_i\over {\cal E}_i}\Big)|12P_{34{\pmb\beta}_1}|^2{\bf Pf}'\Psi_{3+|{\pmb\alpha}_1|}{\bf Pf}'\W\Psi_{3+|{\pmb\alpha}_1|}\Big){1\over P^2_{34{\pmb\beta}_1}}\nn
& &\Big(\int\Big(\prod_{j\in{\pmb\beta_1}}\,{d\sigma_j\over {\cal E}_j}\Big)|34P_{12{\pmb\alpha}_1}|^2{\bf Pf}'\Psi'_{3+|{\pmb\beta}_1|}{\bf Pf}'\W\Psi'_{3+|{\pmb\beta}_1|}\Big)\nn
&=&{1\over2}\sum_{\epsilon^M\W\epsilon^{M'}}\,{\cal A}^{\epsilon,\W\epsilon}_{\rm GR}(\{\underline{\hat{1}},\underline{2},{\pmb\alpha}_1,\underline{\hat{P}}_{34{\pmb\beta}_1}^{\epsilon^M,\W\epsilon^{M'}}\}){1\over P^2_{34{\pmb\beta}_1}}
{\cal A}^{\epsilon,\W\epsilon}_{\rm GR}(\{\underline{\hat{3}},\underline{4},{\pmb\beta}_1,\underline{\hat{P}}_{12{\pmb\alpha}_1}^{\epsilon^M,\W\epsilon^{M'}}\})\,.
\eea

For the mirrored configuration
\bea
& &(y_1=+\sqrt{\sigma_1^2-\Lambda^2}\,,~\sigma_1)\,,~~~~(y_2=+\sqrt{\sigma_2^2-\Lambda^2}\,,~\sigma_2)\,,~~~~
(y_{\alpha_l}=+\sqrt{\sigma_{\alpha_l}^2-\Lambda^2}\,,~\sigma_{\alpha_l})\,,\nn
& &(y_3=-\sqrt{\sigma_3^2-\Lambda^2}\,,~\sigma_3)\,,~~~~(y_4=-\sqrt{\sigma_4^2-\Lambda^2}\,,~\sigma_4)\,,~~~~
(y_{\beta_m}=-\sqrt{\sigma_{\beta_m}^2-\Lambda^2}\,,~\sigma_{\beta_m})\,,
\eea
the treatment is extremely similar. Summing over two configurations gives
\bea
& &{\cal A}^{\epsilon,\W\epsilon}_{\rm GR}(\{1,2,\cdots,n\})\Big|^{1,2,{\pmb\alpha}_1}_{3,4,{\pmb\beta}_1}+
{\cal A}^{\epsilon,\W\epsilon}_{\rm GR}(\{1,2,\cdots,n\})\Big|^{3,4,{\pmb\beta}_1}_{1,2,{\pmb\alpha}_1}\nn
&=&\sum_{\epsilon^M\W\epsilon^{M'}}\,{\cal A}^{\epsilon,\W\epsilon}_{\rm GR}(\{\underline{\hat{1}},\underline{2},{\pmb\alpha}_1,\underline{\hat{P}}_{34{\pmb\beta}_1}^{\epsilon^M,\W\epsilon^{M'}}\}){1\over P^2_{12{\pmb\alpha}_1}}
{\cal A}^{\epsilon,\W\epsilon}_{\rm GR}(\{\underline{\hat{3}},\underline{4},{\pmb\beta}_1,\underline{\hat{P}}_{12{\pmb\alpha}_1}^{\epsilon^M,\W\epsilon^{M'}}\})\,,~~~~\label{phy-pole-gen-1}
\eea
which is the first line of \eref{phy-pole-YM-gen}.

Applying the same method to the separation $\{\{1,4,{\pmb\alpha}_2\},\{2,3,{\pmb\beta}_2\}\}$, we obtain the factorized formula
\bea
& &{\cal A}^{\epsilon,\W\epsilon}_{\rm GR}(\{1,2,\cdots,n\})\Big|^{1,4,{\pmb\alpha}_2}_{2,3,{\pmb\beta}_2}+
{\cal A}^{\epsilon,\W\epsilon}_{\rm GR}(\{1,2,\cdots,n\})\Big|^{2,3,{\pmb\beta}_2}_{1,4,{\pmb\alpha}_2}\nn
&=&\sum_{\epsilon^M\W\epsilon^{M'}}\,{\cal A}^{\epsilon,\W\epsilon}_{\rm GR}(\{\underline{\hat{1}},\underline{4},{\pmb\alpha}_2,\underline{\hat{P}}_{23{\pmb\beta}_2}^{\epsilon^M,\W\epsilon^{M'}}\}){1\over P^2_{14{\pmb\alpha}_2}}
{\cal A}^{\epsilon,\W\epsilon}_{\rm GR}(\{\underline{2},\underline{\hat{3}},{\pmb\beta}_2,\underline{\hat{P}}_{14{\pmb\alpha}_2}^{\epsilon^M,\W\epsilon^{M'}}\})\,,~~~~\label{phy-pole-gen-2}
\eea
which is the second line of \eref{phy-pole-YM-gen}.

%%%%%%%%%%%%%%%%%%%%%%%%%%%%%%%%%%%%%%%%%%%

\end{document}